\documentclass[12pt]{article}
\usepackage{epsfig,amsfonts,amssymb}
\usepackage{cite}
\input epsf.sty
\usepackage{cite}

\newcommand{\hab}{}

\def\ZZZ{{\hbox{ Z\kern-1.6mm Z}}}
\def\RRR{{\hbox{ R\kern-2.4mm R}}}
\def\CCC{{\hbox{ C\kern-2.0mm C}}}
\def\zzz{{\hbox{z\kern-1mm z}}}

\newcommand{\nn}{\nonumber \\}

\newcommand{\qeq}{{\hbox{=\kern-2.3mm ? \kern.5mm }}}
\renewcommand{\qeq}{=}

\newcommand{\eps}{\epsilon}
\newcommand{\vareps}{\varepsilon}

\newcommand{\vp}{\varphi}
\newcommand{\ve}{\varepsilon}

\newcommand{\DD}{{\cal D}}

\newcommand{\AAA}{{\cal A}}

\newcommand{\KK}{{\cal K}}

\newcommand{\MM}{{\cal M}}

\newcommand{\OO}{{\cal O}}

\newcommand{\LL}{{\cal L}}
 
\newcommand{\half}{{1\over 2}}
\newcommand{\wt}{\widetilde}
\newcommand{\wh}{\widehat}

\newcommand{\NN}{{\cal N}}

\newcommand{\SSS}{{\cal S}}

\newcommand{\be}{\begin{equation}}
\newcommand{\ee}{\end{equation}}
\newcommand{\ben}{\begin{eqnarray}\displaystyle}
\newcommand{\een}{\end{eqnarray}}

\newcommand{\refb}[1]{(\ref{#1})}
\newcommand{\p}{\partial}
\newcommand{\sectiono}[1]{\section{#1}\setcounter{equation}{0}}

\newcommand{\zet}{\zeta}

\def\one{{\hbox{ 1\kern-.8mm l}}}
\def\zero{{\hbox{ 0\kern-1.5mm 0}}}

\begin{document}

\baselineskip 24pt

\begin{center}
{\Large \bf  Logarithmic Corrections to $\NN=2$ Black Hole
Entropy: An Infrared Window into the Microstates}

\end{center}

\vskip .6cm
\medskip

\vspace*{4.0ex}

\baselineskip=18pt

\centerline{\large \rm Ashoke Sen}

\vspace*{4.0ex}

\centerline{\large \it Harish-Chandra Research Institute}
\centerline{\large \it  Chhatnag Road, Jhusi,
Allahabad 211019, India}

\vspace*{1.0ex}
\centerline{\small E-mail:  sen@mri.ernet.in}

\vspace*{5.0ex}

\centerline{\bf Abstract} \bigskip

Logarithmic corrections to the extremal black hole entropy
can be computed purely in terms of the low energy data
-- the spectrum of massless fields and their interaction.
The demand of reproducing these corrections 
provides a strong constraint on any
microscopic theory of quantum gravity
that attempts to explain the black hole
entropy. Using quantum entropy function formalism
we compute logarithmic corrections to the entropy
of half BPS black holes
in $\NN=2$ supersymmetric string theories.
Our results allow us to test various proposals
for the measure in the OSV formula, and
we find agreement with the measure
proposed by Denef and Moore if we assume their result
to be valid at weak topological string coupling.
Our analysis also gives the logarithmic corrections to
the entropy of extremal Reissner-Nordstrom black holes
in ordinary Einstein-Maxwell theory.

\vfill \eject

\baselineskip=18pt

\tableofcontents

\sectiono{Introduction and summary}  \label{sintro}

Recent years have seen considerable progress towards an
understanding of the black hole entropy beyond the original
formula due to Bekenstein and Hawking relating the entropy
to the area of the event horizon. In particular
Wald's formula gives a prescription for computing the
black hole entropy
in a classical theory of gravity with higher derivative 
terms, possibly coupled to other matter 
fields\cite{9307038,9312023,9403028,9502009}.
In the extremal limit this leads to a simple algebraic 
procedure for determining the near horizon field
configurations and the entropy\cite{0506177,0508042}. 
A proposal for computing quantum
corrections to this formula was suggested in
\cite{0809.3304,0805.0095} by exploiting the presence of 
$AdS_2$ factors in the near horizon geometry of extremal
black holes. In this formulation, 
called the quantum entropy function
formalism,
the degeneracy associated with the black hole
horizon is given by the string theory
partition function $Z_{AdS_2}$
in the near horizon geometry of the black hole.  
Such a partition function is divergent
due to the infinite volume of $AdS_2$, but the rules of 
$AdS_2/CFT_1$ correspondence gives a precise procedure
for removing this divergence.
While in the classical limit this prescription gives us
back the exponential of the Wald entropy, it can in principle be
used to systematically calculate the quantum corrections to the
entropy of an extremal black hole.

In this paper our main focus will be on logarithmic
corrections to the black hole entropy.
These arise from one loop quantum corrections to $Z_{AdS_2}$
involving massless
fields and are insensitive to the details of the ultraviolet properties
of the theory. On the other hand, being corrections to the
black hole entropy, they give us non-trivial information about
the microstates of the black hole. For this reason they can be
regarded as an infrared window into the microphysics of black
holes. 
In two previous papers\cite{1005.3044,1106.0080} 
we used the quantum entropy function
to compute  logarithmic corrections to the entropy
of 1/8 BPS and 1/4 BPS black holes in
$\NN=8$ and $\NN=4$ supersymmetric string 
theories respectively and found
results in perfect agreement with the microscopic 
results of 
\cite{9607026,0412287,0505094,0506249,
0508174,0510147,
0602254,0603066,0605210,0607155,0609109,0612011,
0708.1270,0802.0544,0802.1556,0803.2692}. 
In this paper we use this formalism to compute logarithmic correction
to the entropy of 
half BPS black holes in $\NN=2$ 
supersymmetric string theories. As in 
\cite{1005.3044,1106.0080} we consider the limit in
which all components of the charge become large at the
same rate. In this limit
we find that for a theory with $n_V$ massless vector multiplets
and $n_H$ massless hypermultiplets, the entropy
including logarithmic
correction is given by
\be \label{elogres}
{A_H\over 4 G_N} + 
{1\over 12}(23+n_H - n_V)\ln {A_H\over G_N} 
+ \OO(1)\, ,
\ee
where $A_H$ is the area of the event horizon and $G_N$ is
the Newton's constant.
The $\OO(1)$ terms include functions of ratios of charges,
and also contains terms carrying inverse powers of 
charges.\footnote{Thus if we take another limit in
which some ratios of charges become large then we may
get additional logarithmic corrections.}
Note that while the result depends on the number of vector
and hypermultiplet fields, it does not depend on the details
of the interaction involving these fields through the prepotential
and the metric on the hypermultiplet moduli space.
Eq.\refb{elogres} is consistent with
the version of the OSV formula\cite{0405146} 
given in \cite{0702146}
if we take their result to be valid at weak topological
string coupling. However \refb{elogres} is in apparent
disagreement with the
measure proposed in \cite{0808.2627,0810.1233}.
A detailed discussion on this can be found in \S\ref{sosv}.
For STU model\cite{9508064,9901117} we 
have $n_H=4$ and $n_V=3$, leading
to a logarithmic correction of $2\ln (A_H/G_N)$ to the entropy. This
agrees with the result of \cite{1106.0080}.

We also
give for comparison the result of 
\cite{1005.3044,1106.0080} for 
supersymmetric black hole entropies
in $\NN=4$ and $\NN=8$ supersymmetric theories:
\ben\label{en=48}
\NN=4 &:& {A_H\over 4 G_N} + \OO(1) \nn
\NN=8 &:& {A_H\over 4 G_N}-4\ln {A_H\over G_N} + \OO(1)\, .
\een

Note that the coefficient given in 
\refb{elogres} is proportional to the gravitational
$\beta$-function in $\NN=2$ supergravity / string theory
given in \cite{duffroc,9203071,9204030,9212045}.
However this
relation does not hold universally. For example in the $\NN=8$
supersymmetric theory the gravitational $\beta$-function 
vanishes\cite{duffroc}
but the logarithmic correction to the entropy given in \refb{en=48}
does not vanish.
The precise relation will be discussed in
\S\ref{slocal}.

Our analysis also gives the result for the logarithmic correction
to the entropy of an extremal Reissner-Nordstrom black
hole in ordinary non-supersymmetric Einstein-Maxwell theory.
The result is $-{241\over 45}\ln {A_H\over G_N}$. If the theory
in addition contains 
$n_S$ massless scalars, $n_F$ massless Dirac fermions
and $n_V$ additional Maxwell fields, all minimally coupled to
background gravitational field but not to the background 
electromagnetic flux, then the net
entropy is given by
\be \label{elognonsusyrn}
{A_H\over 4G_N} -{1\over 180} ({ 964} 
+ n_S + 62 n_V +11n_F
) \, \ln \, {A_H\over G_N} + \OO(1)\, .
\ee
We emphasize that in this formula $n_V$ is the number of
{\it additional} minimally coupled Maxwell field. Thus if we
just had an extremal Reissner-Nordstrom black hole in
Einstein gravity coupled to a single
Maxwell field then $n_V=0$ in our
convention.

Various other 
earlier approaches to computing
logarithmic corrections to black hole entropy can be found
in \cite{9407001,9408068,9412161,9604118,
9709064,0002040,0005017,
0104010,0112041,0406044,0409024,0805.2220,
0808.3688,0809.1508,0911.4379,1003.1083,1008.4314}.
Of these the method advocated in \cite{9709064}, and
subsequently developed further in \cite{1008.4314}
and reviewed in \cite{1104.3712},
is closest to the one we are following; so we have given 
a detailed comparison between the two methods below 
eq.\refb{ebir2}. For now we would like to mention that
the method of \cite{9709064,1008.4314,1104.3712} would
correctly reproduce the dependence on $n_S$, $n_V$ and $n_F$
for  extremal Reissner-Nordstrom black holes
in \refb{elognonsusyrn} 
but will fail to give the constant term 964 correctly.
This is due to the fact that the constant term comes from 
fluctuations of the
metric and the gauge field under which the black hole
is charged, and for these fields
the analysis of \cite{9709064,1008.4314,1104.3712} 
would be insufficient
on two counts: first it does not take into account correctly the
mixing between these fields due to the presence of the
gauge field flux in the near horizon geometry
of the black hole, and second it fails to take into account correctly
the effect of integration over the zero modes of these fields.
The naive application of the analysis of 
\cite{9709064,1008.4314,1104.3712} would also fail to
get the result \refb{elogres} or \refb{en=48} for
the supersymmetric black holes in $\NN=2,4,8$ supergravity
for which
both the mixing between the fields and the 
integration over the
zero modes play a crucial role. As we discuss in \S\ref{slocal},
the effect of mixing between the fields can be incorporated by
augmenting the analysis of 
\cite{9709064,1008.4314,1104.3712} by supersymmetry, -- a
fact already anticipated in \cite{duffroc}. However the
effect of zero mode integration still needs to be taken into account separately.

Refs.\cite{0905.2686,1012.0265} attempted an exact 
evaluation
of $Z_{AdS_2}$ using localization methods. The general formula
for the logarithmic correction to the half BPS black hole
entropy in these
theories, described in \refb{elogres}, 
shows that $Z_{AdS_2}$ receives
non-trivial contribution from not only the vector multiplets but also
the gravity multiplet and the hypermultiplets. This makes the evaluation
of this partition function a much more challenging problem, 
but also a more
interesting one.

Before concluding the introduction we would like to discuss the
region of validity of our formul\ae. There are two independent
questions: for which range of charges is our analysis valid and
in which region of the moduli space is our analysis valid?
As we have already mentioned, our analysis will be valid in the
limit when all components of the charge are scaled uniformly, so
that the four dimensional near horizon geometry becomes weakly curved
and the internal space remains at a fixed shape and size as we scale
the charges. The precise limit may be taken as follows. First we
take a black hole solution  in the $\NN=2$ supergravity
with finite area event horizon and regular attractor values
of the vector multiplet moduli, 
but do not require the charges to be quantized. We
then scale all
the charges carried by this black hole by some large
number
$\Lambda$ and at the end shift the charges
by finite amounts to nearby integers
in such a way that the final charge vector is
primitive. In this limit the area of the horizon and hence the
entropy scale as $\Lambda^2$ and 
the vector multiplet moduli
remain fixed at regular values. 
To determine the chamber in the moduli space where our
results are valid, note that 
our result is based on the analysis of the near
horizon geometry of a single black hole.
Thus we need to work in the
attractor domain where the enigmatic configurations discussed in
\cite{0702146} are absent. Even in this case the total index receives
contribution from multicentered scaling solutions besides the single
centered black hole. 
In order that our result for single centered black hole
entropy gives the dominant contribution we
need to ensure that the contribution to the index from
the scaling solutions are either absent or subleading.
We discuss this point in detail in \S\ref{smulti}.

A related issue arises for extremal Reissner-Nordstrom black holes
whose entropy is given by \refb{elognonsusyrn}. Due to the
existence of multicentered black holes with each center
carrying a fraction of the total charge, 
 the index receives contribution not only
from single centered black holes but also from multi-centered
configurations.
This can be
avoided by considering a dyonic configuration carrying a primitive
charge vector instead of a purely electrically charged configuration.
Since the Einstein-Maxwell theory is duality invariant, our result
\refb{elognonsusyrn} will continue to be valid in such a situation,
but the multicentered configurations are avoided 
since the total charge vector, being
primitive, can no longer be
split into multiple charge vectors which are proportional to
each other. (A complete proof of this is still lacking however;
see the discussion in \S\ref{smulti}.)

A final point about notation: while in the macroscopic
description we compute the entropy of the black hole, on
the microscopic side we always compute an appropriate
index. It was argued in \cite{0903.1477,1009.3226} 
that the entropy of the
single centered black hole also represents the logarithm
of the index carried by the same black hole.
For this reason we shall use the word entropy and
logarithm of the index interchangeably throughout
our discussion.

The rest of the paper is organized as follows. \S\ref{sstrategy}
and \S\ref{sminimal} contains mostly review of known material.
In \S\ref{sstrategy} we describe the general strategy for computing
logarithmic corrections to the entropy of an extremal black
hole. In \S\ref{sminimal} we illustrate this by calculating the
logarithmic corrections to the entropy due to masless scalar,
fermion and vector fields, {\it assuming that they only couple
minimally to the background metric and is not affected by the
any other background field if present.} In particular for the extremal
Reissner-Nordstrom black hole this analysis
does not apply to the gauge field which has non-zero background field
strength since due to the Maxwell term in the action such gauge fields
will be affected by the background flux. In \S\ref{s1} we apply the
method reviewed in \S\ref{sstrategy}
to compute the logarithmic correction to the
entropy of an extremal Reissner-Nordstrom
black hole. This is important for our analysis since 
the bosonic sector of pure $\NN=2$ supergravity is described
by ordinary Einstein-Maxwell theory and consequently the
results of this section describe the
bosonic contribution to the logarithmic correction to
BPS black hole entropy in pure $\NN=2$
supergravity.
In \S\ref{sfermion} we augment this result by computing the logarithmic
correction to BPS black hole entropy due to the fermionic fields of
$\NN=2$ supergravity. Adding the results of \S\ref{s1} and
\S\ref{sfermion} we arrive at the result given in \refb{elogres} for
$n_H=n_V=0$. In \S\ref{sgeneral} we apply the same method to
compute the logarithmic correction to the entropy of a BPS black
hole in a general supergravity theory with arbitrary number of vector
and hypermultiplets. This leads to \refb{elogres}. In \S\ref{slocal} we
discuss an alternative but equivalent method for deriving the same results,
making use of the supersymmetry of the theory. This method is simpler,
but requires certain assumption about possible supersymmetric one loop
counterterms in $\NN=2$ supergravity theory. One could in principle
elevate this into a rigorous analysis -- at the same level as that in 
\S\ref{s1}-\S\ref{sgeneral} -- by classifying all possible four derivative
supersymmetric terms in the action that could be generated as
one loop correction in $\NN=2$ supergravity. In \S\ref{smulti} we
explore if multi-centered scaling solutions could invalidate our result by
generating new configurations whose entropy is of the same order or
larger than the single center black hole entropy we analyze. Although
we do not have any rigorous result we argue that it is extremely unlikely.
In \S\ref{sosv} we carry out a detailed comparison of our results with
different versions of the OSV formula for black hole entropy which have
been proposed in the literature. 
While our result agrees with that of \cite{0702146} assuming its validity
in the scaling limit we are studying, it disagrees with the proposal of
\cite{0808.2627}. 
We argue however  this disagreement
can be rectified by certain changes in
the proposed formul\ae\ of \cite{0808.2627} 
without violating any basic principle used in arriving
at these results. In appendix \ref{sbasis} we collect the results for
eigenfunctions and eigenvalues of the laplacian on $AdS_2\times S^2$
for various fields. In appendix \ref{suse} we collect some useful
mathematical identities used in our analysis.  Finally in appendix
\ref{ssymplectic} we demonstrate that in a general $\NN=2$
supergravity theory
coupled to a set of vector and hypermultiplet fields, the action that
describes the fluctuations of various fields around the BPS black
hole background to quadratic order
has a universal form that depends only on the
number of vector and hypermultiplet fields but not on the details of
their interaction {\it e.g.} the prepotential for the vector multiplet and
the moduli space metric for the hypermultiplet. This has been used
in the analysis of \S\ref{sgeneral} and is responsible
for the universal form of \refb{elogres} that does not depend on the details
of the interaction.

\sectiono{General strategy} \label{sstrategy}

In this section we shall review the general strategy for computing
logarithmic corrections to the entropy of extremal black holes.
We shall focus on spherically symmetric
extremal black holes in four dimensions, 
but the method we describe
is easily generalizable to non-spherical (rotating) black holes.

Suppose
we have an extremal black hole with near horizon geometry
$AdS_2\times S^2$, with equal radius of
curvature 
$a$ of $AdS_2$ and $S^2$. 
Then
the Euclidean near horizon metric takes the form
\be \label{emets}
ds^2 = a^2 (d\eta^2 +\sinh^2\eta \, d\theta^2) + a^2 (d\psi^2 +
\sin^2\psi d\phi^2)\, .
\ee
We shall denote by $x^\mu$ all four 
coordinates on $AdS_2\times
S^2$, by
$x^m$ the coordinates $(\eta,\theta)$
on $AdS_2$ and by $x^\alpha$ the
coordinates $(\psi,\phi)$ on $S^2$
and introduce
the invariant antisymmetric tensors $\vareps_{\alpha\beta}$
on $S^2$ and $\vareps_{mn}$ on $AdS_2$
respectively, computed with the background metric
\refb{emets}:
\be \label{edefve}
\ve_{\psi\phi} = a^{2}\, \sin\psi, \qquad \ve_{\eta\theta}
= a^{2} \, \sinh\eta\, .
\ee
All indices will be raised and lowered with the background metric $g_{\mu\nu}$ defined in
\refb{emets}.

Let $Z_{AdS_2}$ denote the partition function of string
theory in the near horizon geometry, evaluated by
carrying out functional integral over all the string fields
weighted by the exponential of the Euclidean action
$\SSS$, 
with boundary conditions such that asymptotically the field
configuration approaches the near horizon geometry
of the black hole.\footnote{Our definition of the euclidean action
includes a minus sign so that the path integral is weighted by
$e^{\SSS}$ instead of $e^{-\SSS}$.}
Since in $AdS_2$ the asymptotic
boundary conditions fix the electric fields, or equivalently
the charges carried by the black hole\cite{0809.3304}, 
and allow the
constant modes of the gauge fields to fluctuate, we need
to include in the path integral a boundary term
$\exp(-i\ointop \sum_k q_k A^{(k)}_\mu dx^\mu)$ where
$A^{(k)}_\mu$ are the gauge fields and $q_k$ are the
corresponding electric charges carried by the black 
hole\cite{0809.3304}.
Thus we have
\be \label{ezads2}
Z_{AdS_2} = \int d\Psi \exp(\SSS
-i\ointop \sum_k q_k A^{(k)}_\mu dx^\mu)\, ,
\ee
where $\Psi$ stands for all the string fields.
$AdS_2/CFT_1$ correspondence
tells us that the full quantum corrected entropy $S_{BH}$
is related to $Z_{AdS_2}$ via\cite{0809.3304}:
\be \label{eads2cft1}
e^{S_{BH} - E_0 L} =  Z_{AdS_2}\, ,
\ee
where $E_0$ is the energy of the ground state
of the black hole carrying a given set of charges,
and $L$ denotes the length of the boundary of $AdS_2$
in a regularization scheme that renders the
volume of $AdS_2$ finite by putting an infrared
cut-off $\eta\le \eta_0$.
Eq.\refb{eads2cft1} is valid in the limit of large $L$
and allows us to compute $S_{BH}$
from the knowledge of $Z_{AdS_2}$.

Let $\Delta\LL_{eff}$ denote the one loop correction to the four
dimensional effective lagrangian density evaluated in the background
geometry \refb{emets}. Then the one loop correction to 
$Z_{AdS_2}$ is given by
\be \label{e2}
\exp\left[
 \int_0^{\eta_0} d\eta \, \int_0^{2\pi} d\theta  \, 
 \int_0^\pi d\psi\, \int_0^{2\pi} d\phi\, \sqrt{\det g} \, 
 \Delta\LL_{eff}\right]
= \exp\left[8\pi^2 \, a^4 \, (\cosh\eta_0-1) \, 
\Delta\LL_{eff}\right]\, .
\ee
Here we have used the fact that due to the 
$SO(2,1)\times SO(3)$ isometry of $AdS_2\times S^2$,
$\Delta\LL_{eff}$ is independent of the coordinates of
$AdS_2$ and $S^2$. Since the length of the boundary, 
situated at $\eta=\eta_0$, is 
given by $L=2\pi a \sinh\eta_0$, the
term proportional to
$\cosh\eta_0$ in the exponent of \refb{e2}
can be written as $-L \Delta E_0
+\OO\left(L^{-1}\right)$
where $\Delta E_0=-4\pi a^3
\Delta\LL_{eff}$ has the interpretation of
the shift in the
ground state energy.
The $L$-independent contribution in the exponent
can be interpreted as the
one loop correction to the black hole 
entropy\cite{0809.3304}. Thus we have
\be \label{e3}
\Delta S_{BH} = -8\pi^2 a^4\, \Delta\, \LL_{eff}\, .
\ee
While the term in the exponent proportional to $L$
and hence $\Delta E_0$ can get further corrections
from boundary terms in the action, the $L$-independent part
$\Delta S_{BH}$ is defined unambiguously.
This reduces the problem of computing one loop correction to
the black hole entropy to that of computing one
loop correction to $\LL_{eff}$.
We shall now describe the general procedure for calculating
$\Delta\LL_{eff}$.

Suppose we have a set of massless 
fields\footnote{Here by massless field we mean any
field whose mass is of order $a^{-1}$ or less.} 
$\{\phi^i\}$ where
the index $i$ could run over several scalar fields, or the space-time
indices of tensor fields. Let $\{f_n^{(i)}(x)\}$ denote an
orthonormal  basis of eigenfunctions of the kinetic operator
expanded around the near horizon geometry, with 
eigenvalues
$\{\kappa_n\}$:
\be \label{eortho}
\int d^4 x \, \sqrt{\det g} \, G_{ij} \, f_n^{(i)}(x)\,
f_m^{(j)}(x) = \delta_{mn}\, ,
\ee
where $g_{\mu\nu}$ is the $AdS_2\times S^2$ metric and
$G_{ij}$ is a metric in the space of fields induced by the
metric on $AdS_2\times S^2$, {\it e.g.} for a vector
field $A_\mu$, $G^{\mu\nu}=g^{\mu\nu}$.
 Then the heat kernel $K^{ij}(x,x')$ is defined as
\be \label{eh1}
K^{ij}(x,x';s) = \sum_n \, e^{-\kappa_n\, s} \, f_n^{(i)}(x)\,
f_n^{(j)}(x')\, .
\ee
In \refb{eortho}, \refb{eh1} we have assumed that we are working in a basis
in which the eigenfunctions are real; if this is not the case then we
need to replace one of the $f_n^{(i)}$'s by $f_n^{(i)*}$.
Among the $f_n^{(i)}$'s there may be a special set of
modes for which $\kappa_n$ vanishes. We shall denote these
zero modes by the special symbol $g_\ell^{(i)}(x)$, and define
\be \label{ezma}
\bar K^{ij} (x, x') = \sum_\ell \,  g_\ell^{(i)}(x)\,
g_\ell^{(j)}(x')\, .
\ee
Defining
\be \label{edefk0}
K(0;s) = G_{ij}\, K^{ij}(x,x; s)\, , \qquad \bar K(0) = G_{ij}\, 
\bar K^{ij}(x,x; s)\, ,
\ee
and using orthonormality of the wave-functions, we get
\be \label{eext23}
 \int d^4 x \, \sqrt{\det g}\,   
\left(K(0;s) - \bar K(0)\right) = {\sum_{n}}' e^{-\kappa_n\, s} \, ,
\ee
where $\sum_n'$ denotes sum over the non-zero modes only.
Note that due to homogeneity of $AdS_2\times S^2$ the right hand sides
of \refb{edefk0} do not depend on $x$.
The contribution of the non-zero modes of the
massless fields to the one loop effective
action can now be expressed as
\be \label{e4}
\Delta\SSS =-{1\over 2}\, {\sum_n}' \ln\kappa_n =
{1\over 2} \int_\eps^\infty {ds\over s} {\sum_n}' e^{-\kappa_n s}
=
{1\over 2}\, \int_{\eps}^\infty\, {ds\over s} \, 
\int d^4 x \, \sqrt{\det g}\,   \left(K(0;s) - \bar K(0)\right)\, ,
\ee
where $\eps$
is an ultraviolet cut-off which we shall take to be of order one, \i.e.\
string scale.\footnote{Throughout this paper we shall assume that the
horizon values of all the moduli fields are of order unity so that
string scale and Planck scale are of the same order. This sets
$G_N\sim 1$.}
Identifying this as the contribution to
$\int d^4 x \sqrt{\det g}\, \Delta\LL_{eff}$ we get the contribution
to $\Delta \LL_{eff}$ from the non-zero modes:
\be \label{e3ab}
\Delta\LL_{eff}^{(nz)} = {1\over 2} \, \int_{\eps}^\infty\, {ds\over s} \, 
\left(K(0;s) - \bar K(0)\right)\, .
\ee
The logarithmic contribution to the entropy -- term proportional to
$\ln a$ -- arises from the $1<< s << a^2$ region in the $s$
integral. If we expand $K(0;s)$ in a Laurent series expansion
in $\bar s =s/a^2$ around $\bar s =0$, and if $K_0$ denotes
the coefficient of the constant mode in this expansion, then 
using \refb{e3} and \refb{e3ab} we see that the
net logarithmic correction to the entropy from the non-zero modes
will be given by
\be \label{enetlognz}
-8\pi^2 a^4 \, \left(K_0 -  \bar K(0)\right) \ln a
= -4\pi^2 a^4 \, \left(K_0 -  \bar K(0)\right) \ln A_H\, ,
\ee
where $A_H=4\pi a^2$ is the area of the event horizon.

The contribution to $Z_{AdS_2}$ from integration over the zero
modes can be evaluated as follows.\footnote{Some discussion
on the effect of zero modes on the ultraviolet divergent contribution
to the black hole entropy can be found in 
\cite{9512047,9605153}.}
First note that we can use
\refb{ezma}, \refb{edefk0} to define the number of zero modes
$N_{zm}$:
\be \label{enz1}
\int d^4 x \, \sqrt{\det g}\,   
\bar K(0) = \sum_\ell 1 = N_{zm}\, .
\ee
In fact often the matrix $\bar K^{ij}$ takes a block diagonal form
in the field space,
with different blocks representing zero modes  of different sets of
fields. In that
case we can use the analog of \refb{enz1} to define the number
of zero modes of each block. If these different blocks are labelled
by different sets
$\{A_r\}$ then the number of zero modes belonging to
the set $A_r$ will be given by
\ben \label{ediffblock}
&& N^{(r)}_{zm} = \int d^4 x \, \sqrt{\det g}\,   
\bar K^r(0)=8\pi^2 a^4 \, \bar K^r(0) \, \left(\cosh\eta_0 - 1
\right), \nn
&& \quad \bar K^r(0) \equiv \sum_{\ell\in A_r} G_{ij}
g_\ell^{(i)}(x)\,
g_\ell^{(j)}(x)\, .
\een
Typically these zero modes are
associated with certain asymptotic symmetries, -- gauge transformation
with parameters which do not vanish at infinity. In this case we
can evaluate the integration over the zero modes by making a change
of variables from the coefficients of the zero modes to the parameters
labelling the (super-)group of asymptotic symmetries. 
Suppose for the zero modes in the $r$'th block the Jacobian for the
change of variables from the fields to supergroup parameters gives a
factor of $a^{\beta_r}$ for each zero mode. 
Then the net $a$ dependent contribution
to $Z_{AdS_2}$ from the zero mode integration will be 
given by
\be \label{enetzmc}
a^{\sum_r \beta_r N^{(r)}_{zm}}
= \exp\left[8\pi^2 a^4 \, \left(\cosh\eta_0 - 1
\right)\ln a \sum_r \beta_r \bar K^r(0)\right]\, .
\ee
Again the coefficient of $\cosh\eta_0$ can be interpreted
as due to a shift in the energy $E_0$, whereas the
$\eta_0$ independent term has the interpretation of a
contribution to the black hole entropy.
This gives the 
following expression for the 
logarithmic correction to the entropy from the zero modes:
\be \label{enetlonzm}
-8\pi^2 a^4\, \ln a \, \sum_r \beta_r  
\bar K^r(0)\, .
\ee
Adding this to \refb{enetlognz} we get
\be \label{edeltasbh}
\Delta S_{BH} = -4\pi^2 a^4\, \ln A_H \,
\left( K_0 + \sum_r (\beta_r  -1)
\bar K^r(0)\right)\, .
\ee
We shall refer to the term proportional to 
$\sum_r (\beta_r  -1)
\bar K^r(0)$ as the zero mode contribution although it should
be kept in mind that only the term proportional to
$\beta_r$ arises from integration over the zero modes,
and the $-1$ term is the result of subtracting the zero
mode contribution from the heat kernel to correctly
compute the result of integration over the non-zero modes.

The contribution from the fermionic fields can be included in
the above analysis as follows.
Let $\{\psi^i\}$ denote the set of fermion fields in the theory.
Here $i$ labels the internal indices or space-time vector index
(for the gravitino fields) but the spinor indices are suppressed.
Without any loss of generality we can take the
$\psi^i$'s to be Majorana
spinors satisfying $\bar \psi^i =(\psi^i)^T \wt C$ where $\wt C$
is the charge conjugation operator. Then the kinetic term for the
fermions have the form
\be \label{eferkin}
-{1\over 2} \, \bar\psi^i \DD_{ij} \psi^j = -{1\over 2} \, 
(\psi^i)^T \wt C \DD_{ij} \psi^j\, ,
\ee
for some appropriate operator $\DD$.
We can now proceed to define the heat kernel of the fermions
in terms of eigenvalues of $\DD$ in the usual manner, but  
with the following simple changes. Since the
integration over the fermions produce $(\det \DD)^{1/2}$
instead of $(\det\DD)^{-1/2}$, we need
to include an extra minus sign in the definition of the heat kernel.
Also since the fermionic kinetic operator is linear in derivative,
it will be convenient to first compute the determinant of 
$\DD^2$ 
and then take an additional square root of the determinant. 
This is implemented by including an extra factor 
of $1/2$ in
the definition of the heat kernel.\footnote{For this it is
important to work with Majorana or Dirac fermions but
not Weyl fermions since the action of $\DD$ changes the
chirality of the state. Thus $\det (\DD^2)\ne (\det \DD)^2$
acting on a Weyl fermion
if the action of $\DD$ on the left and the right moving
fermions are different.}
We shall denote
by $K_0^f$ the constant part of the fermionic heat kernel
in the small $s$ expansion
after taking into account this factor of $-1/2$.
For analysis of the zero 
modes however
we need to work with the kinetic operator and not its square since
the zero mode structure may get modified upon taking the square
{\it e.g.} the kinetic operator may have
blocks in the Jordan canonical form
which squares to zero,  but the matrix itself may be 
non-zero.\footnote{This problem would not arise if we work
with $\wt C\DD$ instead of $\DD$ since $\wt C \DD$ is 
represented by an
anti-symmetric matrix. However for other reasons it is convenient
to work with $\DD$ instead of $\wt C\DD$.}
Let us denote by $\bar K^f(0)$ the total fermion zero mode 
contribution
to the heat kernel. This must be subtracted from the
total heat kernel. 
Thus we arrive at an expression similar to \refb{enetlognz}
for the fermionic non-zero mode contribution to the entropy:
\be \label{enetlognzfer}
 -4\pi^2 a^4 \, \left(K^f_0 -  \bar K^f(0)\right) \ln A_H\, ,
\ee
Next we need to carry out
the integration over the zero modes. 
Taking into account the extra factor of $-1/2$ in the definition
of the fermionic heat kernel we see that the analog of
\refb{ediffblock} for the total number of fermion zero modes 
$N^{(f)}_{zm}$ now
takes the form
\be \label{enfzm}
N^{(f)}_{zm} = -16\pi^2 a^4 \, \bar K^f(0) \, \left(\cosh\eta_0 - 1
\right)\, .
\ee
Let us further assume that integration over each fermion zero
modes gives a factor of $a^{-\beta_f/2}$ for some constant
$\beta_f$.
Then the total $a$-dependent contribution from integration
over the fermion zero modes is given by
\be \label{eadep}
\exp\left[8\pi^2 a^4 \, \left(\cosh\eta_0 - 1
\right)\, \beta_f \bar K^f(0) \, \ln a \right]\, .
\ee
As usual the coefficient of $\cosh\eta_0$ can be interpreted
as due to a shift in the energy $E_0$, whereas the
$\eta_0$ independent term has the interpretation of a
contribution to the black hole entropy.
Combining this with the contribution \refb{enetlognzfer}
from the non-zero modes
we arive at the 
following expression for the 
logarithmic correction to the entropy from the fermion
zero modes:
\be \label{edeltasbhfermi}
\Delta S_{BH} = -4\pi^2 a^4\, \ln A_H \,
\left( K^f_0 +  (\beta_f  -1)
\bar K^f(0)\right)\, .
\ee
In later sections we shall describe the
computation of $K(0;s)$ and $\bar K^r(0)$ for various
fields, as well as of the coefficients $\beta_r$ for 
gauge fields,
metric and the gravitinos.

\sectiono{Simple examples with minimally coupled massless fields}
\label{sminimal}

We shall now review some simple applications of the
results of the previous section
by computing logarithmic corrections to the black hole
entropy due to minimally coupled scalar, vector and
fermion fields.\footnote{Analysis of logarithmic correction
to the black hole entropy due to massless scalars with
non-minimal coupling to background gravity can be found
in \cite{9504022}.  However for our analysis we also
need to deal with the case where the fluctuations in various fields
are coupled to background fluxes. These will be discussed in later
sections.}
First consider the example of a massless scalar whose
only interaction with other fields is a coupling to gravity
via minimal coupling. Let us denote by
$K^s(x,x';s)$ the heat kernel associated with such a scalar.
It follows from
\refb{eh1} and the fact that
$\square_{AdS_2\times S^2}=\square_{AdS_2}+\square_{S^2}$
that the heat kernel of a massless scalar field
on $AdS_2\times S^2$ 
is given by the product
of the heat kernels on $AdS_2$ and $S^2$, and in the
$x'\to x$ limit takes the form\cite{campo}
\be \label{e4a}
 K^s(0;s) = K^s_{AdS_2}(0;s) K^s_{S^2}(0;s)\, .
\ee
$K^s_{S^2}$ and $K^s_{AdS_2}$ in turn can be calculated using
\refb{eh1} since we know the eigenfunctions and the eigenvalues of the
Laplace operator on these respective spaces. 
The eigenfunctions $f_{\lambda,\ell}$ 
on $AdS_2$ are described in \refb{e5p}.
Since $f_{\lambda,\ell}$  vanishes at
$\eta=0$ for $\ell\ne 0$, only the $\ell=0$ 
eigenfunctions will contribute to
$K^s_{AdS_2}(0;s)$. At $\eta=0$ $f_{\lambda,0}$
has the value
$\sqrt{\lambda\tanh(\pi\lambda)}/\sqrt{2\pi a^2}$.
The corresponding eigenvalue of $-\square_{AdS_2}$ is
$(\lambda^2 +{1\over 4})/a^2$.
Thus \refb{eh1} gives
\be \label{e5q}
K^s_{AdS_2}(0;s) =  {1\over 2\pi\, a^2}
\int_0^\infty \, d\lambda \, \lambda
\tanh(\pi\lambda) \, 
\exp\left[- s\left(\lambda^2 +{1\over 4}\right)/a^2\right]\, 
\, .
\ee
On $S^2$ the eigenfunctions are $Y_{lm}(\psi,\phi)/a$
and the corresponding eigenvalues are $-l(l+1)/a^2$.
Since $Y_{lm}$
vanishes at $\psi=0$ for $m\ne 0$, and $Y_{l0}=\sqrt{2l+1}/\sqrt{4\pi}$ 
at $\psi=0$ we have
\be \label{e6p}
K^s_{S^2}(0;s) = {1\over 4\pi a^2} 
\sum_{l} e^{-sl(l+1)/a^2} (2l+1)\, .
\ee
We can bring this to a form similar to
\refb{e5q} by expressing it as
\be \label{ess1}
{1\over 4\pi i\, a^2} \, e^{s/4a^2}\, 
\ointop \, d\wt\lambda \, \wt\lambda \, 
\tan(\pi\wt\lambda)\, e^{- s\wt\lambda^2/a^2}\, ,
\ee
where $\ointop$ denotes integration along a contour that travels
from  $\infty$ to 0 staying below the real axis and returns to
$\infty$ staying above the real axis. By deforming the integration
contour to a pair of straight lines through the origin --
one at an angle $\kappa$ below the positive real
axis and the other at an angle $\kappa$ above the positive
real axis -- we get
\be \label{ess2}
K^s_{S^2}(0;s) ={1\over 2\pi a^2} e^{s/4a^2}
\, {\rm Im} \, \int_0^{e^{i\kappa}\times \infty}
\, \wt\lambda \, d\wt\lambda\, \tan(\pi\wt\lambda) \, e^{-s\wt\lambda^2/a^2}
\, , \qquad 0<\kappa<< 1\, .
\ee
Combining \refb{e6p} and \refb{e5q} we get the heat kernel of a
scalar field on $AdS_2\times S^2$:
\ben \label{ecomb1}
K^s(0;s) 
&=& {1\over 8\pi^2 a^4}\,
\sum_{l=0}^\infty (2l+1) \int_0^\infty \, d\lambda \, \lambda
\tanh(\pi\lambda) \, \exp\left[ -\bar s \lambda^2 - \bar 
s \left(l+{1\over 2}
\right)^2
\right]\nn
&=& {1\over 4\pi^2 a^4}\,
\int_0^\infty \, d\lambda \, \lambda
\tanh(\pi\lambda) \, 
\, {\rm Im} \, \int_0^{e^{i\kappa}\times \infty}
\, \wt\lambda \, d\wt\lambda\, \tan(\pi\wt\lambda)\,
\exp\left[ -\bar s \lambda^2 - \bar 
s \wt\lambda^2
\right]\, , \nn
\een
where
\be \label{e7}
\bar s = s/a^2\, .
\ee

In order to find the logarithmic correction to the entropy we need to
expand $K^s(0;s)$ in a power series expansion in $\bar s$ and pick
the coefficient $K^s_0$ of the constant term in this expansion. 
With the help of \refb{esa1}, \refb{esa2}
we get:
\ben \label{e8}
K^s_{AdS_2}(0;s) &=& {1\over 4\pi a^2\, \bar s}\,
e^{-\bar s / 4} \, \left[ 1 + 
\sum_{n=0}^\infty {(-1)^n\over n!} (2n+1)! {\bar s^{n+1}
\over \pi^{2n+2}} {1\over 2^{2n}} \left(2^{-2n-1}-1\right)
\zeta(2n+2)\right]\nn
&=& {1\over 4\pi a^2\, \bar s}\,
e^{-\bar s / 4} \, \left(1 -{1\over 12} \bar s+ 
{7\over 480} \bar s^2 + \OO(\bar s^3) \right)\, ,
\een
\ben \label{e9}
K^s_{S^2}(0;s) &=& {1\over 4\pi a^2\, \bar s}\,
e^{\bar s / 4} \, \left[ 1 -
\sum_{n=0}^\infty {1\over n!} (2n+1)! {\bar s^{n+1}
\over \pi^{2n+2}} {1\over 2^{2n}} \left(2^{-2n-1}-1\right)
\zeta(2n+2)\right]\nn
&=& {1\over 4\pi a^2\, \bar s}\,
e^{\bar s / 4} \, \left(1 +{1\over 12} \bar s+ 
{7\over 480} \bar s^2 + \OO(\bar s^3) \right)\, .
\een
Substituting \refb{e8} and \refb{e9} into \refb{e4a} we get
\be \label{e10}
K^s(0;s) =  
{1\over 16\pi^2 a^4\, \bar s^2} 
\left( 1 +{1\over 45} \bar s^2 +\OO(s^4)\right)\, .
\ee
This gives $K^s_0 = 1/720\pi^2 a^4$.
Eq.\refb{eigen1} shows that for the scalar all the eigenvalues of the
kinetic operator
$-\square$ are
positive and hence there are no zero modes.
Hence, using 
\refb{edeltasbh} we get the logarithmic
contribution to the entropy from a minimally coupled scalar
to be
\be \label{e14}
\Delta S_{BH} = -{1\over 180} \ln A_H\, .
\ee

Next we consider the case of a Maxwell field $A_\mu$ whose only
coupling is via the minimal coupling to the background metric. The
action of such a field is given by
\be \label{ea1}
\SSS_A = -{1\over 4} \int d^4 x \sqrt{\det g}
\, F_{\mu\nu} F^{\mu\nu}\, ,
\ee
where $F_{\mu\nu} \equiv \p_\mu A_\nu - \p_\nu A_\mu$ is the
gauge field strength. Adding a gauge fixing term
\be \label{ea2}
S_{gf} = -{1\over 2} \int d^4 x \sqrt{\det g} \, (D_\mu A^\mu)^2\, ,
\ee
we can express the action as
\be \label{ea3}
\SSS_A +\SSS_{gf}
= -{1\over 2} \int d^4 x \sqrt{\det g} A_\mu
(\Delta A)^\mu\, ,
\ee
where
\be \label{eddelta}
(\Delta \, A)_\mu \equiv -\square \, A_\mu + R_{\mu\nu}
A^\nu\, ,\qquad \square A_\mu \equiv g^{\rho\sigma} D_\rho D_\sigma
A_\mu\, .
\ee
A vector in
$AdS_2\times S^2$ decomposes into a (vector, scalar) plus
a (scalar, vector), with the first and the second factors representing
tensorial properties in $AdS_2$ and $S^2$ respectively.
Furthermore, on any of these
components the action of the kinetic operator
can be expressed as $\Delta_{AdS_2}+\Delta_{S^2}$, with $\Delta$
defined as in \refb{eddelta} for vectors and as $-\square$ for
scalars. Thus
we can construct the eigenfunctions of $\Delta$
by taking the product of appropriate eigenfunctions of
$\Delta_{AdS_2}$ and $\Delta_{S^2}$, and the corresponding
eigenvalue of $\Delta$ on $AdS_2\times S^2$ will be given by the
sum of the eigenvalues of $\Delta_{AdS_2}$ and $\Delta_{S^2}$.
This gives
\be \label{e20}
K^{v}(0;s)=K^{v}_{AdS_2}(0,s) K^s_{S^2}(0;s)
+ K^s_{AdS_2}(0,s) K^{v}_{S^2}(0;s)\, .
\ee
Thus we need to compute $K^{v}_{AdS_2}(0,s)$ and
$K^{v}_{S^2}(0;s)$.  Finally,
quantization of gauge fields also requires us to introduce
two anticommuting scalar ghosts whose kinetic operator is given by
the standard laplacian $-\square$ in the harmonic gauge.
They give a net contribution of $-2 K^s(0;s)$ to the heat kernel.

To find $K^v_{S^2}$ we use the basis functions given in
\refb{ebasis}. These have $\Delta$ eigenvalue $\kappa^{(k)}_1$
and hence 
the  contribution from any of these two
eigenfunctions to the
vector heat kernel $K^v_{S^2}(x,x;s)$ is given by $(\kappa_1^{(k)})^{-1}\,
e^{-\kappa_1^{(k)} s}\, 
g^{\mu\nu}\p_\mu U_k(x)
\p_\nu U_k(x)$. Now since $K^v_{S^2}(x,x;s)$ is independent
of $x$ after summing over the contribution from all the states, we could
compute it by taking the volume average of each term.
Taking a volume average 
allows us to integrate by parts and gives
the same result as the volume average of
$(\kappa_1^{(k)})^{-1} \,
e^{-\kappa_1^{(k)} s}\, U_k(x) (-\square)  U_k(x)
= e^{-\kappa_1^{(k)} s}\, U_k(x)^2$. 
This is the same as the contribution
from $U_k$ to the scalar heat kernel. Thus the net contribution to
$K^v_{S^2}(0,s)$ from the pair of basis states given in \refb{ebasis}
is given by $2 K^s_{S^2}(0;s) - 1/2\pi a^2$, where the subtraction
term $-1/2\pi a^2$ accounts for the absence of the contribution
from the $l=0$ modes.
Similarly the contribution from the basis states \refb{ebasistwo} to 
$K^v_{AdS_2}(0;s)$ is given by $2 K^s_{AdS_2}(0;s)$. We must
add to this the contribution from the discrete modes given in
\refb{e24p}. Using \refb{esumvector} we see that this 
contribution is given by $1/2\pi a^2$, leading to
$K^v_{AdS_2}(0;s)=2 K^s_{AdS_2}(0;s) + 1/2\pi a^2$.
Thus we get the net contribution to the $K(0;s)$ from the vector
field, including the ghosts, to be:
\ben \label{enetv}
K^v(0,s) &=& \left(2 K^s_{S^2}(0;s) -{1\over 2\pi a^2}\right)
K^s_{AdS_2}(0;s) + 
\left(2 K^s_{AdS_2}(0;s)  +{1\over 2\pi a^2}\right)
K^s_{S^2}(0;s) \nn
&& - 2 K^s_{S^2}(0;s) K^s_{AdS_2}(0;s)\, .
\een
Using \refb{e8}, \refb{e9} we get
\be \label{enetv1}
K^v(0,s) = {1\over 8\pi^2 a^4}\left({1\over \bar s^2} 
+ {31\over 45}  +\OO(\bar s^4)\right)\, ,
\ee
leading to $K^v_0 = 31/360\pi^2 a^4$.

Gauge fields also have zero modes arising from the product of
$a^{-1} Y_{00}(\psi,\phi)$ with the discrete modes 
$\p_m\Phi^{(\ell)}$ given in \refb{e24p}.
Using \refb{edefk0} and \refb{esumvector}
we get the contribution to $\bar K$ from
these zero modes to be
\be \label{evzm1}
\bar K^v(0) = a^{-2} \, \sum_\ell \left(Y_{00}(\psi,\phi)\right)^2
g^{mn}\p_m \Phi^{(\ell)}(x) \p_n\Phi^{(\ell)}(x)
= {1\over 8\pi^2 a^4}\, .
\ee
We could also derive the expression for 
as follows. It follows from \refb{ediffblock} that 
$8\pi^2 a^4 \bar K^v(0)(\cosh\eta_0-1)$ 
has the interpretation of the total
number of gauge field zero modes. This in turn is given by the
number of discrete modes $N_1$ on $AdS_2$ given in \refb{etotalvz}
since the gauge field zero modes are obtained by taking the
product of the unique $l=0$ mode of a scalar in $S^2$ and the
discrete modes of the vector field in $AdS_2$.  Thus we
have $8\pi^2 a^4 \bar K^v(0)=1$.

We now need to compute the coefficient $\beta_v$ appearing in
\refb{edeltasbh} for the zero modes of the vector fields.
This computation proceeds as follows.
First we express the metric $g_{\mu\nu}$
on $AdS_2\times S^2$ as
$a^2 g^{(0)}_{\mu\nu}$ where 
$g^{(0)}_{\mu\nu}$
is independent of $a$. The path integral over 
$A_\mu$ is normalized such that
\be \label{eap2}
\int [DA_\mu] \exp\left[- \int d^4 x \, \sqrt{\det g} \, g^{\mu\nu} A_\mu A_\nu
\right] = 1\, ,
\ee
\i.e.\
\be \label{eap3}
\int [DA_\mu] \exp\left[- a^2 \int d^4 x \, \sqrt{\det g^{(0)}} \, 
g^{(0)\mu\nu} A_\mu A_\nu
\right] = 1\, .
\ee
{}From this we see that up to an $a$ independent 
normalization
constant, $[DA_\mu]$  actually corresponds to integration
with measure $\prod_{\mu,x} d(a A_\mu(x))$. 
On the other hand  the gauge field
zero modes are associated with
deformations produced by the gauge transformations
with non-normalizable parameters: $\delta A_\mu\propto
\p_\mu \Lambda(x)$ for some functions $\Lambda(x)$ with 
$a$-independent
integration range. Thus
the result of integration over the gauge field zero
modes can be found by first changing the integration over the
zero modes of
$(aA_\mu)$ to integration over $\Lambda$ and then
picking up the contribution from the Jacobian in this
change of variables. This gives a factor of $a$ from integration
over each zero mode of $A_\mu$.
It now follows from the definition of $\beta_r$ given 
in the paragraph below \refb{ediffblock} that we have
\be \label{ebetav}
\beta_v = 1\, .
\ee
Eq.\refb{edeltasbh} now gives the net logarithmic contribution
to $S_{BH}$ from the minimally coupled vector field to be
\be \label{eminimalvector}
-4\pi^2 a^4\, \ln A_H \,
\left( K^v_0 + (\beta_v  -1)
\bar K^v(0)\right) = -{31\over 90}\ln A_H\, .
\ee
Note that the term proportional to $\bar K^v(0)$ does not
contribute since $\beta_v=1$.

Next we consider the case of a massless
Dirac fermion, again with only
interaction being minimal coupling to the metric on $AdS_2\times S^2$.
The eigenfunctions and eigenvalues 
of the square of the Dirac operator are given by the direct product of
$(\chi_{l.m}^\pm, \eta_{l,m}^\pm)$ given in \refb{ed2} with
$(\chi_k^\pm(\lambda), \eta_k^\pm(\lambda))$ given in \refb{ed2a}.
We can compute the heat kernel for the fermion using the relations:
\ben \label{erelchi}
&& \sum_{m} \left((\chi^+_{l,m})^\dagger \chi^+_{l,m}
+ (\chi^-_{l,m})^\dagger \chi^-_{l,m} +
(\eta^+_{l,m})^\dagger \eta^+_{l,m}
+ (\eta^-_{l,m})^\dagger \eta^-_{l,m}\right) = {1\over \pi a^2}(l+1)\, , \nn
&& \sum_k \left((\chi^+_{k}(\lambda))^\dagger \chi^+_{k}(\lambda)
+ (\chi^-_{k}(\lambda))^\dagger \chi^-_{k}(\lambda)
+ (\eta^+_{k}(\lambda))^\dagger \eta^+_{k}(\lambda)
+ (\eta^-_{k}(\lambda))^\dagger \eta^-_{k}(\lambda)\right)
= {1\over \pi a^2} \lambda \coth(\pi\lambda)\, . \nn
\een
The first of these relations is derived by evaluating it at $\psi=0$
where only the $m=0$ terms contribute whereas the second relation
is derived by evaluating it at $\eta=0$ where only the $k=0$ terms
contribute. Using this we get the contribution to $K(0;s)$ from the
fermion fields to be
\ben \label{ediracf}
K^f(0;s)&=& -{1\over \pi^2 a^4} 
\int_0^\infty d\lambda e^{-\bar s\lambda^2}
\, \lambda \, \coth(\pi\lambda)
 \sum_{l=0}^\infty
(l+1)
\, e^{-\bar s\left(l+1\right)^2}\nn
&=& -{1\over \pi^2 a^4} \int_0^\infty d\lambda 
\, \lambda \, \coth(\pi\lambda)
\int_0^{e^{i\kappa}\times \infty} 
d\wt\lambda \, \wt\lambda \, 
\cot(\pi\wt\lambda) \, e^{-\bar s\wt\lambda^2-\bar s\lambda^2}\, .
\een
Note that we have included a minus sign 
in the heat kernel to
account for the fermionic nature of the fields.
Since we are squaring the kinetic operator we should
have also gotten a factor of 1/2, but this is compensated
for by a factor of 2 arising out of the complex nature of
the fields. In other words when we 
expand a Dirac fermion in the
basis $(\chi^\pm_{l,m}, \eta^\pm_{l,m})\otimes
(\chi_k^\pm(\lambda), \eta_k^\pm(\lambda))$, the
coefficients of expansion are arbitrary
complex numbers, and hence we double the number
of integration variables.
Using \refb{esa1a}, \refb{esa2a} we now get
\be \label{ekffin}
K^f(0;s) = -{1\over 4\pi^2 a^4 \bar s^2} \left( 1 - {11\over 180}\bar s^2
+\OO(\bar s^4)\right)\, ,
\ee
leading to $K^f_0=11/720\pi^2 a^4$. Since there are no zero modes
for the fermions, \refb{edeltasbh} leads to the following contribution to the
black hole entropy due to a minimally coupled
massless Dirac fermion:
\be \label{efermisbh}
-{11\over 180} \ln A_H\, .
\ee
If instead we choose to work with Majorana fermions
then \refb{efermisbh} is replaced by 
$-{11\over 360} \ln A_H$. 

Our analysis shows that if we have a set of $n_S$ minimally
coupled massless scalar fields, $n_V$ minimally coupled
Maxwell fields and $n_F$ minimally coupled massless
Dirac fields, then they lead to a net logarithmic contribution
of
\be \label{enetlogminimal}
\Delta S_{BH} = -{1\over 180} \ln \, A_H 
(n_S + 62 n_V + 11 n_F)\, 
\ee
to the black hole entropy.
We shall now describe an alternative method for arriving at this
result. First note that in all the cases discussed above
only the $K_0$ term in \refb{edeltasbh} is responsible for the
logarithmic correction; the contribution proportional to
$(\beta_r -1) \bar K^r$ vanishes either due to the
vanishing of $\bar K^r$ due to absence of zero modes
(as in the case of scalars and fermions) or due to
the vanishing of $\beta_r-1$ (as in the case of gauge fields).
On the other hand one can show 
that\cite{duffobs,christ-duff1,christ-duff2,duffnieu,birrel,gilkey,0306138} 
the
contribution to $K_0$ -- the constant term in the small
$\bar s$ expansion of the heat kernel -- is given by
\be \label{ebirrel}
K_0 = -{1\over 90\pi^2} (n_S + 62 n_V + 11 n_F) E
- {1\over 30\pi^2} (n_S + 12 n_V + 6 n_F) I\, ,
\ee
where
\ben \label{edefie}
E &=& {1\over 64} \left(R_{\mu\nu\rho\sigma} 
R^{\mu\nu\rho\sigma} - 4 R_{\mu\nu} R^{\mu\nu}
+ R^2\right) \nn
I &=&  -{1\over 64} \left(R_{\mu\nu\rho\sigma} 
R^{\mu\nu\rho\sigma} - 2 R_{\mu\nu} R^{\mu\nu}
+ {1\over 3} R^2\right)\, .
\een
For the metric \refb{emets} we have $I=0$ and $E=-1/8a^2$.
Thus we get
\be \label{ebir2}
K_0 = {1\over 720\pi^2 a^4} (n_S + 62 n_V + 11 n_F)\, .
\ee
Substituting this into \refb{edeltasbh} we recover
\refb{enetlogminimal}.

The result \refb{enetlogminimal} agrees with earlier
results on logarithmic corrections to the extremal black
hole entropy computed {\it e.g.} in 
\cite{9709064,1008.4314,1104.3712}.
This will not be the case for the results derived in later
sections, so it is important to understand the relation
between the two computations. 
First \cite{9709064,1008.4314,1104.3712} do not use the quantum
entropy function for their
computation, but use 
the relation between the entanglement entropy and the
partition function in the presence of a conical defect. But
as argued in \cite{9407001,9412020} the entropy 
computed by this method gives the same result computed
using the $K_0$ given in \refb{ebirrel} -- so this is
not a coincidence. Second,
 as we have seen in the analysis described above
 the zero modes
conspire in such a way that the result is controlled
completely by the coefficient $K_0$ arising in the
small $\bar s$ expansion of the heat kernel. 
If this had not been the case then we would have to
account for the extra contribution proportional to
$\bar K^r(0) (\beta_r-1)$ which is absent in the analysis
of \cite{9709064,1008.4314,1104.3712}.
As we shall
see in the next few sections, $\bar K^r(0) (\beta_r-1)$
will be non-vanishing
when we are considering fluctuations of the metric
or gravitino
degrees of freedom. 
Third, in arriving at \refb{enetlogminimal} 
we have analyzed fields which couple to gravity minimally
without any coupling to any background flux. 
This however is not always the case, {\it e.g.}
whenever there is
any background flux, {\it e.g.} for Reissner-Nordstrom
black holes, the kinetic term of the 
metric and some gauge fields get additional contribution
due to the background flux which is not captured in
the simple formula given in \refb{ebirrel}. 
A similar effect occurs the the fermionic sector.
It may be possible to generalize \refb{ebirrel} and hence the
analysis of \cite{9709064,1008.4314,1104.3712} to such cases,
but the results currently available in 
\cite{9709064,1008.4314,1104.3712} 
are not sufficient to
compute correctly the logarithmic correction to the extremal
black hole entropy due to metric and gravitino
fluctuations, and other fields with non-trivial coupling to the
background flux.
It will be interesting to generalize the earlier analysis of
\cite{9709064,1008.4314,1104.3712} to incorporate the effect of the
zero modes and the 
background flux, and see if the results for 
logarithmic correction
to the entropy agree with those given in 
\S\ref{s1}-\ref{sgeneral}.

\sectiono{Extremal Reissner-Nordstrom black holes} 
\label{s1}

We now consider the Einstein-Maxwell theory with
the action
\be \label{ernk1}
\SSS = \int \sqrt{\det g}\,  \LL_b, \qquad
\LL_b= \left[ R - F_{\mu\nu} F^{\mu\nu}\right]\, ,
\ee
where $R$ is the scalar curvature computed with
the metric $g_{\mu\nu}$ and $F_{\mu\nu}
=\p_\mu A_\nu - \p_\nu A_\mu$ is the gauge field
strength. 
Note that we have set $G_N = 1/16\pi$.
The near horizon geometry of an extremal electrically
charged
Reissner-Nordstrom solution in this theory is given by
(see {\it e.g.} \cite{0708.1270})
\be \label{ernk1.5}
ds^2 \equiv \bar g_{\mu\nu} dx^\mu dx^\nu =
a^2(d\eta^2 +\sinh^2\eta d\theta^2) + a^2 (d\psi^2 +
\sin^2 \psi d\phi^2), \qquad \bar F_{mn}
= {i} \, a^{-1}\, \ve_{mn}\, .
\ee
The parameter $a$ is related to the electric charge $q$ via
the relation $q=a$.
The classical Bekenstein-Hawking entropy of this black
hole is given by
\be \label{ebeken}
S_{BH} = 4\pi A_H=16 \pi^2 a^2=16\pi^2 q^2\, .
\ee
Since this theory possesses an electric-magnetic duality
symmetry, the result for the entropy of a dyonic black
hole carrying electric charge $q$ and magnetic charge $p$
can be found from that of an electrically charged black
hole by replacing $q$ by $\sqrt{q^2+p^2}$. This holds
for the classical entropy given in \refb{ebeken} as well
as the logarithmic correction that will be discussed below.

To compute logarithmic corrections to the entropy of this
black hole we consider fluctuations of the metric and
gauge fields of form
\be \label{ernk2}
g_{\mu\nu} = \bar g_{\mu\nu} + h_{\mu\nu}, \qquad
A_\mu = \bar A_\mu + {1\over 2}\, 
\AAA_\mu, \quad
F_{\mu\nu} = \bar F_{\mu\nu} +{1\over 2}
(\p_\mu \AAA_\nu - \p_\nu\AAA_\mu)
\equiv \bar F_{\mu\nu} +
{1\over 2} \, f_{\mu\nu}\, .
\ee
In subsequent discussions 
all indices will be raised and lowered by the 
background metric
$\bar g$.
Substituting \refb{ernk2} into \refb{ernk1}, adding to this a
gauge fixing term
\be \label{egauge}
\LL_{gf} = -{1\over 2} 
g^{\rho\sigma}\,
\left(D^\mu h_{\mu\rho} -{1\over 2} D_\rho \, 
h^\mu_{~\mu}\right)
\left( D^\nu \, h_{\nu\sigma} -{1\over 2} 
D_\sigma h^\nu_{~\nu}\right)
 - {1\over 2} D^\mu \AAA_\mu D^\nu \AAA_\nu
\, ,
\ee
and throwing away total derivative terms, 
we get the total Lagrangian
density for the fluctuating fields:
\ben \label{sboson}
\LL_b + \LL_{gf} &=& \hbox{constant} 
-{1\over 4} h_{\mu\nu} 
\left(\wt \Delta
h\right)^{\mu\nu} + {1\over 2} 
\AAA_\mu (\bar g^{\mu\nu}\square  - R^{\mu\nu}) 
\AAA_\nu\nn
&& + a^{-2}\, \left({1\over 2} \, h^{mn} h_{mn}
- {1\over 2} \, h^{\alpha\beta} h_{\alpha\beta}
+ h^{m\alpha} h_{m\alpha}
+{1\over  4} (h^\alpha_{~\alpha} - h^m_{~m})^2 \right)\nn
&& - { 2i a^{-1}}
\vareps^{mn}\, f_{\alpha m} h^\alpha_{~n}
-  {i\over 2} \, a^{-1}\,  \vareps^{mn} f_{mn} 
\left( h^\gamma_{~\gamma} - h^p_{~p}\right)
\, ,
\een
where
\ben \label{e35}
\left(\wt\Delta h\right)_{\mu\nu} 
&=& -\square h_{\mu\nu} - R_{\mu\tau} h^\tau_{~\nu}
- R_{\nu\tau} h_\mu^{~\tau} - 2 R_{\mu\rho\nu\tau} h^{\rho\tau}
+{1\over 2} \, \bar g_{\mu\nu} \, 
\bar g^{\rho\sigma} \, \square\, h_{\rho\sigma}
\nn && + R\, h_{\mu\nu} 
+ \left(\bar g_{\mu\nu} R^{\rho\sigma} + R_{\mu\nu}
\bar g^{\rho\sigma}\right) h_{\rho\sigma} -{1\over 2}\, R\, 
\bar g_{\mu\nu} \, \bar g^{\rho\sigma}\, h_{\rho\sigma}
\, .
\een
In this formula
all components of the Riemann and Ricci tensor and the
curvature scalar are computed with the background metric
$\bar g_{\mu\nu}$.
To this we must also add the 
Lagrangian density for the ghost fields\cite{1005.3044}:
\be \label{emghost}
\LL_{ghost} = 
\left[b^\mu \left(\bar g_{\mu\nu}\square + R_{\mu\nu}\right) 
 c^\nu + b\square c
- 2 \, b \bar F_{\mu\nu} \, D^\mu c^\nu\right]
\, .
\ee

We now need to find the eigenmodes and eigenvalues
of the kinetic operator and then calculate the determinant.
We follow the same strategy as in 
\cite{1005.3044,1106.0080}, \i.e.\ first expand the 
various fields as linear combinations
of the eigenmodes described in appendix \ref{sbasis},
substitute them into the action \refb{sboson}, \refb{emghost},
and then find the eigenvalues of the kinetic
operator. For this we can work at
fixed $l$ and $\lambda$ values since at the quadratic level
the modes carrying different $l$ and $\lambda$
values do not mix. 
This simplifies the problem enormously since at
fixed values of $l$ and $\lambda$ the kinetic operator
reduces to a finite dimensional matrix 
$\MM(l+{1\over 2},\lambda)$.
The net contribution to $K(0;s)$ can then be computed
using the formula
\ben \label{k0smaster}
K(0;s) &=& {1\over 8\pi^2 a^4} \, 
\sum_{l=0}^\infty (2l+1) \int_0^\infty d\lambda \,
\lambda\, 
\tanh(\pi\lambda)\, Tr e^{s \MM(l+{1\over 2},\lambda)} \nn
&=& {1\over 4\pi^2 a^4} \, {\rm Im}
\int_0^{e^{i\kappa}\times \infty}\, d\wt\lambda \,
\wt\lambda\, 
\tan(\pi\wt\lambda) \int_0^\infty d\lambda \,
\lambda\, 
\tanh(\pi\lambda)\, Tr e^{s \MM(\wt\lambda,\lambda)} \, .
\een
It will be convenient to introduce a new matrix $M$ via
the relation:
\be \label{enewmatrix}
\MM
= \left\{ -(\kappa_1 + \kappa_2) \, I + a^{-2} \, M\right\}
\, ,
\ee
where $I$ is the identity matrix and
 \be \label{ernparam}
\kappa_1 = a^{-2} l(l+1) = a^{-2}
\left(\wt\lambda^2 -{1\over 4}\right), \qquad
\kappa_2 = a^{-2} \left(\lambda^2 +{1\over 4}\right)\, .
\ee
Substituting this into \refb{k0smaster} we get the
first contribution to $K(0;s)$ which we denote by
$\wt K^B_{(1)}(0;s)$:\footnote{The superscript $B$ stands for
bosonic fields. Of course in the Einstein-Maxwell theory all
physical fields are bosonic and hence this symbol is redundant,
but eventually we shall regard this as the bosonic sector of
$\NN=2$ supergravity. The `tilde' on $K$ stands for the fact that
we have overcounted the contribution from the
$l=0$ and $l=1$ sectors by ignoring the constraints 
mentioned below
\refb{ebasis}, \refb{ebasisgrav}. Again this notation has been used
keeping in mind a similar notation to be used in
\S\ref{sfermion} for the fermionic sector of $\NN=2$ supergravity.}
\be\label{ek0snext}
\wt K^B_{(1)}(0;s) = {1\over 4\pi^2 a^4} \, 
\int_0^{e^{i\kappa}\times \infty}\, d\wt\lambda \,
\wt\lambda\, 
\tan(\pi\wt\lambda) \int_0^\infty d\lambda \,
\lambda\, 
\tanh(\pi\lambda)\, e^{-\bar s (\lambda^2 +\wt\lambda^2)} \,
Tr (e^{\bar s M})\, .
\ee
We can now carry out the small $\bar s$ expansion by
expanding the last term as
\be\label{eexpandlast}
Tr (e^{\bar s M}) = \sum_{n=0}^\infty 
{1\over n!} \bar s^n \, Tr(M^n)\, 
\ee
and using \refb{esa1}, \refb{esa2} to evaluate the integrals.
\refb{ek0snext} is not the complete contribution however, since
for $l=0$ and $1$ some modes will be absent 
due to the
constraints on the modes mentioned below
\refb{ebasis}, \refb{ebasisgrav}. This requires a subtraction
term which we shall call $\wt K^B_{(2)}$. Finally we also
have to include the contribution from the discrete modes
given in \refb{e24p}, \refb{ehmn} which we shall denote by
$K^B_{(3)}$. 

Our first task will be to find the matrix $M$. For this we
expand the various fields as 
\ben \label{ewr2}
&& \AAA_\alpha = {1\over \sqrt {\kappa_1}}
\left( C _1 \p_\alpha \, u + C _2 \ve_{\alpha\beta}
\p^\beta \, u\right)\, , \qquad
\AAA_m =  {1\over \sqrt {\kappa_2 }}
\left( C _3 \p_m      \, u + C _4 \ve_{mn}
\p^n \, u\right)\, , \nn
&& h_{m\alpha} = {1\over \sqrt {\kappa_1\kappa_2}}
\left(B_1 \, \p_\alpha \p_m \, u + B_2 \, \ve_{mn}  \p_\alpha \p^n u
+ B_3 \, \ve_{\alpha\beta} \, \p^\beta \p_m u +
B_4 \, \ve_{\alpha\beta} \, \ve_{mn} \, \p^\beta \p^n u\right)\, , \nn
&& h_{\alpha\beta} = {1\over \sqrt 2} \, (i\, B_5 
+ B_6)\, g_{\alpha\beta} \, u + {1\over \sqrt{
\kappa_1 - 2 a^{-2}}}
\left(D_\alpha\xi_\beta + D_\beta \xi_\alpha 
- g_{\alpha\beta} \, D^\gamma\xi_\gamma\right),
 \nn
&& h_{mn} = {1\over \sqrt 2} \, (i\, B_5 
- B_6)\, g_{mn} \, u + {1\over 
\sqrt{\kappa_2 + 2 a^{-2}}}
\left(D_m\wh\xi_n + D_n \wh\xi_m - g_{mn} \, D^p\wh\xi_p\right),
\nn
&& \qquad  \quad \xi_\alpha = {1\over \sqrt{\kappa_1}}\left(
B_7 \p_\alpha \, u 
+ B_8 \, \ve_{\alpha\beta}
\p^\beta \, u\right)\, ,\qquad \wh\xi_m = 
{1\over \sqrt{\kappa_2}}\, \left(B_9 \p_m      \, u + B_{0} 
\, \ve_{mn}
\p^n \, u\right)\, . \nn
\een
Here $u$ denotes the product of $Y_{lm}(\psi,\phi)/a$ and
a basis vector $f_{\lambda,\ell}(\eta,\theta)$
given in \refb{e5p} for some fixed $(l,\lambda)$. 
$B_i$'s  and $C_i$'s are constants labelling the
fluctuations. Substituting this into the action we can
compute the matrix $\MM$
of the kinetic operator. The result is
\ben \label{eget1}
&& {1\over 2} \pmatrix{\vec B & \vec C} \,
\MM\, 
\pmatrix{\vec B\cr \vec C} \nn
&=& 
 - {1\over 2} (\kappa_1 +\kappa_2) \, \left[\sum_{i=1}^4 C_i^2
+ \sum_{i=1}^6 B_i^2 \right] -{1\over 2} (\kappa_1+\kappa_2
- 4\, a^{-2}) (B_7^2 + B_8^2) \nn &&
 -{1\over 2} (\kappa_1+\kappa_2 +4a^{-2})
(B_9^2 + B_0^2) \nonumber \\ 
&& + a^{-2} \sum_{i=1}^4 B_i^2 - 2 i a^{-2} B_5 B_6
- a^{-2} (B_7^2 + B_8^2) + a^{-2} (B_9^2 + B_0^2) 
+ 2 a^{-2} B_6^2 \nn
&& - 2 i a^{-1} \left[ -\sqrt{\kappa_1} C_3 B_2 + \sqrt{\kappa_1}
C_4 B_1 + \sqrt{\kappa_2} C_1 B_2 + \sqrt{\kappa_2} C_2 B_4
+ \sqrt{2\kappa_2} B_6 C_4\right] \, .\nn
\een
The matrix $\MM$ and hence the matrix
$M$ defined via \refb{enewmatrix}, \refb{eget1}
has block diagonal form and
is easy to diagonalize.
First of all we note the the modes labelled by
$B_3$, $B_7$, $B_8$, $B_9$ and $B_0$ do not mix
with any other mode and the modes $B_4$ and $C_2$
only mix with each other but not with any other mode. 
The modes $B_2$, $C_1$ and
$C_3$ mix with each other but not with any other
mode. Finally the modes $B_5$, $B_6$, $C_4$ and 
$B_1$ mix with each other but not with any other mode.
The eigenvalues of $M$ in these different sectors
are given by
\ben\label{erneigen}
&& B_3:2, \quad 
B_7: 2,  \quad B_8: 2, \quad B_9:-2, \quad B_0: -2,
\quad B_4, C_2: 1\pm i\sqrt{4\kappa_2 a^2 -1},
 \nn &&
B_2, C_1, C_3: 0, 1\pm i\sqrt{4 a^2 (\kappa_1+
\kappa_2) - 1}\nn &&
B_5, B_6, C_4, B_1: \hbox{Eigenvalues of} \pmatrix{
0 & -2 i  & 0 & 0\cr
-2i & 4 & -2ia\sqrt{2\kappa_2} & 0\cr
0 & -2ia\sqrt{2\kappa_2} & 0 & -2ia\sqrt{\kappa_1}\cr
0 & 0 & -2 i a \sqrt{\kappa_1} & 2}\, .
\een
{}From this we get
\ben \label{ern2}
Tr (M) &=& 12,\nn
Tr(M^2) &=& 36 - 32 \lambda^2 - 16\wt\lambda^2,\nn
Tr(M^3) &=& 24 - 144\lambda^2 - 48\wt\lambda^2,\nn
Tr(M^4) &=& 68 - 464 \lambda^2 + 192 \lambda^4
- 112 \wt\lambda^2 + 192 \lambda^2 \wt\lambda^2 
+ 64 \wt\lambda^4\, .
\een
Substituting this into \refb{ek0snext} and carrying our the
$\lambda$, $\wt\lambda$ integrals using
\refb{esa1}, \refb{esa2} we get the constant term
in the small $\bar s$ expansion of $\wt K^B_{(1)}(0;s)$
to be
\be \label{ekb1value}
\wt K^B_{(1)}(0;s): {337\over 360\pi^2 a^4}\, .
\ee

We now need to remove the contribution due to the modes
which are absent for $l=0$ and $l=1$.
For $l=1$ the modes $B_7$ and $B_8$ are absent
due to the constraint mentioned below \refb{ebasisgrav}.
The removed  eigenvalues of $M$ are 2 and 2, and so
those of $\MM$ are $-a^{-2}(\lambda^2+{1\over 4})$ and 
$-a^{-2}(\lambda^2+{1\over 4})$.
For $l=0$ the modes $C_1$, $C_2$, $B_1$, $B_2$,
$B_3$, $B_4$, $B_7$, $B_8$ are absent due to the
constraint mentioned below \refb{ebasis}. The removed
eigenvalues of $M$ are:
\ben \label{ernremoved}
B_1:2, \quad 
B_3:2, \quad B_7: 2, \quad B_8:2, \quad B_4, C_2: 1\pm
2i\lambda, \quad B_2, C_1: 1\pm 2i\lambda\, . \een
This gives a net subtraction term 
\be \label{ern4}
\wt K^B_{(2)}(0;s)= -{1\over 8\pi^2 a^4} \int_0^\infty
d\lambda \, \lambda\, \tanh\pi\lambda\, e^{-\bar s \lambda^2}
\, e^{-\bar s/4}\, 
\left[6 + 2 e^{\bar s(1+2i\lambda)} + 2 e^{\bar s(1-2i\lambda)}
+ 4 e^{2\bar s}\right]\, .
\ee
The first term inside the square bracket is the contribution from
the $l=1$ modes while the other terms represent contribution from
the $l=0$ modes.
Again by expanding the term inside 
the square bracket in a power series
expansion in $\bar s$ and using \refb{esa1} we get the
$\bar s$ independent contribution to $\wt K^B_{(2)}$ in the
small $\bar s$ expansion to be
\be \label{ekb2value}
\wt K^B_{(2)}(0;s): {1\over 24\pi^2 a^4}\, .
\ee

Next we need to include the contribution due to the
discrete modes. For this we expand the fields as
\ben \label{edefo1}
&& \AAA_m = E_1 v_m + E_2 \ve_{mn} v^n, 
 \nn
&& h_{m\alpha} = {1\over \sqrt{\kappa_1}}\,
\left(E_3 \p_\alpha v_m + \wt E_3 \ve_{mn} \p_\alpha v^n
+
E_4 \ve_{\alpha\beta} \p^\beta v_m
+ \wt E_4 \ve_{\alpha\beta} \ve_{mn} \p^\beta v^n \right)
\nn
&& h_{mn} ={a\over \sqrt 2}\,
 \left(D_m \wh\xi_n + D_n \wh\xi_m 
- g_{mn} D^p \wh\xi_p\right) \, , \quad 
\wh\xi_m = E_5 v_m + \wt E_5 \ve_{mn} v^n
\, ,
\een
and
\be \label{edefo2.5}
h_{mn} = E_6 w_{mn}\, .
\ee
Here $v_m$ is the product of a spherical harmonic with one of the
vectors in \refb{ebasistwo} and $w_{mn}$ is the
product of a spherical harmonic 
with one of the basis vectors given in
\refb{ehmn}.
Following the strategy of \cite{1005.3044,1106.0080},
we have taken $v_m$ to be a real basis vector, and regarded
$v_m$ and $\ve^{mn}v_n$ as independent. 
This effectively
doubles the number of modes and hence we need to
halve the
contribution from each mode. Thus for example the
contribution to the heat kernel
on $AdS_2$ from each of these basis vectors is now
given by  a half of \refb{esumvector}, \i.e.\
$1/4\pi a^2$ since the net contribution is shared between
$v_m$ and $\ve^{mn} v_n$.
There is no mixing between the modes described in
\refb{edefo1} and \refb{edefo2.5}; hence we can 
compute their contributions separately.
Substituting \refb{edefo1} into the action 
we get the kinetic
term to be
\ben \label{ern4.5}
&& -{1\over 2}\, \kappa_1 \, (E_1^2 + E_2^2) 
- {1\over 2}\, \sum_{i=3}^4 (\kappa_1 - 2 a^{-2})
(E_i^2 + \wt E_i^2) 
- {1\over 2}\, (\kappa_1 + 2 a^{-2}) (E_5^2 +\wt E_5^2)
\nn &&
+ 2 i a^{-1}\sqrt{\kappa_1} 
\left(E_1 \wt E_3 - E_2 E_3
\right)\nn
&\equiv& 
-{1\over 2} \kappa_1 \left(E_1^2 + E_2^2 +\sum_{i=3}^5
(E_i^2 +\wt E_i^2)\right) + {1\over 2} a^{-2}
\pmatrix{\vec E & \vec {\wt E}} \wh M \pmatrix{\vec E\cr
\vec{\wt E}}
\een
Eigenvalues of $\wh M$ defined through \refb{ern4.5}
are given by
\ben \label{erneigendis}
&& E_4: 2, \quad  \wt E_4:2, \quad E_5: -2, \quad \wt E_5: -2,
\quad (E_1, \wt E_3): 1\pm  i \sqrt{4l^2+4l-1}, \nn
&& (E_2, E_3): 1\pm i \sqrt{4l^2+4l-1}\, .
\een
For $l=0$ however the modes $E_3$, $\wt E_3$, $E_4$,
$\wt E_4$ carrying $\wh M$ eigenvalues
$2$, $2$, $2$, $2$ are absent due to the condition
mentioned below \refb{ebasis}. Finally the mode
\refb{edefo2.5} gives a kinetic term
\be \label{ern4.6}
-{1\over 2}\kappa_1 E_6^2\, .
\ee
Combining these results we get the net
contribution to the heat kernel from the discrete modes
to be 
\ben \label{ern5}
K^B_{(3)}(0;s) &=& {1\over 16\pi^2 a^4} \bigg[
\sum_{l=0}
^\infty \bigg\{ (2l+1) e^{-\bar s l (l+1)} \, 
\bigg( 2 e^{2\bar s} + 2 e^{-2\bar s}
+2 e^{\bar s(1 + i \sqrt{4l^2 + 4 l -1})}
+ 2 e^{\bar s(1 - i \sqrt{4l^2 + 4 l -1})} \bigg)\bigg\} 
\nn &&
  \qquad \qquad   \qquad \qquad - 4 \, e^{2\bar s}
\nn
&&  \qquad \qquad
+ 6\sum_{l=0}
^\infty  (2l+1) e^{-\bar s l (l+1)}  
\bigg]
\nn &=& {1\over 4\pi^2 a^4}
{\rm Im}\, 
\int_0^{e^{i\kappa}\times\infty} d\wt\lambda\, \wt\lambda\,
\tan\pi\wt\lambda\, e^{-\bar s \wt\lambda^2} e^{\bar s/4}
\bigg( e^{2\bar s} + e^{-2\bar s} 
+ e^{\bar s(1 + i \sqrt{4\wt\lambda^2 - 2})}
\nn &&  
\qquad \qquad \qquad \qquad 
+ e^{\bar s(1 - i \sqrt{4\wt\lambda^2 - 2})}+ 3\bigg)
- {1\over 4\pi^2 a^4} e^{2\bar s} 
\een
The first line represents the contribution from the
eigenvalues \refb{erneigendis} and the second line
represents the effect of removing the four $l=0$
modes. The third line represents the contribution
from the mode $E_6$ with kinetic term given in
\refb{ern4.6}. We can evaluate the integral by 
expanding the terms inside $(~)$ in the fourth and
fifth lines in a power series expansion in $\bar s$
and using \refb{esa2}. The result for the constant term in the
small $\bar s$ expansion of $K^B_{(3)}$ is:
\be \label{ekb3value}
K^B_{(3)}: -{5\over 24\pi^2 a^4}\, .
\ee

Next we turn to the ghost fields. 
The last term in \refb{emghost} describes mixing between the 
fields $b$
and $c^\nu$, but this has no effect on the determinant since
the mixing matrix has an upper triangular form. Thus we can
separately evaluate the contribution from the $(b,c)$ fields
and $(b^\mu, c^\nu)$ fields.
The contribution from
the $b$, $c$ ghosts associated with the $U(1)$ gauge field
is negative of that of two scalars. This gives the first
contribution from the ghosts:
\be \label{ern5.5}
K^{ghost}_{(1)} 
= -{1\over 2\pi^2 a^4} \, {\rm Im}\, 
\int_0^{e^{i\kappa}\times\infty} d\wt\lambda\, \wt\lambda\,
\tan\pi\wt\lambda\, 
\int_0^\infty d\lambda \, \lambda\, \tanh\pi\lambda\, 
e^{-\bar s\, (\lambda^2 +\wt\lambda^2)}\, .
\ee
For finding the contribution due to the $b_\mu$, $c_\mu$
ghosts associated with general coordinate invariance,
we expand them in modes:
\ben \label{eghdecomp}
b_\alpha &=& A {1\over \sqrt{\kappa_1}}\,
\p_\alpha u + B {1\over \sqrt{\kappa_1}}\,
\ve_{\alpha\beta}\p^\beta u, \nn
b_m &=& C {1\over \sqrt{\kappa_2}}\,
\p_m u + D {1\over \sqrt{\kappa_2}}\,
\ve_{mn} \p^n u, \nn
c_\alpha &=& E {1\over \sqrt{\kappa_1}}\,
\p_\alpha u + F {1\over \sqrt{\kappa_1}}\,
\ve_{\alpha\beta}\p^\beta u, \nn
c_m &=& G {1\over \sqrt{\kappa_2}}\,
\p_m u + H {1\over \sqrt{\kappa_2}}\, \ve_{mn} 
\p^n u\, .
\een
Substituting this into the first term in
\refb{emghost} we get the ghost kinetic term:
\be \label{eghac}
(\kappa_1 + \kappa_2 - 2 a^{-2}) (AE + BF)
+ (\kappa_1 + \kappa_2 + 2 a^{-2})(CG + DH)\, .
\ee
This gives the second contribution to the heat kernel
of the ghosts
\ben \label{ern6}
K^{ghost}_{(2)} &=& -{1\over 8\pi^2 a^4} 
\sum_{l=0}^\infty (2l+1) \int_0^\infty d\lambda \, 
\lambda\, \tanh\pi\lambda\, e^{-\bar s\, \lambda^2 -
{1\over 4} \bar s 
-\bar s l(l+1))}\bigg[4 e^{-2\bar s} +  4 e^{2\bar s}
\bigg] 
\nn
&=&  -{1\over 4\pi^2 a^4} 
\, {\rm Im}\, 
\int_0^{e^{i\kappa}\times\infty} d\wt\lambda\, \wt\lambda\,
\tan\pi\wt\lambda\, 
\int_0^\infty d\lambda \, \lambda\, \tanh\pi\lambda\, 
e^{-\bar s\, (\lambda^2 +\wt\lambda^2)}
\bigg[4 e^{-2\bar s} +  4 e^{2\bar s}
\bigg] \, .\nn
\een
We need to subtract from this the contribution due to
the absent modes $A$, $B$, $E$, $F$ for $l=0$.
This is given by
\be \label{ern7}
K^{ghost}_{(3)} = {1\over 2\pi^2 a^4} \int_0^\infty 
d\lambda \, \lambda\, \tanh\pi\lambda\, e^{-\bar s\, \lambda^2}
e^{2\bar s - {1\over 4} \bar s}\, .
\ee
Finally we need to include the contribution due to the
discrete modes where we take $b_m$ and $c_m$ to be
proportional to $v_m$. This gives the final contribution
to the ghost heat kernel:
\be \label{ern8}
K^{ghost}_{(4)}=
-{1\over 2\pi^2 a^4}  \, {\rm Im}\, 
\int_0^{e^{i\kappa}\times\infty} d\wt\lambda\, \wt\lambda\,
\tan\pi\wt\lambda\, e^{-\bar s\,\wt\lambda^2}
e^{-2\bar s + {1\over 4} \bar s}
\ee
The small $s$ expansion of \refb{ern5.5},
\refb{ern6}-\refb{ern8} can be found by standard method 
described above and we get
the following constant terms in the small $s$ expansion:
\ben \label{eghostvalue}
K^{ghost}_{(1)} &:& -{1\over 360\pi^2 a^4} \nn
K^{ghost}_{(2)} &:& -{91\over 90\pi^2 a^4} \nn
K^{ghost}_{(3)} &:& {5\over 12\pi^2 a^4} \nn
K^{ghost}_{(4)} &:& {5\over 12\pi^2 a^4} \, . \nn
\een
Adding all the contributions in \refb{ekb1value},
\refb{ekb2value}, \refb{ekb3value}
and \refb{eghostvalue} we get the
total contribution to the constant term in the small $\bar s$ 
expansion of the heat kernel
\be \label{ekbtot}
K^B_0 =  {53\over 90 \pi^2 a^4}\, .
\ee

Next we turn to the contribution due to the zero modes.
We first need to remove from $K^B_0$ the contribution due to the
zero modes and then compute the contribution to $Z_{AdS_2}$ from
integration over the zero modes. The combined effect of these is
encoded in the $\sum_r (\beta_r-1) \bar K^r(0)$ term in
\refb{edeltasbh}. Thus we need to compute $\beta_r$ and 
$\bar K^r(0)$
due to various zero modes. The relevent zero modes come
from the gauge field $A_\mu$  and the
metric $h_{\mu\nu}$  which we shall label by
$r=v$ and $r=m$ respectively.
We can identify these zero modes by examining 
the discrete mode contribution \refb{ern5} to $K(0;s)$. 
First of all note that for $l=0$
the $(2l+1) 
e^{\bar s (-l(l+1)+ 1 + i\sqrt{4l^2+4l-1})}$ term becomes a constant
signalling the presence of a zero mode. Working backwards we can
identify them as due to the modes $E_1$, $E_2$ of the gauge field
$A_\mu$. Since this term gives a contribution of $1/8\pi^2 a^4$
to $K(0;s)$ we have $\bar K^v(0)=1/8\pi^2 a^4$. But we have seen
that $\beta_v=1$ for the gauge fields and hence these zero modes
do not contribute to $\sum_r (\beta_r-1) \bar K^r(0)$. The other
zero modes come from the $3 (2l+1) e^{-l(l+1)\bar s}$ term in 
\refb{ern5} in the $l=0$ sector
and the $(2l+1) e^{-l(l+1)\bar s + 
2\bar s}$ term in the $l=1$ sector. The former
corresponds to the modes represented by $E_6$ while the latter correspond
to the modes represented by $E_5$, $\wt E_5$. By examining
\refb{edefo1}, \refb{edefo2.5} 
we see that both are modes of the metric. Physically
the former represent deformations 
associated with the asymptotic Virasoro
symmetries of the $AdS_2$ metric, while the latter are the zero
modes of the $SU(2)$ gauge fields obtained from the dimensional
reduction on $S^2$. The total
contribution from these modes to $K(0,s)$ is given by $6/8\pi^2a^4$
and hence we have $\bar K^m(0)= 3/4\pi^2 a^4$.

To complete the analysis we need to compute $\beta_m$ associated
with the metric deformation. For this we proceed as in
\refb{eap2}, \refb{eap3}. The path integral over the metric fluctuation
$h_{\mu\nu}$ is normalized as
\be \label{ebp2}
\int [Dh_{\mu\nu}] \exp\left[- \int d^4 x \, \sqrt{\det g} \, 
g^{\mu\nu} g^{\rho\sigma} h_{\mu\rho} h_{\nu\sigma}
\right] = 1\, ,
\ee
\i.e.\
\be \label{ebp3}
\int [Dh_{\mu\nu}] \exp\left[- \int d^4 x \, 
\sqrt{\det g^{(0)}} \, 
g^{(0)\mu\nu} g^{(0)\rho\sigma}h_{\mu\rho} h_{\nu\sigma}
\right] = 1\, .
\ee
Thus the correctly normalized integration
measure, up to an $a$ independent constant, is 
$\prod_{x,(\mu\nu)} dh_{\mu\nu}(x)$.
We now note that the zero modes are associated
with diffeomorphisms with non-normalizable parameters:
$h_{\mu\nu}\propto D_\mu\xi_\nu + D_\nu\xi_\mu$, with
the diffeomorphism parameter $\xi^\mu(x)$
having $a$ independent integration range. Thus the $a$
dependence of the integral over the metric zero modes
can be found by finding the Jacobian from the change
of variables from $h_{\mu\nu}$ to $\xi^\mu$. Lowering
of the index of $\xi^\mu$ gives a factor of $a^2$, leading to
a factor of $a^2$ per zero mode. Thus we have $\beta_m=2$
and hence the contribution to $\sum_r (\beta_r-1) \bar K^r(0)$
from the zero modes of the metric is given by
\be \label{emetzero}
(2-1) {3\over 4\pi^2 a^4} = {3\over 4\pi^2 a^4} \, .
\ee
Adding \refb{emetzero} to \refb{ekbtot} and substituting this into
\refb{edeltasbh}, we get the net contribution
to the logarithmic correction to the entropy of an extremal
Reissner-Nordstrom black hole:
\be \label{ekblog}
\Delta S_{BH} = - {241\over 45} \ln A_H\, .
\ee

If in addition the theory contains $n_S$ minimally coupled
massless scalar, $n_F$ minimally coupled massless Dirac
fermion and $n_V$ minimally coupled Maxwell fields, then
the total logarithmic correction to $S_{BH}$ is given
by the sum of \refb{enetlogminimal} and \refb{ekblog}:
\be \label{ebhplus}
\Delta S_{BH} = -{1\over 180}(964 +
n_S + 62 n_V + 11 n_F)\, \ln A_H\, .
\ee

\sectiono{Half BPS black holes in
pure $\NN=2$ supergravity} \label{sfermion}

We shall now consider half BPS black holes in pure
$\NN=2$ supergravity\cite{ferraranieu}. This requires adding to the
Einstein-Maxwell action described in the previous section
the fermionic action of a pair of Majorana spinors 
$\psi_\mu$ and $\vp_\mu$ satsifying
\be \label{ecccond}
\bar\psi_\mu = \psi_\mu^T \wt C, \qquad \bar\vp_\mu =
\vp_\mu^T \wt C\, ,
\ee
for each $\mu$. Here $\wt C$ is the charge conjugation
operator defined in \refb{echargec}. 
The quadratic part of the
fermionic action is given by
\ben \label{sfermiaction}
S_f &=& \int d^4 x \, \sqrt{\det g} \, \LL_f, \nn
\LL_f &=& -{1\over 2}
\bar \psi_{\mu}\gamma^{\mu\nu\rho} D_\nu \psi_\rho 
-{1\over 2}
\bar \vp_{\mu}\gamma^{\mu\nu\rho} D_\nu \vp_\rho 
+\half F^{\mu\nu}\bar \psi_\mu \vp_\nu + {1\over 4} 
F_{\rho\sigma} 
\bar \psi_\mu \gamma^{\mu\nu\rho\sigma} \vp_\nu \nn &&
- \half F^{\mu\nu}\bar \vp_\mu \psi_\nu - {1\over 4}  
F_{\rho\sigma} 
\bar \vp_\mu \gamma^{\mu\nu\rho\sigma} \psi_\nu\, .
\een
For quantization we need to add to this a gauge fixing term
\be \label{efergauge}
\LL_{gf} = {1\over 4} \bar\psi_\mu \gamma^\mu \gamma^\nu D_\nu 
\gamma^\rho \psi_\rho
+ {1\over 4} \bar\vp_\mu \gamma^\mu \gamma^\nu D_\nu 
\gamma^\rho \vp_\rho\, ,
\ee
and a ghost action
\be \label{emghostfermi}
\LL_{ghost} = 
\sum_{r=1}^2
\left[\bar{\tilde b_r} \, \Gamma^\mu D_\mu \tilde c_r
+ \bar{\tilde e_r} \, \Gamma^\mu D_\mu \tilde e_r\right]\, .
\ee
Here for each $r$ ($r=1,2$) $\tilde b_r$, $\tilde c_r$ and
$\tilde e_r$ represent spin half bosonic ghosts. The two values
of $r$ correspond to two local supersymmetries which the theory
possesses, $\tilde b_r$ and $\tilde c_r$ are the standard Fadeev-Popov
ghosts, and $\tilde e_r$ is a special ghost originating due to the unusual
nature of the gauge fixing terms we have used\cite{1005.3044}.

The sum of $\LL_f$ and $\LL_{gf}$, evaluated in the background
\refb{ernk1.5}, can be expressed as
\be \label{ermf1}
\LL_f +\LL_{gf} = -{1\over 2} \left[\bar \psi^\alpha \KK^{(1)}_\alpha 
+ \bar \psi^m \KK^{(2)}_m 
+ \bar \vp^\alpha \KK^{(3)}_\alpha 
+ \bar \vp^m \KK^{(4)}_m \right]
\ee
where
\ben \label{edefrnk}
\KK^{(1)}_\alpha &=&  - {1\over 2} \gamma^n 
(\not \hskip -4pt D_{S^2} +\sigma_3 \, \not \hskip -4pt D_
{AdS_2}) \gamma_\alpha \psi_n
 - {1\over 2}\gamma^\beta\left(\not \hskip -4pt D_{S^2} 
+\sigma_3 \, \not \hskip -4pt D_{AdS_2}\right) \gamma_\alpha
\psi_\beta + \hab \, i\, 
a^{-1}\ve_\alpha^{~\beta} \sigma_3 \tau_3 \vp_\beta
\nn
\KK^{(2)}_m &=& 
- {1\over 2} \gamma^\beta (\not \hskip -4pt D_{S^2} 
+\sigma_3 \, \not \hskip -4pt D_{AdS_2}) \gamma_m\psi_\beta
- {1\over 2}\gamma^n\left(\not \hskip -4pt D_{S^2} 
+\sigma_3 \, \not \hskip -4pt D_{AdS_2}\right) \gamma_m
\psi_n
-  \hab \, i\,   a^{-1}\ve_m^{~n} \vp_n \, \nn
\KK^{(3)}_\alpha &=&  - {1\over 2} \gamma^n 
(\not \hskip -4pt D_{S^2} +\sigma_3 \, \not \hskip -4pt D_
{AdS_2}) \gamma_\alpha \vp_n
 - {1\over 2}\gamma^\beta\left(\not \hskip -4pt D_{S^2} 
+\sigma_3 \, \not \hskip -4pt D_{AdS_2}\right) \gamma_\alpha
\vp_\beta - \hab \, i\, 
a^{-1}\ve_\alpha^{~\beta} \sigma_3 \tau_3 \psi_\beta
\nn
\KK^{(4)}_m &=& 
- {1\over 2} \gamma^\beta (\not \hskip -4pt D_{S^2} 
+\sigma_3 \, \not \hskip -4pt D_{AdS_2}) \gamma_m\vp_\beta
- {1\over 2}\gamma^n\left(\not \hskip -4pt D_{S^2} 
+\sigma_3 \, \not \hskip -4pt D_{AdS_2}\right) \gamma_m
\vp_n
+  \hab \, i\,   a^{-1}\ve_m^{~n} \psi_n \, .\nn
\een
We now expand the fermion fields in the basis described in
appendix \ref{sbasis}. As in the case of the bosonic fields
we can work at fixed values of $l$ and $\lambda$. Let 
$\chi$ denote the product of $\chi^+_{l,m}$ or $\eta^+_{l,m}$
defined in \refb{ed2} and $\chi_k^+(\lambda)$ or $\eta_k^+(\lambda)$
defined in \refb{ed2a}. Then $\chi$ satisfies
\be \label{es90}
\not \hskip-4pt D_{S^2} \chi= i\zet_1 \, \chi,
\quad \not \hskip-4pt D_{AdS_2}\chi= i\zet_2 \, \chi,
\quad \zet_1=(l+1)/a \ge 1/a, \quad \zet_2=\lambda/a \ge 0 \, .
\ee
Furthermore, using eqs.\refb{es90} and
the representation of the $\gamma$-matrices
given in \refb{egam1} we get
\ben \label{erelations}
&& \ve_\alpha^{~\beta}\gamma_\beta = i \sigma_3\gamma_\alpha,
\quad 
\ve_{\alpha\beta} D^\beta\chi = -i \sigma_3 D_\alpha\chi
-\zet_1 \sigma_3 \gamma_\alpha\chi\, , \nn &&
\ve_m^{~n}\gamma_n = i \tau_3\gamma_m,
\quad  
\ve_{mn} D^n\chi = -i\tau_3 D_m\chi - \zet_2 
\tau_3 \sigma_3 \gamma_m\chi\, .
\een
The basis functions involving $\chi^{-}_{l,m}$ and $\eta^-_{l,m}$
will be represented as $\sigma_3\chi^{+}_{l,m}$ and 
$\sigma_3 \eta^+_{l,m}$ respectively; thus we shall not include
them separately. Similarly the basis functions 
$\chi^-_k(\lambda)$ and $\eta^-_k(\lambda)$ will be
represented as $\tau_3\chi^+_k(\lambda)$ and 
$\tau_3\eta^+_k(\lambda)$.
We now introduce the modes of $\psi_\mu$ and $\vp_\mu$
via the expansion
\ben \label{es89}
\psi_\alpha &=& b_1 \gamma_\alpha\chi + b_2 \sigma_3
\gamma_\alpha\chi + b_3 D_\alpha\chi + b_4 \sigma_3  D_\alpha\chi 
\nn &&
+ b_1' \gamma_\alpha \tau_3 \chi + b_2' \sigma_3
\gamma_\alpha \tau_3 \chi + b_3' \tau_3 D_\alpha \chi 
+ b_4' \sigma_3  \tau_3  D_\alpha\chi 
\nn
\psi_m &=& c_1 \gamma_m\chi + c_2 \sigma_3
\gamma_m\chi + c_3 \sigma_3 D_m\chi + c_4  
D_m\chi 
\nn &&
+ c_1' \gamma_m \tau_3 \chi + c_2' \sigma_3
\gamma_m  \tau_3\chi + c_3' \sigma_3  \tau_3
D_m\chi + c_4'  \tau_3  
D_m\chi \nn
\vp_\alpha &=& g_1 \gamma_\alpha\chi + g_2 \sigma_3
\gamma_\alpha\chi + g_3 D_\alpha\chi + g_4 \sigma_3  D_\alpha\chi 
\nn &&
+ g_1' \gamma_\alpha \tau_3 \chi + g_2' \sigma_3
\gamma_\alpha \tau_3 \chi + g_3' \tau_3 D_\alpha \chi 
+ g_4' \sigma_3  \tau_3  D_\alpha\chi 
\nn
\vp_m &=& h_1 \gamma_m\chi + h_2 \sigma_3
\gamma_m\chi + h_3 \sigma_3 D_m\chi + h_4  
D_m\chi 
\nn &&
+ h_1' \gamma_m \tau_3 \chi + h_2' \sigma_3
\gamma_m  \tau_3\chi + h_3' \sigma_3  \tau_3
D_m\chi + h_4'  \tau_3  
D_m\chi
\een
where $b_i, b_i',c_i,c_i',g_i, g_i',h_i,h_i'$ are constants.
Substituting this into \refb{edefrnk} we get
\ben \label{ekkans}
\KK^{(1)}_\alpha &=& B_1 \gamma_\alpha\chi + B_2 \sigma_3
\gamma_\alpha\chi + B_3 D_\alpha\chi + B_4 \sigma_3  D_\alpha\chi 
\nn &&
+ B_1' \gamma_\alpha \tau_3 \chi + B_2' \sigma_3
\gamma_\alpha \tau_3 \chi + B_3' \tau_3 D_\alpha \chi 
+ B_4' \sigma_3  \tau_3  D_\alpha\chi 
\nn
\KK^{(2)}_m &=& C_1 \gamma_m\chi + C_2 \sigma_3
\gamma_m\chi + C_3 \sigma_3 D_m\chi + C_4  
D_m\chi 
\nn &&
+ C_1' \gamma_m \tau_3 \chi + C_2' \sigma_3
\gamma_m  \tau_3\chi + C_3' \sigma_3  \tau_3
D_m\chi + C_4'  \tau_3  
D_m\chi\nn
\KK^{(3)}_\alpha &=& G_1 \gamma_\alpha\chi + G_2 \sigma_3
\gamma_\alpha\chi + G_3 D_\alpha\chi + G_4 \sigma_3  D_\alpha\chi 
\nn &&
+ G_1' \gamma_\alpha \tau_3 \chi + G_2' \sigma_3
\gamma_\alpha \tau_3 \chi + G_3' \tau_3 D_\alpha \chi 
+ G_4' \sigma_3  \tau_3  D_\alpha\chi 
\nn
\KK^{(4)}_m &=& H_1 \gamma_m\chi + H_2 \sigma_3
\gamma_m\chi + H_3 \sigma_3 D_m\chi + H_4  
D_m\chi 
\nn &&
+ H_1' \gamma_m \tau_3 \chi + H_2' \sigma_3
\gamma_m  \tau_3\chi + H_3' \sigma_3  \tau_3
D_m\chi + H_4'  \tau_3  
D_m\chi
\een
where
\ben \label{es91}
B_1 &=& -i\zeta_1 b_1 + {1\over 2}\zeta_1^2 b_3 
+ {1\over 2}\zeta_1 \zeta_2 b_4 
+ i\zeta_1 c_1 -{1\over 2}\zeta_1 \zeta_2 c_3 + 
{1\over 2}\left(\zeta_2^2 + {1\over a^2} \right) c_4 
- \hab a^{-1} g_1' -\hab \, i \, \zeta_1 a^{-1} g_3'
 \nn
B_2 &=& i\zeta_1 b_2 
+{1\over 2}\zet_1 \zet_2 b_3
-{1\over 2}\zeta_1^2 b_4 
+ i\zet_1 c_2- {1\over 2}\left(\zeta_2^2 
+ {1\over a^2} \right) c_3
-{1\over 2}\zet_1 \zet_2 c_4 
-\hab\,  a^{-1} g_2' -  \hab\, i\, \zeta_1 a^{-1} g_4' \nn
B_3 &=& i\zet_2 b_4  - 2c_1 - i\zet_2 c_3 
+\hab\, a^{-1} g_3' \nn
B_4 &=& i\zet_2 b_3 -2 c_2 -i\zet_2 c_4 
+\hab\, a^{-1} g_4' \nn
C_1 &=& -i\zet_2 b_2 + {1\over 2}\left(\zet_1^2 -{1\over a^2}
\right) b_3 +{1\over 2}\zet_1\zet_2 b_4 
-i\zet_2 c_2 -{1\over 2}\zet_1 \zet_2 c_3 +
{1\over 2}\zet_2^2 c_4 
-\hab\, a^{-1} h_1' + \hab\, i\, \zeta_2 a^{-1} h_3'
 \nn
C_2 &=& i\zet_2 b_1 -{1\over 2}\zet_1\zet_2 b_3 
+{1\over 2}\left( \zet_1^2 -{1\over a^2}\right) b_4
-i\zet_2 c_1 +{1\over 2}\zet_2^2 c_3 +{1\over 2} \zet_1
\zet_2 c_4 
-\hab\, a^{-1} h_2' +\hab\, i\, \zeta_2 a^{-1} h_4'
\nn
C_3 &=& 2 b_2 + i\zet_1 b_4 - i\zet_1 c_3 
-\hab\, a^{-1} h_3'\nn
C_4 &=& -2 b_1 -i\zet_1 b_3 + i\zet_1 c_4 
-\hab\, a^{-1} h_4'\nn
B'_1 &=& -i\zeta_1 b'_1 + {1\over 2}\zeta_1^2 b'_3 
- {1\over 2}\zeta_1 \zeta_2 b'_4 
+ i\zeta_1 c'_1 +{1\over 2}\zeta_1 \zeta_2 c'_3 + 
{1\over 2}\left(\zeta_2^2 + {1\over a^2} \right) c'_4 
-\hab\,  a^{-1} g_1 -  \hab\, i\,  \zeta_1 a^{-1} g_3
 \nn
B'_2 &=& i\zeta_1 b'_2 
-{1\over 2}\zet_1 \zet_2 b'_3
-{1\over 2}\zeta_1^2 b'_4 
+ i\zet_1 c'_2 -{1\over 2}\left(\zeta_2^2 
+ {1\over a^2} \right) c'_3
+{1\over 2}\zet_1 \zet_2 c'_4 
-\hab\,  a^{-1} g_2 -  \hab\, i\,  \zeta_1 a^{-1} g_4 \nn
B'_3 &=& -i\zet_2 b'_4  - 2c'_1 + i\zet_2 c'_3 
+\hab\,  a^{-1} g_3 \nn
B'_4 &=& -i\zet_2 b'_3 -2 c'_2 +i\zet_2 c'_4 
+\hab\,  a^{-1} g_4 \nn
C'_1 &=& i\zet_2 b'_2 + {1\over 2}\left(\zet_1^2 -{1\over a^2}
\right) b'_3 -{1\over 2}\zet_1\zet_2 b'_4 
+i\zet_2 c'_2 +{1\over 2}\zet_1 \zet_2 c'_3 +
{1\over 2}\zet_2^2 c'_4 
-\hab\,  a^{-1} h_1 - \hab\, i\, \zeta_2 a^{-1} h_3
 \nn
C'_2 &=& -i\zet_2 b'_1 +{1\over 2}\zet_1\zet_2 b'_3 
+{1\over 2}\left( \zet_1^2 -{1\over a^2}\right) b'_4
+i\zet_2 c'_1 +{1\over 2}\zet_2^2 c'_3 -{1\over 2} \zet_1
\zet_2 c'_4 
-\hab\,  a^{-1} h_2 -  \hab\, i \zeta_2 a^{-1} h_4
\nn
C'_3 &=& 2 b'_2 + i\zet_1 b'_4 - i\zet_1 c'_3 
-\hab\,  a^{-1} h_3\nn
C'_4 &=& -2 b'_1 -i\zet_1 b'_3 + i\zet_1 c'_4 
-\hab\,  a^{-1} h_4\nn
G_1 &=& -i\zeta_1 g_1 + {1\over 2}\zeta_1^2 g_3 
+ {1\over 2}\zeta_1 \zeta_2 g_4 
+ i\zeta_1 h_1 -{1\over 2}\zeta_1 \zeta_2 h_3 + 
{1\over 2}\left(\zeta_2^2 + {1\over a^2} \right) h_4 
+\hab\, a^{-1} b_1' +\hab\, i\,  \zeta_1 a^{-1} b_3'
 \nn
G_2 &=& i\zeta_1 g_2 
+{1\over 2}\zet_1 \zet_2 g_3
-{1\over 2}\zeta_1^2 g_4 
+ i\zet_1 h_2- {1\over 2}\left(\zeta_2^2 
+ {1\over a^2} \right) h_3
-{1\over 2}\zet_1 \zet_2 h_4 
+  \hab\,  a^{-1} b_2' +\hab\, i\, \zeta_1 a^{-1} b_4' \nn
G_3 &=& i\zet_2 g_4  - 2h_1 - i\zet_2 h_3 
-\hab\,  a^{-1} b_3' \nn
G_4 &=& i\zet_2 g_3 -2 h_2 -i\zet_2 h_4 
-\hab\,  a^{-1} b_4' \nn
H_1 &=& -i\zet_2 g_2 + {1\over 2}\left(\zet_1^2 -{1\over a^2}
\right) g_3 +{1\over 2}\zet_1\zet_2 g_4 
-i\zet_2 h_2 -{1\over 2}\zet_1 \zet_2 h_3 +
{1\over 2}\zet_2^2 h_4 
+  \hab\, a^{-1} c_1' -\hab\, i\,  \zeta_2 a^{-1} c_3'
 \nn
H_2 &=& i\zet_2 g_1 -{1\over 2}\zet_1\zet_2 g_3 
+{1\over 2}\left( \zet_1^2 -{1\over a^2}\right) g_4
-i\zet_2 h_1 +{1\over 2}\zet_2^2 h_3 +{1\over 2} \zet_1
\zet_2 h_4 
+  \hab\, a^{-1} c_2' -\hab\, i\,  \zeta_2 a^{-1} c_4'
\nn
H_3 &=& 2 g_2 + i\zet_1 g_4 - i\zet_1 h_3 +  
\hab\, a^{-1} c_3'\nn
H_4 &=& -2 g_1 -i\zet_1 g_3 + i\zet_1 h_4 +  
\hab\,  a^{-1} c_4'\nn
G'_1 &=& -i\zeta_1 g'_1 + {1\over 2}\zeta_1^2 g'_3 
- {1\over 2}\zeta_1 \zeta_2 g'_4 
+ i\zeta_1 h'_1 +{1\over 2}\zeta_1 \zeta_2 h'_3 + 
{1\over 2}\left(\zeta_2^2 + {1\over a^2} \right) h'_4 
+  \hab\, a^{-1} b_1 +\hab\, i\, \zeta_1 a^{-1} b_3
 \nn
G'_2 &=& i\zeta_1 g'_2 
-{1\over 2}\zet_1 \zet_2 g'_3
-{1\over 2}\zeta_1^2 g'_4 
+ i\zet_1 h'_2 -{1\over 2}\left(\zeta_2^2 
+ {1\over a^2} \right) h'_3
+{1\over 2}\zet_1 \zet_2 h'_4 
+ \hab\,  a^{-1} b_2 +\hab\, i\,  \zeta_1 a^{-1} b_4 \nn
G'_3 &=& -i\zet_2 g'_4  - 2h'_1 + i\zet_2 h'_3 
-  \hab \, a^{-1} b_3 \nn
G'_4 &=& -i\zet_2 g'_3 -2 h'_2 +i\zet_2 h'_4 
-  \hab\,  a^{-1} b_4 \nn
H'_1 &=& i\zet_2 g'_2 + {1\over 2}\left(\zet_1^2 -{1\over a^2}
\right) g'_3 -{1\over 2}\zet_1\zet_2 g'_4 
+i\zet_2 h'_2 +{1\over 2}\zet_1 \zet_2 h'_3 +
{1\over 2}\zet_2^2 h'_4 
+  \hab\,  a^{-1} c_1 +\hab\, i\,  \zeta_2 a^{-1} c_3
 \nn
H'_2 &=& -i\zet_2 g'_1 +{1\over 2}\zet_1\zet_2 g'_3 
+{1\over 2}\left( \zet_1^2 -{1\over a^2}\right) g'_4
+i\zet_2 h'_1 +{1\over 2}\zet_2^2 h'_3 -{1\over 2} \zet_1
\zet_2 h'_4 
+ \hab\,  a^{-1} c_2 +\hab\, i\,  \zeta_2 a^{-1} c_4
\nn
H'_3 &=& 2 g'_2 + i\zet_1 g'_4 - i\zet_1 h'_3 
+\hab\,  a^{-1} c_3\nn
H'_4 &=& -2 g'_1 -i\zet_1 g'_3 + i\zet_1 h'_4 
+  \hab\,  a^{-1} c_4 \, . \nn
\een
We can express this as
\be \label{epm1}
\pmatrix{\vec B\cr \vec C\cr \vec G\cr \vec H\cr 
\vec B'\cr \vec C'
\cr \vec G'\cr \vec H'}
= \MM \pmatrix{\vec b\cr \vec c\cr \vec g\cr \vec h\cr
\vec b'\cr \vec c'\cr \vec g' \cr \vec h'}\, ,
\ee
where $\MM$ is a $32\times 32$ matrix. If we introduce the
matrix $\MM_1$ through
\be \label{epm2}
\MM^2 = -(\zeta_1^2 + \zeta_2^2) I_{32} + a^{-2} \MM_1\, ,
\ee
then the
fermionic contribution to the heat kernel from the $l\ge 1$,
\i.e.\ $\zeta_1\ge 2/a$ modes will be given by
\be \label{ekf1}
K^f_{(1)}(0;s) =  -{1\over 8\pi^2 a^4} \, 
\sum_{l=1}^\infty (l+1) \, 
\int_0^\infty d\lambda \, \lambda\, \coth\pi\lambda\, 
e^{-\bar s\, \left(\left(l+1\right)^2 +\lambda^2
\right)}
\sum_{n=0}^\infty \, {\bar s^n \over n!} \, Tr(\MM_1^n)\, .
\ee
Note the normalization factor 1/8 instead of 1 as in
\refb{ediracf}. A factor of 1/4 can be traced
to the fact that in the analog of
\refb{erelchi} we should no longer include the 
$\chi^-$'s or $\eta^-$'s in the
sum since in the basis of expansion \refb{es89},
\refb{ekkans} we have included, besides $\chi$, the
states $\sigma_3\chi$, $\tau_3\chi$ and $\sigma_3
\tau_3\chi$. Another factor of 1/2 arises from the fact that 
we are dealing
with Majorana fermions instead of Dirac fermions.

In \refb{ekf1} we have not included the $l=0$
contribution. 
This is due to the fact that
for $l=0$, \i.e.\ $\zeta_1=a^{-1}$ the modes 
$D_\alpha\chi$
and $\gamma_\alpha\chi$ are related by \refb{eindep}.
Thus we can set $b_3=b_4=b_3'=b_4'=
g_3=g_4=g_3'=g_4'=0$ and replace
the expressions for $B_1$, $B_2$, $B'_1$, $B'_2$,
$G_1$, $G_2$, $G'_1$, $G'_2$
by those of $B_1 + i B_3/2a$, $B_2+iB_4/2a$, 
$B_1' + i B_3'/2a$, $B_2'+iB_4'/2a$,
$G_1 + i G_3/2a$, $G_2+iG_4/2a$, 
$G_1' + i G_3'/2a$, $G_2'+iG_4'/2a$ respectively.
This gives a $24\times 24$ matrix $\wt \MM$ 
relating
$(B_1, B_2, B_1', B_2', C_1,\cdots C_4, C_1',\cdots C_4',
G_1, G_2, G_1', G_2', H_1,\cdots H_4, H_1',\cdots H_4')$
to 
$(b_1, b_2, b_1', b_2', c_1,\cdots c_4, c_1',\cdots c_4',
g_1, g_2, g_1', g_2', h_1,\cdots h_4, h_1',\cdots h_4')$.
Let us introduce the matrix $\wt \MM_1$ via:
\be \label{epm2a}
\wt\MM^2 = -(a^{-2} + \zeta_2^2) I_{24} + 
a^{-2} \wt\MM_1\, .
\ee
Then the contribution from the $l=0$ modes will be
given by
\be \label{ekf2}
K^f_{(2)}(0;s) =  -{1\over 8\pi^2 a^4} \, 
\int_0^\infty d\lambda \, \lambda\, \coth\pi\lambda\, 
e^{-\bar s\, \left(1 +\lambda^2
\right)}
\sum_{n=0}^\infty \, {\bar s^n \over n!} \, Tr(\wt\MM_1^n)\, .
\ee

We can now write
\be \label{ekf1kf2}
K^f_{(1)}(0;s) + K^f_{(2)}(0;s) = \wt 
K^f_{(1)}(0;s) + \wt K^f_{(2)}(0;s)\, ,
\ee
where
\ben \label{epm4}
\wt K^f_{(1)}(0;s) &=&  -{1\over 8\pi^2 a^4} \,
\sum_{l=0}^\infty (l+1) \, 
\int_0^\infty d\lambda \, \lambda\, \coth\pi\lambda\, 
e^{-\bar s\, \left(\left(l+1\right)^2 +\lambda^2
\right)}
\sum_{n=0}^\infty \, {\bar s^n \over n!} \, Tr(\MM_1^n)\nn
&=& -{1\over 8\pi^2 a^4} \, {\rm Im}\, 
\int_0^{e^{i\kappa}\times\infty} d\wt\lambda\, \wt\lambda\,
\cot\pi\wt\lambda\, 
\int_0^\infty d\lambda \, \lambda\, \coth\pi\lambda\, 
e^{-\bar s\, (\lambda^2 +\wt\lambda^2)}
\sum_{n=0}^\infty \, {\bar s^n \over n!} \, Tr(\MM_1^n)\, ,
\nn
\een
and
\be \label{ekf2new}
\wt K^f_{(2)}(0;s) =  -{1\over 8\pi^2 a^4} \, 
\int_0^\infty d\lambda \, \lambda\, \coth\pi\lambda\, 
e^{-\bar s\, \left(1 +\lambda^2
\right)}
\sum_{n=0}^\infty \, {\bar s^n \over n!} \, \left[
Tr(\wt\MM_1^n) - Tr(\MM_1^n)|_{l=0}\right]\, .
\ee

Finally we have to include the contribution from the
discrete modes obtained by taking $\psi_m$ to be a
linear combination of the product of the
modes given in
\refb{eadd1} and \refb{ed2}. 
The contribution from these modes may be
analyzed by setting $\lambda=i$ \i.e.\ $\zeta_2=i/a$,
$b_i=b_i'=g_i=g_i'=0$ for $1\le i\le 4$,
and $c_{i+2}=2ac_i$, $c'_{i+2}=2ac'_i$, 
$h_{i+2}=2ah_i$, $h'_{i+2}=2ah'_i$ for $i=1,2$
in
\refb{es91}. Eq.\refb{es91} now gives $B_i=B_i'=G_i=G_i'=0$ 
for $1\le i\le 4$,
and $C_{i+2}=2aC_i$, $C'_{i+2}=2aC'_i$, 
$H_{i+2}=2aH_i$, $H'_{i+2}=2aH'_i$ for $i=1,2$, and we
get a $8\times 8$ matrix $\wh\MM$ that relates the constants
$C_i, C_i'$, $H_i, H_i'$ to $c_i$, $c_i'$, $h_i$, $h_i'$ for
$i=1,2$. 
We again introduce the matrix $\wh\MM_1$ via
\be \label{epm2ahat}
\wh\MM^2 = -(-a^{-2} + \zeta_1^2) I_{8} + 
a^{-2} \wh\MM_1\, .
\ee
Then the contribution to the heat kernel from the fermionic
discrete modes will be given by:
\ben \label{ekf3}
K^f_{(3)}(0;s) &=&  
-{1\over 8\pi^2 a^4} \, \sum_{l=0}^\infty (l+1) \,
e^{-\bar s(l+1)^2 + \bar s} 
\sum_{n=0}^\infty \, {\bar s^n \over n!} \, Tr(\wh\MM_1^n)\nn
&=& -{1\over 8\pi^2 a^4} \, 
{\rm Im}\, 
\int_0^{e^{i\kappa}\times\infty} d\wt\lambda\, \wt\lambda\,
\cot\pi\wt\lambda\, 
e^{\bar s\, \left(1 -\wt\lambda^2
\right)}
\sum_{n=0}^\infty \, {\bar s^n \over n!} \, Tr(\wh\MM_1^n)
\, .
\een

Explicit computation gives
\ben \label{epm3}
Tr(\MM_1) &=& -32\nn
Tr(\MM_1^2) &=& 128 + 64 (l+1)^2 + 64 \lambda^2\nn
Tr(\MM_1^3) &=& -512 - 384 (l+1)^2 - 384 \lambda^2\nn
Tr(\MM_1^4) &=& 2048 + 2048 (l+1)^2 + 256 (l+1)^4 
+ 2048 \lambda^2 + 512 (l+1)^2 \lambda^2 + 256 \lambda^4\, ,
\nn
\een
\ben \label{epm3a}
Tr(\wt\MM_1) &=& -32\nn
Tr(\wt\MM_1^2) &=& 192 + 64 \lambda^2\nn
Tr(\wt\MM_1^3) &=& -896 - 384 \lambda^2\nn
Tr(\wt\MM_1^4) &=& 4352 + 2560 \lambda^2 + 256 \lambda^4\, ,
\nn
\een
and
\ben \label{epm3b}
Tr(\wh\MM_1) &=& -16\nn
Tr(\wh\MM_1^2) &=& 32 + 32 (l+1)^2\nn
Tr(\wh\MM_1^3) &=& -64 - 192 (l+1)^2 \nn
Tr(\wh\MM_1^4) &=&128 + 768 (l+1)^2 + 128 (l+1)^4 \, .
\nn
\een
Substituting these into eqs.\refb{epm4}, \refb{ekf2new} and 
\refb{ekf3}
and using eqs.\refb{esa1a}, \refb{esa2a} we get the following
constant terms in the small $\bar s$ expansions of the heat
kernels:
\be \label{epm5}
\wt K_{(1)}^f: {11\over 180\pi^2 a^4}\, ,
\ee
\be \label{epm5a}
\wt K_{(2)}^f: -{5\over 12\pi^2 a^4}\, ,
\ee
and
\be \label{epm5ab}
K_{(3)}^f: -{5\over 12\pi^2 a^4}\, .
\ee
Finally the six Majorana
ghost fields give a contribution equal to that of
three minimally coupled Dirac fermions but with opposite sign.
Thus using \refb{ekffin} we get the constant term in the heat kernel
from the ghost fields to be:
\be \label{epm6}
K^f_{ghost} : -{11\over 240\pi^2 a^4}
\ee
Adding up the contributions \refb{epm5}-\refb{epm6} 
we get the total fermionic
contribution to the constant term in $K(0;s)$:
\be \label{ekftot}
K^f_0 = -{589\over 720\pi^2 a^4}\, .
\ee

To this we have to add the extra contribution due to the zero modes.
These modes arise in the sector containing the discrete
modes with $l=0$. The kinetic operator 
in this sector is represented by the matrix $\wh\MM$
defined above \refb{epm2ahat}. Explicit computation
shows that for $l=0$, \i.e.\ $\zeta_1=1$ this matrix has
the form:
\be \label{epm}
\left(
\begin{array}{cccccccc}
 -i & 0 & 0 & 0 & 0 & 0 & 1 & 0 \\
 0 & i & 0 & 0 & 0 & 0 & 0 & 1 \\
 0 & 0 & -i & 0 & 1 & 0 & 0 & 0 \\
 0 & 0 & 0 & i & 0 & 1 & 0 & 0 \\
 0 & 0 & -1 & 0 & -i & 0 & 0 & 0 \\
 0 & 0 & 0 & -1 & 0 & i & 0 & 0 \\
 -1 & 0 & 0 & 0 & 0 & 0 & -i & 0 \\
 0 & -1 & 0 & 0 & 0 & 0 & 0 & i
\end{array}
\right)\, .
\ee
This has four zero eigenvalues, representing four zero
modes. Eq.\refb{ekf3} now
shows that the net contribution
to $K(0;s)$ from these zero modes is given by
$-1/2\pi^2 a^4$. This is to be identified as the
contribution $\bar K^f(0)$ in \refb{enetlognzfer}
that must be subtracted 
from the
heat kernel. 

It remains to calculate the constant $\beta_f$ that
appears in \refb{edeltasbhfermi}. 
It was shown in \cite{1106.0080}
that the effect of fermion zero mode integration is to
add back to $K^f_0$ three times the contribution that we
subtract, \i.e.\ we have $\beta_f=3$.
For completeness we shall briefly recall the argument.
First following an argument similar to the one given below
\refb{eap3} for the gauge fields, one can show that the
path integral measure for the gravitino fields $\psi_\mu$
corresponds to $\prod_{\mu,x} d(a \psi_\mu(x))$.
To evaluate the integral we note that the fermion zero
mode deformations correspond to local supersymmetry
transformation ($\delta\psi_\mu\propto D_\mu\eps$) with
supersymmetry transformation parameters $\eps$ which
do not vanish at infinity. Now since the anti-commutator of
two supersymmetry transformations correspond to a
general coordinate transformation with parameter
$\xi^\mu =\bar \eps \gamma^\mu \eps$, and since
$\gamma^\mu\sim a^{-1}$, we conclude that
$\eps_0=a^{-1/2}\eps$
provides a parametrization of the asymptotic supergroup
in an $a$-independent manner. Writing $\delta(a\psi_\mu)
\propto a^{3/2}D_\mu\eps_0$, and using the fact that the
integration over the supergroup parameter $\eps_0$ produces
an $a$ independent result, we now see that each
fermion zero mode integration produces a factor of
$a^{-3/2}$. 
Comparing this with the definition of $\beta_f$ given below
\refb{enfzm} we get $\beta_f=3$.

Using \refb{edeltasbhfermi} we now see that the net 
logarithmic contribution to the entropy from the gravitino
fields is given by
\be \label{efermicont}
-4\pi^2 a^4 \, \ln A_H\, \left(-{589\over 720\pi^2 a^4}
-{1\over 2\pi^2 a^4} (3-1)
\right) = (1309/180) \ln A_H\, .
\ee
Adding this to the bosonic contribution given in
\refb{ekblog} we get 
a net contribution
of
\be \label{en=2fin}
{23\over 12} \ln A_H\, ,
\ee
to the black hole entropy.

\sectiono{Half BPS black holes in
$\NN=2$ supergravity coupled to matter fields}
\label{sgeneral}

We shall now consider a more general $\NN=2$ supergravity theory
containing $n_V$ vector multiplets and $n_H$ hypermultiplets.
Since at quadratic order in the expansion around the near horizon
background the fluctuations in the vector multiplet fields do not mix
with the fluctuations in the hypermultiplet fields, we
can evaluate separately the logarithmic correction to the entropy due
to the vector multiplets and the hypermultiplets. 
The action involving these fields can be found in
\cite{r5}.

Let us begin with the vector multiplet fields.
Suppose we have an $\NN=2$ 
supergravity theory coupled to 
$n_V$
vector multiplets. The coupling of the vector multiplet
fields to supergravity
will be described by the
prepotential $F(\vec X)$ which is a homogeneous function of degree
2 in $n_V+1$ complex variables  $X^0,\cdots X^{n_V}$, 
with $X^k/X^0$ having
the interpretation of the $n_V$ complex scalars in the $n_V$ 
vector multiplets. Now it has been shown in appendix \ref{ssymplectic}
that with the help of a symplectic transformation we can introduce
new special coordinates $Z^A$ ($0\le A\le n_V$)
in the vector multiplet moduli space
such that 
\begin{enumerate}
\item In the near horizon geometry $Z^k=0$ 
for $k=1,\cdots n_V$.
\item The prepotential in the new coordinate system has the form:
\be \label{enewpre}
F = -{i\over 2} \left((Z^0)^2 - \sum_{k=1}^{n_V} (Z^k)^2\right) +\cdots\, ,
\ee
where $\cdots$ denotes terms which are cubic and higher order in the
$Z^k$'s and hence do not effect the action up to quadratic order in the
fluctuations around the near horizon geometry.
\item The only non-vanishing background electromagnetic
field in the near horizon geometry is
$F^0_{mn}$ of the form:
\be \label{ef0munu}
F^0_{mn} 
= -2 i a^{-1} \ve_{mn}, \quad m,n\in AdS_2\, ,
\ee
in the gauge $Z^0=1$. Here
$a$ denotes the radii of the near horizon $AdS_2$ and $S^2$.
\end{enumerate}
With this choice of the prepotential, the relevant part of the bosonic action
can be computed using the general formul\ae\ given {\it e.g.} in
\cite{r5}. 
We work in the gauge $Z^0=1$ and 
define a set of complex scalar
fields $\phi^k$ through the equation:
\be \label{ecomsca}
Z^k = {1\over 2}\phi^k = {1\over 2} (\phi^k_R + i \phi^k_I)\, .
\ee
Up to quadratic order in the fluctuations in the near 
horizon geometry
the action given in \cite{r5} takes the form:
\ben \label{equad}
&&\int d^4 x\sqrt{\det g}\, \bigg[  R -{1\over 2} \p_\mu \phi^k_R
\p^\mu \phi^k_R - {1\over 2} \p_\mu \phi^k_I
\p^\mu \phi^k_I-{1\over 4} \left\{1 + {1\over 2} \sum_k \left(
(\phi^k_R)^2 - (\phi^k_I)^2\right)\right\}
F^{0\mu\nu} F_{0\mu\nu}\nn
&& \qquad \qquad
-{1\over 4} F^{k\mu\nu} F^k_{\mu\nu} -{1\over 2}\phi^k_R
F^{0\mu\nu} F^k_{\mu\nu} +{1\over 2}\phi^k_I \wt F^{0\mu\nu}
F^k_{\mu\nu} + \cdots\bigg]\, ,
\een
where $\cdots$ denotes terms cubic and higher order in the
fluctuations and
\be \label{edefftilde}
\wt F^{0\mu\nu} = {1\over 2} \,
i \eps^{\mu\nu\rho\sigma}F^0_{\rho\sigma},
\qquad \eps^{mn\alpha\beta}=\ve^{mn}
\ve^{\alpha\beta}\, .
\ee

Comparing \refb{ef0munu} with \refb{ernk1.5}
(or \refb{equad} with \refb{ernk1})  we see that
$F^0_{\mu\nu}$ can be identified as 
$-2 F_{\mu\nu}$ where $F_{\mu\nu}$
is the graviphoton field strength appearing in \S\ref{s1}.
The bosonic fields in the vector multiplet are the real
scalar fields
$\phi^k_{R,I}$ and the vector fields $A^k_\mu$ whose
field strengths are given by $F^k_{\mu\nu}$.
In the background \refb{ef0munu} the action involving these 
fields to quadratic order is given by:
\ben \label{qvecquad}
&& 
\int d^4 x\sqrt{\det g}\, \bigg[-{1\over 4} F^{k\mu\nu} F^k_{\mu\nu}-{1\over 2} \p_\mu \phi^k_R
\p^\mu \phi^k_R - {1\over 2} \p_\mu \phi^k_I
\p^\mu \phi^k_I
+ a^{-2} \sum_{k=1}^{n_V}
((\phi^k_R)^2 -( \phi^k_I)^2) \nn
&& \qquad \qquad \qquad \qquad + i \,
a^{-1} \phi^k_R \ve^{mn} F^k_{mn}
+ a^{-1}\, \phi^k_I \ve_{\alpha\beta} F^k_{\alpha\beta}\bigg]\, .
\een
Note the mass terms for the scalars and
the mixing between the vector and the scalar fields
appearing in the last two terms.
This has exactly the same structure as the one which appeared in the
analysis of the matter multiplet fields in $\NN=4$ supergravity in
\cite{1005.3044}. Thus we can borrow the
result of \cite{1005.3044}, which shows 
that the net contribution to the
heat kernel from these fields, after taking into account the effect
of the ghost fields, is given by $4 K^s(0;s)$ for each vector
multiplet, with $K^s$ given in \refb{e10}.
Since $\beta_v=1$ we do not need to give any special
treatment to the zero modes of the vector fields.

Let us now turn to the contribution from the fermions in the vector
multiplet. Each vector multiplet contains two Majorana fermions
or equivalently one Dirac fermion. It can be shown using the results
of \cite{r5} that for quadratic prepotential of the
type we have, the kinetic operator 
of the vector multiplet fermions is the standard Dirac operator
in the $AdS_2\times S^2$ background metric.
Thus the heat Kernel is given by $K^f(0;s)$
given in \refb{ediracf}. As a result the net
contribution to the heat Kernel from each vector multiplet field is given
by
\be \label{eeach}
4 K^s(0;s) + K^f(0;s) = {4\over 720\pi^2 a^4} 
+ {11\over 720\pi^2 a^4}
+\cdots = {1\over 48\pi^2 a^4}+\cdots \, ,
\ee
where as usual $\cdots$ represent terms containing other
powers of $s$.
This corresponds to a contribution to the entropy of 
$-{1\over 12} \ln A_H$ per
vector multiplet.

Let us now turn to the hypermultiplet fields consisting
of four real scalars and a pair of Weyl fermions.
The four scalars are minimally coupled to the
background gravitational field without any
coupling to the graviphoton flux, and 
give a contribution of $4 K^s(0;s)$. 
Each hypermultiplet contains a pair of Weyl fermions
$\zeta_a$ ($a=1,2$)
whose action in the Lorentzian theory, 
to quadratic order, is
given by\cite{r5}
\be \label{ehy1}
-{1\over 2} \bar\zeta^a \not \hskip-4pt D \zeta_a
+{1\over 4} \, \bar\zeta^a \ve_{ab} \Sigma_{\mu\nu}
F^{0\mu\nu} \, \zeta^b + \hbox{h.c.}\, ,
\ee
where $\ve=\pmatrix{0 & 1\cr -1 & 0}$,
$\Sigma^{\mu\nu} = {1\over 4} 
[\gamma^\mu,\gamma^\nu]$, and $\zeta_a$ and $\bar\zeta^a$
are related as
\be \label{ehy2}
\bar\zeta^a = (\zeta_a)^\dagger \, \gamma^0
= (\zeta^a)^T \wt C\, ,
\ee
$\wt C$ being the charge conjugation operator.
In writing down \refb{ehy1} we have already used the
fact that for the background we are considering 
$F^0_{\mu\nu}$ is the only non-vanishing field strength.
\refb{ehy2} can be taken as the definition of $\zeta^a$
in terms of $\zeta_a$.
Since $\zeta^a$ defined via \refb{ehy2} has
opposite chirality of $\zeta_a$, we can define a
Majorana spinor $\chi^a$ via
\be \label{ehy3}
\chi^a = \zeta_a + \zeta^a\, ,
\ee
and express the action as
\be \label{ehy3a}
-{1\over 2} \bar\chi^a \not \hskip-4pt D \chi^a
+{1\over 4} \, \bar\chi^a \ve_{ab} \Sigma_{\mu\nu}
F^{0\mu\nu} \, \chi^b \, .
\ee
This can now be continued to Euclidean space with
$\bar \chi^a \equiv (\chi^a)^T \wt C$.
Using the explicit form of the $\gamma$ matrices given in
\refb{egam1} and the background value of
$F^0_{\mu\nu}$ given in \refb{ef0munu} we get
\be \label{ehy4}
-{1\over 2} \bar\chi^a \not \hskip-4pt D \chi^a
-{1\over 2} \, a^{-1} \bar\chi^a \ve_{ab} \tau_3 \chi^b \, .
\ee
Thus the kinetic operator is given by
\be \label{ekinhy}
\delta_{ab}\, \not \hskip-4pt D + a^{-1} \ve_{ab} \tau_3 
= \DD_1+\DD_2,\quad
\DD_1\equiv 
\delta_{ab}\,\not \hskip-4pt D_{S^2} 
+ a^{-1} \ve_{ab} \tau_3, \quad
\DD_2\equiv \delta_{ab}\, \sigma_3 \, 
\not \hskip-4pt D_{AdS_2}\, .
\ee
Since $\DD_1$ and $\DD_2$ anti-commute
we have $(\DD_1+\DD_2)^2 = (\DD_1)^2 +
(\DD_2)^2$. The eigenvalues of
$\DD_2^2$ are given by $-\lambda^2/a^2$. On the other
hand since $\not \hskip-4pt D_{S^2}$ has eigenvalues
$\pm i(l+1)a^{-1}$, and $- a^{-1} \ve_{ab} \tau_3$ has eigenvalues
$\pm i\, a^{-1}$, and these operators act on different
spaces, the eigenvalues of $\DD_1$ are given by
$\pm i(l+1\pm 1)a^{-1}$. Thus 
$(\DD_1)^2+(\DD_2)^2$ has eigenvalues
$-(l+1\pm 1)^2/a^2 - \lambda^2/a^2$ and 
the net contribution to the
heat kernel from the two Majorana fermions 
in the hypermultiplet is
given by
\be \label{ehy5}
-{1\over 2\pi^2 a^4} 
\int_0^\infty d\lambda e^{-\bar s\lambda^2}
\, \lambda \, \coth(\pi\lambda)
 \sum_{l=0}^\infty
(l+1)
\, \left[e^{-\bar s\left(l+2\right)^2}
+ e^{-\bar s\, l^2}\right]\, .
\ee
We can evaluate this in two different ways -- either by
shifting $l\to l\mp 1$ in the two terms as in
\cite{1005.3044}, or by directly expressing this as a
double integral and using eqs.\refb{esa1a}, \refb{esa2a}.
We shall follow the second approach and express
\refb{ehy5} as
\be \label{ehy5a}
-{1\over 2\pi^2 a^4} {\rm Im} \int_0^{e^{i\kappa}\times \infty} 
d\wt\lambda \, \wt\lambda\, \cot(\pi\wt\lambda)\,
\int_0^\infty d\lambda \, \lambda \, \coth(\pi\lambda)
e^{-\bar s\wt \lambda^2-\bar s
\wt\lambda^2}
 \, \left[e^{-2\bar s \wt\lambda -\bar s}
+ e^{2\bar s \wt\lambda -\bar s}\right]\, .
\ee
The terms in the square bracket can now by expanded in
a power series in $\bar s$ and we can evaluate the integrals
using \refb{esa1a}, \refb{esa2a}. The resulting constant term
in the small $\bar s$ expansion of the expression is given
by $-19/720\pi^2 a^4$.
Combining this with the contribution $4/720\pi^2 a^4$
from the bosonic contribution $4K^s(0;s)$,
we get
\be \label{ehypereach}
K^{hyper}(0;s)  = -{1\over 48\pi^2 a^4}+\cdots \, .
\ee
This corresponds to a contribution of 
${1\over 12} \ln A_H$ per
hypermultiplet. Combining \refb{en=2fin} with the results of this
section we see that an $\NN=2$ supergravity 
theory with $n_V$ vector multiplets
and $n_H$ hypermultiplets will have a logarithmic correction
to the entropy given by
\be \label{ehypvec}
{1\over 12} (23 + n_H - n_V) \, \ln \, A_H\, .
\ee

\sectiono{Local method, duality anomaly and
ensemble choice} \label{slocal}

In this section we shall discuss an alternative derivation of the
results for $\NN=2$ supergravity using local methods. Indeed, with
hindsight we could have read out these results from those in
\cite{duffroc} which computed the trace anomalies due to
various fields in gauged supergravity theories.
For this we begin with the generalized version of
\refb{ebirrel} including the effect of $n_{3/2}$
Majorana spin 3/2 field and $n_2$ spin 2 fields.
Then \refb{ebirrel} takes the 
form\cite{christ-duff1,christ-duff2,duffnieu,birrel,gilkey,
0306138}
(for a recenet review see 
\cite{1104.3712})\footnote{For metric and spin 3/2 fields the
individual coefficients multiplying $E$ and $I$ are gauge
dependent\cite{christ-duff1} but the
coefficient of $R_{\mu\nu\rho\sigma} R^{\mu\nu\rho\sigma}$
is gauge independent. As we shall see, this will be the
only relevant coefficient that enters our analysis.}
\be \label{ebirrela}
K_0 = -{1\over 90\pi^2} (n_S + 62 n_V + 11 n_F) E
- {1\over 30\pi^2} (n_S + 12 n_V + 6 n_F - {233\over 6}
n_{3/2} + {424\over 3} n_2) I\, ,
\ee
\ben \label{edefiea}
E &=& {1\over 64} \left(R_{\mu\nu\rho\sigma} 
R^{\mu\nu\rho\sigma} - 4 R_{\mu\nu} R^{\mu\nu}
+ R^2\right)\, , \nn
I &=&  -{1\over 64} \left(R_{\mu\nu\rho\sigma} 
R^{\mu\nu\rho\sigma} - 2 R_{\mu\nu} R^{\mu\nu}
+ {1\over 3} R^2\right)\, .
\een
Now in the near horizon background we are interested
in, we also have background gauge fields besides
the background metric, and so we cannot apply
\refb{ebirrela} directly. But we can try to use
supersymmetry to find the supersymmetric completion
of these terms. Of these since $E$ is a topological term,
it is supersymmetric by itself and does not require the
addition of any other term. On the other hand
supersymmetrization of $I$ has been carried out
in \cite{bergroowit,9602060,9603191}. 
Although the resulting action is quite complicated, it 
is known that 
supersymmetrization
of $I$,
evaluated in the near horizon background of the black 
hole\cite{9603191,9801081,9812082,9904005,0007195,0603149}, takes
the same value as 
$-E$\cite{9711053,0508042} 
even though $I$ itself vanishes
in the near horizon geometry and $E$ does not
vanish.\footnote{This could be 
due to the fact that supersymmetrization
of $I$ and $-E$ are equivalent via a field redefinition since they have 
the
same coefficient of the 
$R_{\mu\nu\rho\sigma}
R^{\mu\nu\rho\sigma}$ term, but we shall not need this
stronger result.} 
Thus for our analysis we can replace the
supersymmetrized $I$ by $-E$ on the 
right hand side of 
\refb{ebirrela}. This gives
\be \label{ebirrelb}
K_0 = -{1\over 90\pi^2} (-2n_S + 26 n_V -7 n_F
+ {233\over 2}
n_{3/2} - {424} n_2) E
\, .
\ee
Using $E=-1/8 a^4$ for the $AdS_2\times S^2$
background, we get
\be \label{ek0local}
K_0  = {1\over 720\pi^2 a^4} (-2n_S + 26 n_V -7 n_F
+ {233\over 2}
n_{3/2} - {424} n_2) \, .
\ee
These coefficients agree with those given in \cite{duffroc}. Using
this result we can
reproduce all the results of the previous sections for $\NN\ge 2$
supergravity theories
correctly.
For example for the hypermultiplet we have
$n_S=4$, $n_F=1$ leading to $K_0
= -1/48\pi^2 a^4$ in agreement with \refb{ehypereach}.
On the other hand for vector multiplets we have
$n_V=1$, $n_S=2$ and $n_F=1$ leading to
$K_0= 1/48\pi^2 a^4$ in agreement with
\refb{eeach}.
For the $\NN=2$ supergravity multiplet we have 
$n_2=1$, $n_{3/2}=2$ and $n_V=1$ leading to
$K_0=-11/48\pi^2 a^4$. This agrees with the sum of 
\refb{ekbtot} and \refb{ekftot}.  For $\NN=4$ supergravity
multiplet we have $n_2=1$, $n_{3/2}=4$, $n_V=6$,
$n_F=2$ and $n_S=2$ leading to
$K_0=1/4\pi^2 a^4$ and for $\NN=8$ supergravity we have
$n_2=1$, $n_{3/2} = 8$, $n_V=28$, $n_F = 28$ and $n_S=70$,
leading to $K_0 = 5/4\pi^2 a^4$. 
These results agree with the corresponding
results in \cite{1106.0080}.
In each of these cases however, the effect of zero modes
needs to be accounted for separately.

Even though this analysis appears to be simpler than the one
carried out in the previous sections, it
requires us to assume that
there are no other local four derivative supersymmetric terms
that could contribute to $K_0$, or, if such terms are
present, they must vanish when evaluated
in the near horizon geometry of the black 
hole.\footnote{For a recent discussion on possible higher
derivative terms in $\NN=2$ supergravity, see
\cite{1010.2150}.}
In contrast
the analysis of the previous sections does not require any
such assumption since we compute the complete contribution
to $K_0$ in the near horizon geometry of the black hole.

\cite{duffnieu} found an ambiguity in computing the coefficient
of $E$ in the trace anomaly: if we replace a field by its dual
field -- {\it e.g.} a scalar field by a 2-form field -- the
coefficient of $E$ changes. 
A recent discussion on this in the context of black hole
entropy can be found in \cite{1009.4439}.
This has been 
understood as due to the contribution to the trace anomaly
from the zero modes\cite{duffroc,0806.3505}. 
Using this ambiguity \cite{duffroc} suggested 
replacing the scalar
field by the 2-form field since that is what appears naturally
in string theory. The resulting contribution to $K_0$ 
agrees with the result of
direct string computation in 
\cite{9203071,9204030,9212045}, and would also produce
correctly the coefficient of the log term in
\refb{elogres} without having to give special treatment
to the zero modes. 
This procedure of replacing a scalar by a 2-form field would
also reproduce correctly the zero result given in
\refb{en=48} for $\NN=4$ supersymmetric theories.
This however is a coincidence; it just
so happens that the extra term we get by first removing the
contribution from the metric and the gravitino
zero modes to the heat kernel
and then carrying out
separately the integration over these zero modes is the same
as the extra term we get in computation of the coefficient of
$E$ if we replace the scalar field by a 2-form field. A similar
replacement for type II string theory on a torus (where several
scalars need to be replaced by 2-form fields and we also need
to include the contribution from some non-dynamical 3-form
fields) will give zero coefficient of the logarithmic 
correction\cite{duffroc}
while the correct coefficient as given in \refb{en=48} 
is $-4$. In contrast the procedure we suggest gives
the correct answer matching the microscopic results
in the $\NN=4$ and 8 supersymmetric theories where
the microscopic results are known.

Also note that 
our procedure for  computing the coefficient of the
logarithmic correction 
does not suffer from
the ambiguity described in the previous paragraph,
 since we remove
the zero mode contribution from the heat kernel completely,
and then integrate separately over the zero modes of
the physical fields. Even in this case one might
have expected an ambiguity depending on which duality
frame we use, since the zero modes over which we integrate
depend on this frame.
This is however fixed by the physical problem at
hand. Let us for example consider adding to the theory a
non-dynamical 3-form field. In this case the non-zero mode
contribution to the heat kernel
vanishes, but integration over the zero modes could
produce non-zero contribution. 
To be more specific, the dimensional reduction of
the 3-form field on $S^2$ gives a gauge field on $AdS_2$
which has a set of zero modes. If we are to integrate
over these zero modes then we would get some additional
logarithmic correction
to the entropy. However in
this case the ensemble that it represents will have the charge
associated with this gauge field fixed. This will correspond
to membrane charge wrapped on $S^2$. This  is not
a physical gauge charge from the point of view of an
asymptotic observer in the
four dimensional Minkowski space-time
and hence should not be fixed in the ensemble. This is turn
shows that we should not be integrating over the zero modes
of the gauge fields sourced by this membrane charge. 
Thus we see that the physical
ensemble we want to calculate the entropy in automatically
fixes the duality frame. This in turn fixes
the relevant zero modes over which we need to
integrate.

\sectiono{Multi-centered black hole solutions} \label{smulti}

Our analysis of logarithmic corrections 
refers to single centered
black hole solutions only. However the microscopic counting
formula does not distinguish between the contributions from
single and multi-centered
contributions, -- it simply counts the total index / degeneracy
for a given total charge. Thus if we are to compare our
results with the result of microscopic counting when such
results become available, we need to either include the
contribution from multi-centered black holes or argue that
such contributions are small compared to that of single
centered black holes.

There are two types of multi-centered black hole solutions
we can consider. If the total charge carried by the black
hole is non-primitive, \i.e.\ can be written as an integral
multiple of another charge vector, then the total charge can
be distributed among multiple centers, carrying parallel
charge vectors. These solutions exist for arbitrary
values of the asymptotic values of the moduli scalar
fields.
Furthermore in this case the positions of the
centers are arbitrary, and the centers can come 
arbitrarily close
to each other producing an intermediate 
$AdS_2\times S^2$ throat associated with the near
horizon geometry of the single centered black hole
carrying the same total charge.
As we go  down the throat, it
splits into multiple $AdS_2\times S^2$ throats
each carrying a fraction of the total flux vector, and
representing the near horizon geometry of 
individual centers. This phenomenon is known as the
anti-de Sitter fragmentation\cite{9812073} via 
Brill instantons\cite{9202037}. This can however
be avoided by taking the total charge vector to be
primitive since in this case it is not possible for the total
charge vector to split into a set of parallel charge vectors.

The second class of multi-centered solutions arise from
the mechanism discussed 
in \cite{0005049,0206072,0304094,0702146}.
In this case the charges carried by the centers are not
parallel and there are certain constraints
 among the relative
distances between the centers.
The (non-)existence of these solutions
depends on the asymptotic values of the moduli scalar
fields,
and most of these solutions cease to exist if we set
the asymptotic values of the moduli fields to be equal to
their attractor values, --
the values they take in the near horizon
geometry of a single centered black hole
carrying the same total charge. Nevertheless \cite{0702146}
pointed out the existence of a class of solutions which
exist even when the asymptotic values of the scalar fields
are set equal to their attractor values. These solutions 
are
known as scaling solutions since in one corner of the
space of parameters labelling these solutions
the distances between the
centers go to zero. This leads to a phenomenon
similar to anti-de Sitter fragmentation\cite{0504221}.

The existence of these scaling solutions could cause potential
problem for comparing our macroscopic results with any
microscopic result since we need to add the contribution from
the scaling solutions to the single centered entropy before
comparing it to the microscopic results. A general
formula for computing the contribution to the index from
these solutions was given in \cite{1103.1887} generalizing
the results of \cite{0807.4556,0906.0011}. It takes 
the form
\be \label{escale1}
f(\{\vec q_{(i)}\}, \{\vec p_{(i)}\}) \prod_{i} d(\vec q_{(i)}, \vec p_{(i)})\, ,
\ee
when the charges carried by the individual centers are
not identical. Here $(\vec q_{(i)}, \vec p_{(i)})$ denote the electric
and the magnetic charge vectors carried by the
$i$th
center, $d(\vec q,\vec p)$ is the contribution to the
index from a single centered black hole carrying
charge $(\vec q, \vec p)$ 
and $f(\{\vec q_{(i)}, \vec p_{(i)}\})$ is a function of
the charges carried by all the centers,
representing the contribution to the index from
the quantum system describing 
the relative motion between the centers. When some of
the centers carry identical charges the result gets
modified\cite{1103.1887}, 
but not in a way that invalidates our discussion
below. 
The contribution from these configurations could dominate
the single centered contribution in two ways: the
number of such multi-centered configurations could be
exponentially large, giving a contribution to the entropy
that is of the same order or larger that that of the single
centered contribution to the entropy, or individual
terms could dominate over the entropy of single centered
black holes. For the special case of D6-$\bar{\rm D}$6-D0
systems the number of configurations was estimated
in \cite{0906.0011}, and although it grows exponentially with
the charge, the power of the charge in the exponent
was found to be smaller than 2. Given the rarity of scaling
solutions to be discussed shortly, we believe that this is
probably a generic features of these solutions. Furthermore
there can also be cancelations between the contributions from
different configurations if they contribute to the index with
opposite signs.
In order to estimate the contribution from the individual
terms we use the result of 
\cite{1103.1887} from which it follows 
that while the index of individual centers
could grow exponentially with the charges, the function
$f(\{\vec q_{(i)}, \vec p_{(i)}\})$ grows polynomially with the charges.
Thus in order for \refb{escale1} to dominate or be of the
same order as the contribution from the single centered black
hole, 
$\sum_i \ln |d(\vec q_{(i)}, \vec p_{(i)})|$ should either 
exceed or be of
the same order as $\ln \left|d\left(\sum_i \vec q_{(i)}, \sum_i 
\vec p_{(i)}\right)\right|$,
-- the latter representing the contribution to the entropy
from a single centered black hole with total charge
$\left(\sum_i \vec q_{(i)}, \sum_i \vec p_{(i)}\right)$.
For this reason it is important to classify all the scaling
solutions carrying a given total charge and examine if 
their contribution could dominate or be of the same 
order as the contribution from a single centered black
hole.\footnote{It has been suggested by Frederik Denef
that the sum of the classical entropies of the individual
centers could not possibly exceed that of the single centered
black hole since this will violate the holographic bound.
Although there is no direct proof of this, some special cases
have been discussed in \cite{gaas}.}

Let us now review the condition under which the scaling
solutions exist. We shall describe the solution in the
limit when all the centers come close to each other
since the (non-)existence of the solution in this
limit will imply (non-)existence of the whole family.
If we define
\be \label{escale2}
\alpha_{ij} = \vec q_{(i)} \cdot \vec p_{(j)} - 
\vec q_{(j)} \cdot \vec p_{(i)}\, ,
\ee
and $\vec x_{(i)}$ denotes the position of the $i$-th center,
then these positions are constrained by the 
requirement\cite{0702146}:
\be \label{escale3}
\sum_{j, j\ne i} {\alpha_{ij} \over |\vec x_{(i)} - \vec x_{(j)}|} = 0 \quad
\forall \, i\, .
\ee
For three centered black hole this translates to the
condition that $\alpha_{12}$, $\alpha_{23}$ and 
$\alpha_{31}$ have the same sign and satisfy the
triangle inequality so that they form three sides of
a triangle.
Another requirement comes from the regularity of the
metric. Let the entropy of a single centered
BPS black hole carrying
charge $(\vec q, \vec p)$ be denoted by $\pi \sqrt{D(\vec q,
\vec p)}$. Then the regularity condition takes the form
\be \label{escale4}
D(\vec h(\vec x), \vec g(\vec x)) > 0 \quad \forall \, \vec x,
\qquad
\vec h(\vec x) \equiv \sum_i { \vec 
q_{(i)}\over |\vec x - \vec x_{(i)}|},
\quad 
\vec g(\vec x) \equiv \sum_i {\vec 
p_{(i)}\over |\vec x - \vec x_{(i)}|}\, .
\ee
Note that while \refb{escale3} is independent of the details
of the theory {\it e.g.} the prepotential, \refb{escale4} is
sensitive to the details of the theory since the function
$D(\vec q, \vec p)$ depends on the prepotential. 
There are further requirements, {\it e.g.} the matrix
multiplying the gauge kinetic term, which is a function of
the vector multiplet scalars, must be positive definite
everywhere in space. These conditions also depend on
the prepotential.

For two centered black holes \refb{escale3} requires
$\alpha_{12}$ to vanish. In this case the function
$f(\{\vec q_{(i)}, \vec p_{(i)}\})$ turns out to be proportional
to $\alpha_{12}$ and as a result two centered scaling solutions
do not contribute to the index. However there are plenty of
solutions to \refb{escale3} involving three or more centers,
giving rise to potential contributors to the index.
The condition \refb{escale4} as well as the
requirement of a positive definite gauge kinetic term  
has been
less studied since this has to be done on a case by case
basis as it depends on the details of the theory. 
\cite{1103.1887} considered a special example of a theory
with a single vector multiplet
with prepotential $-  (X^1)^3 /6X^0$ and found that a 
3-centered 
solution to \refb{escale3}, with each center described by a
regular event horizon, fails to satisfy \refb{escale4}. This
leads us to suspect that the scaling solutions may be rare
and may not be a potential
competitor to the contribution to the index from a single
centered black hole. We shall now
describe the results for some simple systems.

First we consider pure supergravity, or more generally 
supergravity coupled to hypermultiplets but no vector
multiplets. Such theories can arise from type IIB string
theory on Calabi-Yau manifolds which do not admit
any deformation of the
complex structure. In this case 
we do not expect any non-singular multi-centered solutions
with non-parallel charges since the only forces are due
to gravity and electromagnetism, and for non-parallel
charges the gravitational force wins over the electromagnetic
force. This argument 
of course ignores the non-linear effects of gravity
and in order to have a convincing result we need to
analyze the possibility of simultaneous solutions to
\refb{escale3}, \refb{escale4}. In this case
the charge
vectors are one dimensional and
$D(q, p)\propto (q^2 + p^2)$. Thus the only way 
\refb{escale4} can fail is if the functions $h(\vec x)$ and
$g(\vec x)$ both vanish at the same point, \i.e.\ the
surfaces $f(\vec x)=0$ and $g(\vec x)=0$ intersect.
It was shown in \cite{gaas} that for 
three centered solutions these surfaces always
intersect, showing the absence of scaling 
solutions.
For larger number of centers a general proof of absence
does not exist, but none have been found so far in
numerical searches.

We have also examined the solution to \refb{escale4}
in the one vector multiplet model with prepotential 
$-(X^1)^3 /6X^0$. Here we have\cite{9612076}
\be
D(p^0,p^1,   q_1,   q_0)
={1\over 9}\, 
\left[ 3 (   q_1p^1)^2  - 
18   q_0 \, p^0 \,    q_1 \, p^1 
- 9    q_0^2 \,(p^0)^2  -  6 (p^1)^3 \,    q_0  
+ 8   p^0 \, (   q_1)^3 \
\right]\, .
\ee
In this case there are known examples of
scaling solutions satisfying 
\refb{escale4}, {\it e.g.}
the D6-$\rm\bar D6$-D0 system discussed in
\cite{0702146,0807.4556,0906.0011,1103.1887}.
These solutions by themselves have individual centers
carrying zero entropy, but by adding sufficiently small amount
of charges to each center we can ensure that the
each center has non-zero (although small) 
entropy and yet the
solution continues to satisfy the condition
\refb{escale4}.\footnote{I wish to thank Frederik Denef
for suggesting this construction.}
Nevertheless it is instructive to explore
how pervasive these solutions are.
For this
we have
randomly generated the charges carried by the three
centers and
picked among them those sets for which 
$\alpha_{12}$, $\alpha_{23}$ and $\alpha_{31}$
satisfy the triangle
inequality and the discriminant $D$ is
positive for each center as well as for the total charge
carried by all the centers. For each of these sets we then
test the positivity of 
$D(\vec f(\vec x), \vec g(\vec x))$ as a function of 
$\vec x$.
We find that in each of the 30 examples generated
this way,  $D$ fails to be
positive in some region of space.

While we do not have  any rigorous result, the
results reviewed in 
this section indicate that 
scaling solutions satisfying \refb{escale3} and
\refb{escale4} simultaneously are rare. This in turn
gives us reason
to hope that at least in some of the theories the 
contribution from the single centered black holes
dominate the index, and we can
directly compare our results for logarithmic corrections
to the microscopic results. 
It will clearly 
be useful to have a better analytic understanding
of the problem.

\sectiono{Comparison with the OSV formula} \label{sosv}

In this section we shall compare our result with various versions 
of the OSV formula\cite{0405146}. 
In a nutshell an OSV type formula is a proposal for the asymptotic
expansion of the black hole entropy in the large charge limit, 
giving the expression for the entropy as a function of the charges to
all orders in an expansion in inverse powers of charges. In particular
any such formula will give a definite predictions for the logarithmic
corrections to the entropy which are the first subleading corrections
to the Bekenstein-Hawking entropy. Thus it can be compared with
\refb{elogres}.
  
We begin with the
version of the OSV formula proposed in
\cite{0702146}. 
Although this formula was derived for a limit of the charges
different from the one we are considering, we shall go ahead
with the assumption that it is valid also in the limit in which
all the charges are scaled uniformly \i.e.\ for `weak
topological string coupling' and at the attractor point in the
moduli space where single centered black hole gives the dominant
contribution to the entropy.
If the theory has $n_V$ vector multiplets
and is described by the prepotential 
$F(X^0, \cdots X^{n_V})$, then the relevant part of the
formula for the index of a single centered black hole carrying
electric charges $\{q_I\}$ and magnetic charges $\{p^I\}$ 
is given by
\be \label{eosv1}
e^{S_{BH}}= \hbox{constant} \times
\int \prod_{I=0}^{n_V}\, d\phi^I\, e^{-\pi \phi^I q_I}\, 
|g_{top}|^{-2}\, e^{-K} \, |Z_{top}|^2\, ,
\ee
where 
\be \label{eosv2}
e^{-K} = i(\bar X^I F_I - X^I \bar F_I), \quad X^I =
\phi^I + i p^I\, ,
\ee
\be \label{eosv3}
Z_{top} = \left({g_{top}\over 2\pi}\right)^{\chi/24}
\, \exp[-i {\pi\over 2} 
F(X)
+\cdots]
\ee
and 
\be \label{eosv4}
g_{top} = {4\pi \over X^0}\, .
\ee
$\chi$ is the euler character of the Calabi-Yau 3-fold on
which type IIA string theory is compactified to produce
the $\NN=2$ supersymmetric string theory. It 
is related to $n_H$ and $n_V$ via
\be \label{eosv5}
\chi = 2(n_V - n_H + 1)\, .
\ee
The $ \left(g_{top}/ 2\pi\right)^{\chi/24}$ factor was not
present explicitly in the original OSV definition of $Z_{top}$ 
but first made its
appearance in \cite{0507014}.
$\cdots$ in \refb{eosv3} denotes additional terms
containing non-negative powers of $g_{top}$ and non-trivial
functions of $X^k/X^0$ and will
not be relevant for our analysis. Finally it must be mentioned
that the analysis of \cite{0702146} was carried out for
$p^0=0$ \i.e.\ real $g_{top}$.

Let us now consider the limit in which all the charges are
scaled by a large parameter $\Lambda$:
$(q^I, p^I)\to (\Lambda q^I, \Lambda p^I)$. Under this
rescaling $A_H \to \Lambda^2 A_H$. We now try to 
evaluate the integration over $\phi^I$ using saddle point
method. To leading order the 
relevant saddle point lies at the extremum
of 
\be \label{eosv6}
-\pi \phi^I q_I + \pi \, {\rm  Im} \, F\, ,
\ee
and sets $\phi^I$ -- the real parts of $X^I$ -- 
to be equal to the attractor values of
the electric fields given in \refb{eattract} in the $w=8$
gauge. Since $F$ is a homogeneous function of
degree 2 in the $X^I$'s and since $q^I$ and
${\rm Im}(X^I)=p^I$
scale as $\Lambda$, it follows that the saddle point values
of $\phi^I$ also scale as $\Lambda$. Furthermore 
since the second derivatives of ${\rm Im}F$
with respect to $\phi^I$ scale as $\Lambda^0$,
the
determinant from the $\phi$ integral has no $\Lambda$
dependence. Finally $e^{-K}$ scales as $\Lambda^2$
and $g_{top}$ scales as $\Lambda^{-1}$. From 
\refb{eosv1} we now see that in the large $\Lambda$
limit
\ben \label{eosv7}
e^{S_{BH}} &=& C(\vec q,\vec p)\, e^{-\pi \phi^I q_I + \pi {\rm Im} \, F}\, 
\Lambda^{\left(4 - {\chi\over 12}\right)}\nn
&=& C(\vec q, \vec p)\,
 \exp\left[-\pi \phi^I q_I + \pi {\rm Im} \, F
+ {1\over 12} (23 -n_V + n_H) \ln \Lambda^2 
\right]\, ,
\een
where $C(\vec q,\vec p)$ represents
sum of terms which scale as $\Lambda^n$
for $n\le 0$.
The $-\pi \phi^I q_I + \pi {\rm Im} \, F$ term has to be
evaluated at the saddle point and gives the classical
Bekenstein-Hawking entropy $A_H/4 G_N$. 
Since this scales as $\Lambda^2$, we can replace $\ln \Lambda^2$
by $\ln (A_H/G_N)$ at the cost of redefining the order one multiplicative 
factor $C(\vec q, \vec p)$. This 
precisely 
agrees with \refb{elogres}.

There are other proposals for modifying the OSV formula
by introducing an additional measure. For example at the order in
which we are working, the
measure used in \cite{0808.2627,0810.1233} differs from that of
\cite{0702146} by a multiplicative factor of 
$\exp\left[\left(2 -{\chi\over 24}\right)
K\right]$.
This makes the measure 
a homogeneous function of degree zero
in the $X^I$'s and predicts zero coefficient of the
logarithmic correction in contradiction to \refb{elogres}.

Given that the OSV formula has played an important role in our
search for an exact / approximate formula for the black hole entropy
in $\NN=2$ supersymmetric string theories, it will be useful to
explore in some detail the significance of possible agreement and
disagreement between different formul\ae. The 
original proposal of  OSV\cite{0405146} made use
of the observation that the Wald entropy of a
black hole in $\NN=2$ string theory, corrected by higher
derivative terms\cite{9812082,9904005},  
is given by the Legendre transform of
$\ln\, |Z_{top}|^2$ where $Z_{top}$ is the topological string partition
function. OSV then suggested that the exact index is given by the
Laplace transform of $|Z_{top}|^2$, -- this reduces to the
exponential of the Legendre transform of 
$\ln\, |Z_{top}|^2$ in the saddle point approximation. There were
however indications that this cannot be completely 
correct (see {\it e.g.} \cite{0507014,bernard,0508174,0601108}), one needs
to include additional measure factor in the integral while
performing the Laplace transform. If we are allowed to choose
the measure freely then any correction to the leading
entropy can be encoded in an appropriate factor in the measure,
at least order by order in an expansion in inverse powers of charges.
Thus in order to make OSV formula useful one must have an
{\it a priori} description of the measure. \cite{0702146} derived the
measure from an indirect microscopic analysis of the degeneracy of
D4-D2-D0 system wrapped on appropriate cycles of a Calabi-Yau
manifold.\footnote{For other attempts to derive OSV conjecture see 
\cite{0602046,0608059,0608021}.} 
Modular invariance of the partition function
allowed them to use Rademacher expansion and express the
partition function in terms of the index associated with polar states
-- states carrying special charge vectors -- and they then identified the
polar states which give dominant contribution to the entropy.
However since their analysis only keeps a subset of the terms
in the full Rademacher expansion, there are error terms. It was  found
that while the error terms are small for a certain range of charges
(in particular when the D0-brane charge is large), in general there is
no guarantee that they will be small when all the charges are
scaled uniformly. 
Indeed it will require surprising cancellations for their
formula to be valid for this range of charges. Thus while the agreement
of our eq.\refb{elogres} with \cite{0702146} indicates that such
cancellations might be present, at present 
we should treat this agreement
as accidental. It is however encouraging to note that
there have been independent 
indications
that such cancellations might take place\cite{0704.2440}.

In contrast \cite{0601108,0808.2627} started from a different perspective, 
using symplectic invariance as the basic principle.\footnote{Symplectic
invariance does not necessarily 
refer to a symmetry of the OSV formula, but
represents the fact that we could change the electric and
magnetic charges by a symplectic transformation and at the same
time change the prepotential according to the specified rules without
changing the value of the integral. In special cases when the prepotential
remains invariant under such a transformation, the transformation may be
a genuine duality symmetry of the theory.} 
OSV formula
treats electric and magnetic charges differently, and to generalize
this to a symplectic invariant form ref.\cite{0601108,0808.2627} 
had to begin with
an integral that involves double the number of integration variables.
They then recovered the OSV type integral by integrating out half
of the variables using saddle point approximation. However
symplectic invariance by itself 
does not completely 
fix the form of the original integrand -- this 
has to be fixed using the knowledge of the effective
action. 
Using the known  local terms in the one loop effective action and their
effect on the black holes entropy \cite{0808.2627} suggested a
specific measure that differs from the measure of \cite{0702146}
by a factor of 
$\exp\left[\left(2 -{\chi\over 24}\right)
K\right]$ to the order at which we are analyzing the entropy. 
However since $K$ is invariant under a symplectic transformation,
we could multiply the original integrand of \cite{0808.2627}
by a
factor of $\exp\left[-\left(2 -{\chi\over 24}\right)
K\right]$ without violating symplectic invariance. 
Then to this order the results of \cite{0808.2627} and 
\cite{0702146} would agree and will both be consistent with
\refb{elogres}. Multiplying the integrand of \cite{0808.2627} by
$\exp\left[-\left(2 -{\chi\over 24}\right)
K\right]$ corresponds to adding to the effective action a non-local
but symplectic invariant term beyond the local terms
considered in \cite{0808.2627}.

In fact the quantum entropy function formalism that we are using for
computing the entropy
is designed to precisely take into account the contribution to the
black hole entropy from both the local and the non-local terms
in the 1PI 
effective action. The effect of local terms can also be taken into account
using Wald's formula, and for these quantum entropy function will give the
same result as Wald's formula. However Wald's formula is not directly
applicable to the non-local terms in the effective action. Quantum entropy
function takes such corrections
into account by directly evaluating the path integral of
string theory in the near horizon geometry which, by virtue of the
intrinsic curvature of $AdS_2$, 
comes with an automatic infrared cut-off. This allows
us to treat the non-local terms as corrections to the local effective
Lagrangian density. This can be seen from eq.\refb{e3ab}, -- it
describes a correction to 
$\LL_{eff}$ which has logarithmic dependence on the radius of
curvature $a$ of $AdS_2$
but is otherwise infrared finite. The logarithmic dependence on $a$
shows  that these terms are non-analylic in the $a\to\infty$
\i.e.\ flat space limit.
The other ingredient of \cite{0808.2627} -- symplectic invariance -- 
is also
implicitly built in our formalism since quantum entropy function is expressed
as a functional integral over all the fields in the theory. 
Symplectic transformation
can be implemented explicitly at the level of path integral, and using this
we can formally 
transform the expression for the quantum entropy function
written in one duality frame to the expression written in another duality
frame.

Thus we conclude that while our result is in conflict with the explicit form
for the OSV integral that appears in \cite{0808.2627}, there is no
disagreement between the basic principles of \cite{0808.2627} and the
quantum entropy function formalism. The cause of the explicit disagreement
can be traced to certain non-local terms in the one loop effective
action which have been included in our
analysis but were not present in \cite{0808.2627}. On the other hand the
agreement of our result with that of \cite{0702146} seems somewhat 
accidental since the latter was derived for a different scaling limits of
charges instead of the uniform scaling limit used here, and at
a different point in the moduli space where multi-centered black holes
could give dominant contribution to the index. It will be
interesting to explore if due to some underlying miraculous cancellation
the formula given in \cite{0702146} could 
be an exact asymptotic expansion
of the index of a single centered black hole 
in the large charge limit, 
giving the result to all orders in inverse powers of $\Lambda$.
While order by order analysis is not suited for this study, localization
methods discussed in \cite{0905.2686,1012.0265} could help prove
or disprove such a claim.

\bigskip

{\bf Acknowledgement:} I would like to thank Shamik Banerjee,
Gabriel
Lopes Cardoso, 
Atish
Dabholkar, Justin David, Bernard de Wit,
Michael Duff, Joao Gomes, 
Rajesh Gopakumar, Rajesh Gupta, Dileep Jatkar,  Swapna Mahapatra,
Ipsita Mandal,  Jan Manschot, 
Sameer Murthy, Boris Pioline, Sergey Solodukhin
and especially Frederik Denef
for useful discussions. I would also like to thank Gabriel Lopes Cardoso,
Bernard de Wit and Swapna Mahapatra for their detailed comments on
an earlier version of the manuscript.
This work was
supported in part by the J. C. Bose fellowship of 
the Department of Science and Technology, India and the 
project 11-R\&D-HRI-5.02-0304.

\appendix

\sectiono{The basis functions in $AdS_2\times S^2$} \label{sbasis}

In this appendix we shall review the results on 
eigenfunctions and
eigenvalues of the Laplacian operator 
$\square\equiv g^{\mu\nu}D_\mu D_\nu$ 
on $AdS_2$ and $S^2$ for
different tensor and spinor fields following \cite{campo,camhig1,campo2,camhig2}.
First consider
the Laplacian acting on the
scalar fields.
On $S^2$ the normalized
eigenfunctions of $-\square$ are just the usual spherical harmonics
$Y_{lm}(\psi,\phi)/a$ with eigenvalues $l(l+1)/a^2$. 
On the other hand on $AdS_2$ the $\delta$-function
normalized eigenfunctions
of $-\square$ are given by\cite{camhig1}\footnote{Although often
we shall give the basis states in terms of complex functions, we
can always work with a real basis by choosing the real and
imaginary parts of the function.}
\ben \label{e5p}
f_{\lambda,\ell}(\eta,\theta)
&=& {1\over \sqrt{2\pi\, a^2}}\, {1\over 2^{|\ell|} (|\ell|)!}\, \left|
{\Gamma\left(i\lambda +{1\over 2} + |\ell|\right)\over
\Gamma(i\lambda)}\right|\, 
e^{i\ell\theta} \sinh^{|\ell|}\eta\nn
&& F\left(i\lambda +{1\over 2}+|\ell|, -i\lambda 
+{1\over 2}+|\ell|; |\ell|+{1}; -\sinh^2{\eta\over 2}\right), \nn
&& \qquad \qquad \qquad \qquad
\ell\in \ZZZ, \qquad 0<\lambda<\infty\, ,
\een
with eigenvalue $\left({1\over 4}+\lambda^2\right)/a^2$. 
Here $F$ denotes hypergeometric function.

The normalized basis of vector fields
on $S^2$ may be taken as
\be \label{ebasis}
{1\over \sqrt{\kappa_1^{(k)}}} \, \p_\alpha U_k, \qquad
{1\over \sqrt{\kappa_1^{(k)}}} \, \vareps_{\alpha\beta} 
\p^\beta U_k \, ,
\ee
where $\{U_k\}$ denote normalized eigenfunctions of the scalar
Laplacian with eigenvalue $\kappa_1^{(k)}$. The basis states
given in \refb{ebasis} have eigenvalue of $-\square$ equal to
$\kappa_1^{(k)} - a^{-2}$. Note that for $\kappa_1^{(k)}=0$,
\i.e.\ for $l=0$, $U_k$ is a constant and $\p_\alpha U_k$ vanishes.
Hence these modes do not exist for $l=0$.

Similarly 
a normalized basis of vector fields
on $AdS_2$ may be taken as
\be \label{ebasistwo}
{1\over \sqrt{\kappa_2^{(k)}}} \, \p_m W_k, \qquad
{1\over \sqrt{\kappa_2^{(k)}}} \, \vareps_{mn} \p^n W_k \, ,
\ee
where $W_k$ are the $\delta$-function
normalized eigenfunctions of the scalar
Laplacian with eigenvalue $\kappa_2^{(k)}$. The basis states
given in \refb{ebasistwo} have eigenvalues of $-\square$ equal to
$\kappa_2^{(k)} + a^{-2}$. There are also
additional square integrable modes of eigenvalue
$a^{-2}$, given 
by\cite{camhig1}
\be \label{e24p}
A = d\Phi^{(\ell)}, \qquad \Phi^{(\ell)} = {1\over \sqrt{2\pi |\ell|}}\,
\left[ {\sinh\eta \over 1+\cosh\eta}\right]^{|\ell|} e^{i\ell\theta},
\quad \ell = \pm 1, \pm 2, \pm 3, \cdots\, .
\ee
These are not included in \refb{ebasistwo} since the $\Phi^{(\ell)}$ 
given in
\refb{e24p} is not normalizable. 
$d\Phi$ given in \refb{e24p} is
self-dual or anti-self-dual depending on the sign of $\ell$.
Thus we do not get independent eigenfunctions
from $*d\Phi^{(\ell)}$. However we can also work with a real basis
in which we take $d {\rm Re}(\Phi^{(\ell)})$ and $d {\rm Im}(\Phi^{(\ell)})\propto 
*d {\rm Re}(\Phi^{(\ell)})$
as the independent basis states for $\ell>0$. The basis states
\refb{e24p} satisfy
\be \label{esumvector}
\sum_\ell g^{mn} \p_m \Phi^{(\ell)*}(x) \p_n \Phi^{(\ell)}(x)
= {1\over 2\pi a^2}\, .
\ee
We have derived this using the fact that due to homogeneity of
$AdS_2$ this sum is independent of $x$, and that at $\eta=0$
only the $\ell=\pm 1$ terms contribute to the sum. Thus the 
total number
of such discrete modes of spin 1 field on $AdS_2$ is given by
\be \label{etotalvz}
N_1 = \int_{AdS_2}\,  d^2 x \, \sqrt{g_{AdS_2}} \, 
\sum_\ell g^{mn} \p_m \Phi^{(\ell)*}(x) \p_n \Phi^{(\ell)}(x)
= {1\over 2\pi} \int_0^{\eta_0} \sinh\eta\, d\eta \, \int d\theta
= \cosh\eta_0 - 1 \, .
\ee

A similar choice of basis can be made for a symmetric rank two
tensor representing the graviton fluctuation. For example on $S^2$
we can choose a basis of these modes to be
\be \label{ebasisgrav}
{1\over \sqrt 2}\, g_{\alpha\beta} U_k, \qquad 
{1\over  \sqrt{2(\kappa_1^{(k)} - 2 a^{-2})}} \, \left[
D_\alpha\xi_\beta + D_\beta \xi_\alpha - D^\gamma 
\xi_\gamma \, g_{\alpha\beta}\right]\, ,
\ee
where $\xi_\alpha$ denotes one of the two vectors given in
\refb{ebasis}. 
The first set of states have $-\square$ eigenvalue $\kappa_1^{(k)}$
and the second set of states have $-\square$ eigenvalue 
$\kappa_1^{(k)}- 4 a^{-2}$.
Note that for $\kappa_1^{(k)}=2a^{-2}$, \i.e.\ for
$l=1$, the second set of states given in \refb{ebasisgrav}
vanishes since the corresponding $\xi_\alpha$'s label the
conformal Killing vectors of the sphere.

On $AdS_2$ the basis states for a symmetric rank two
tensor may be chosen as
\be \label{ebasgravtwo}
{1\over \sqrt 2}\, g_{mn} W_k, \qquad 
{1\over  \sqrt{2(\kappa_2^{(k)} + 2 a^{-2})}} \, \left[
D_m\wh\xi_n + D_n \wh\xi_m - D^p 
\wh\xi_p \, g_{mn}\right]\, ,
\ee
where $\wh\xi_m$ denotes one of the two vectors given in
\refb{ebasistwo},  or the vector given in
\refb{e24p}. 
The first set of states have $-\square$ eigenvalue $\kappa_2^{(k)}$
and the second set of states have $-\square$ eigenvalue 
$\kappa_2^{(k)}+ 4 a^{-2}$.
Besides these there is another set of
square integrable modes of eigenvalue $2a^{-2}$ of
$-\square$, given by\cite{camhig1}
\ben \label{ehmn}
h_{mn}&=& w^{(\ell)}_{mn}, \nn
w^{(\ell)}_{mn} dx^m dx^n &=&
{a\over  \sqrt{\pi}} \, \left[ {  |\ell| 
(\ell^2-1)\over
2} \right]^{1/2} \, {(\sinh\eta)^{|\ell|-2} \over (1 +\cosh\eta)^{|\ell|}}
\, e^{i\ell\theta}\, (d\eta^2 + 2 \, i\, \sinh\eta\, d\eta d\theta
- \sinh^2\eta \, d\theta^2) \nn
&& \quad \ell\in \ZZZ, \quad |\ell|\ge 2\, .
\een
Locally these can be regarded as deformations generated by
a diffeomorphism on $AdS_2$, 
but these diffeomorphisms 
themselves are not square integrable. 
The basis states
\refb{ehmn} satisfy
\be \label{esummetric}
\sum_\ell g^{mn} g^{pq} w^{(\ell)*}_{mp}(x) w^{(\ell)}_{nq}(x)
= {3\over 2\pi a^2}\, .
\ee
We have derived this using the fact that due to homogeneity of
$AdS_2$ this sum is independent of $x$, and that at $\eta=0$
only the $\ell=\pm 2$ terms contribute to the sum.
Thus as in \refb{etotalvz}
the total number of such discrete modes is given by
\be \label{ediscretemetric}
N_2 = {3 \cosh\eta_0} - 3 \, .
\ee

We can construct
the basis states of various fields on $AdS_2\times S^2$ by taking
the product of the basis states on $S^2$ and $AdS_2$.
For example for a scalar field the basis states will be given by
the product of $Y_{lm}(\psi,\phi)$ with the states
given in \refb{e5p}, and satisfy
\be \label{eigen1}
\square\, f_{\lambda,k}(\eta,\theta) \, Y_{lm}(\psi, \phi)
= -{1\over a^2}\, 
\left\{l(l+1) + \lambda^2 +{1\over 4} \right\}
\, f_{\lambda,k}(\eta,\theta) \, Y_{lm}(\psi, \phi)\, .
\ee
For a vector field
on $AdS_2\times S^2$ the basis
states will contain two sets. One set
will be given by the product of $Y_{lm}(\psi,\phi)$ and
\refb{ebasistwo} or \refb{e24p}. The other set will
contain the product of
the functions \refb{e5p} on $AdS_2$ and the
vector fields \refb{ebasis} on $S^2$.
The basis states for a symmetric rank two tensor field on
$AdS_2\times S^2$ can be constructed in a similar manner.

Finally we turn to the basis states for the fermion fields.
Consider a Dirac spinor
on $AdS_2\times S^2$. It decomposes
into a product of a Dirac spinor on $AdS_2$ and a Dirac spinor
on $S^2$. 
We use the following conventions for the vierbeins and the gamma
matrices
\be \label{evier1}
e^0= a\, \sinh\eta \, d\theta, \quad e^1 = a\, d\eta, \quad
e^2 = a\, \sin\psi\, d\phi, \quad e^3 = a\, d\psi\, ,
\ee
\be \label{egam1}
\gamma^0 = -\sigma_3\otimes \tau_2, \quad \gamma^1 = \sigma_3
\otimes \tau_1, \quad \gamma^2 = -\sigma_2\otimes I_2, \quad 
\gamma^3 = \sigma_1\otimes I_2\, ,
\ee
where $\sigma_i$ and $\tau_i$ are two dimensional Pauli matrices
acting on different spaces and $I_2$ is $2\times 2$ identity
matrix. In this convention 
the Dirac operator on
$AdS_2\times S^2$ can be written as
\be \label{ek1}
\not \hskip-4pt D_{AdS_2\times S^2}
= \not \hskip-4pt D_{S^2} + \sigma_3 \, 
\not \hskip-4pt D_{AdS_2}\, ,
\ee
where
\be \label{ed1}
\not \hskip -4pt D_{S^2} = a^{-1}\left[ -
\sigma^2\, {1\over \sin\psi} \p_\phi
+ \sigma^1 \, \p_\psi +{1\over 2}\, \sigma^1\, \cot\psi\right]\, ,
\ee
and
\be \label{ed1a}
\not \hskip -4pt D_{AdS_2} =
a^{-1}\left[ -\tau^2\, {1\over \sinh\eta} \p_\theta
+ \tau^1 \, \p_\eta +{1\over 2}\, \tau^1\, \coth\eta\right]\, .
\ee

The eigenstates of $\not\hskip -4pt D_{S^2}$
are given by\cite{9505009}
\ben \label{ed2}
\chi_{l,m}^{\pm} &=& {1\over \sqrt{4\pi a^2}}\,
{\sqrt{(l-m)!(l+m+1)!}\over l!}\,
e^{ i\left(m+{1\over 2}\right)\phi} 
\pmatrix{ i\, \sin^{m+1}{\psi\over 2}\cos^m {\psi\over 2}
P^{\left(m+1, m\right)}_{l-m}(\cos\psi)
\cr
\pm \sin^{m}{\psi\over 2}\cos^{m+1} {\psi\over 2}
P^{\left(m, m+1\right)}_{l-m}(\cos\psi)
}, \nn
\eta_{l,m}^{\pm} &=& {1\over \sqrt{4\pi a^2}}\,
{\sqrt{(l-m)!(l+m+1)!}\over l!}\,
e^{ -i\left(m+{1\over 2}\right)\phi} 
\pmatrix{ \sin^{m}{\psi\over 2}\cos^{m+1} {\psi\over 2}
P^{\left(m, m+1\right)}_{l-m}(\cos\psi)
\cr
\pm i \, \sin^{m+1}{\psi\over 2}\cos^m {\psi\over 2}
P^{\left(m+1, m\right)}_{l-m}(\cos\psi)}, 
\nn
&& 
\qquad l,m\in \ZZZ, \quad l\ge 0, \quad 0\le m\le l\, ,
\een
satisfying
\be \label{ed3}
\not \hskip -4pt D_{S^2} \chi_{l,m}^\pm =\pm i\, a^{-1}\,
\left(l +1\right) 
\chi_{l,m}^\pm\, , \qquad
\not \hskip -4pt D_{S^2} \eta_{l,m}^\pm =\pm i\, a^{-1}\,
\left(l +1\right) 
\eta_{l,m}^\pm\, .
\ee
Here $P^{\alpha,\beta}_n(x)$ are the Jacobi Polynomials:
\be \label{ed4}
P_n^{(\alpha,\beta)}(x) = { (-1)^n\over 2^n \, n!} (1-x)^{-\alpha}
(1+x)^{-\beta} {d^n\over dx^n} \left[ (1-x)^{\alpha+n}
(1+x)^{\beta+n}\right]\, .
\ee
$\chi^\pm_{l,m}$ and $\eta^\pm_{l,m}$ provide an
orthonormal set of basis functions, {\it e.g.}
\be \label{edefchinorm}
a^2 \int_{S^2} \left(\chi^{\pm}_{l,m}\right)^\dagger \, 
\chi^{\pm}_{l',m'} \, \sin\psi \,
d\psi\, d\phi = \delta_{ll'}\delta_{mm'}
\ee
etc.

The eigenstates of $\not\hskip -4pt D_{AdS_2}$
are given by\cite{9505009}
\ben \label{ed2a}
\chi_{k}^{\pm}(\lambda) &=& {1\over \sqrt{4\pi a^2}}\,
\left|{\Gamma\left( {1} + k + i\lambda\right)
\over \Gamma(k+1) \Gamma\left({1\over 2}+i\lambda\right)}\right|\,
e^{ i\left(k+{1\over 2}\right)\theta}  \nn
&& \qquad 
\pmatrix{ i \, {\lambda\over k+1}\, 
\cosh^{k}{\eta\over 2}\sinh^{k+1} {\eta\over 2}
F\left(k+1+i\lambda, k+1-i\lambda; k+2;-\sinh^2{\eta\over 2}\right)
\cr
\pm \cosh^{k+1}{\eta\over 2}\sinh^k {\eta\over 2}
F\left(k+1+i\lambda, k+1-i\lambda; k+1;-\sinh^2{\eta\over 2}\right)}, \nn \cr \cr
\eta_{k}^{\pm}(\lambda) &=& {1\over \sqrt{4\pi a^2}}\,
\left|{\Gamma\left( {1} + k + i\lambda\right)
\over \Gamma(k+1) \Gamma\left({1\over 2}+i\lambda\right)}\right|\,
e^{ -i\left(k+{1\over 2}\right)\theta}\nn
&&  \qquad 
\pmatrix{ \cosh^{k+1}{\eta\over 2}\sinh^k {\eta\over 2}
F\left(k+1+i\lambda, k+1-i\lambda; k+1;-\sinh^2{\eta\over 2}\right)
\cr
\pm i \, {\lambda\over k+1}\, 
\cosh^{k}{\eta\over 2}\sinh^{k+1} {\eta\over 2}
F\left(k+1+i\lambda, k+1-i\lambda; k+2;-\sinh^2{\eta\over 2}\right)
}, 
\nn \cr && 
\qquad  k\in \ZZZ, \quad 0\le k<\infty, \quad 0<\lambda<\infty\, ,
\een
satisfying
\be \label{eadsev}
\not \hskip -4pt D_{AdS_2} \chi_{k}^{\pm}(\lambda)=\pm i\, a^{-1}\,
\lambda\,  \chi_{k}^{\pm}(\lambda)
\, , \qquad
\not \hskip -4pt D_{AdS_2} \eta_{k}^{\pm}(\lambda)
=\pm i\, a^{-1}\,
\lambda\,  \eta_{k}^{\pm}(\lambda)
\, .
\ee
$\chi_k^\pm(\lambda)$ and $\eta_k^\pm(\lambda)$
provide an orthonormal set of basis functions on $AdS_2$,
{\it e.g.} 
\be \label{eadsnorm}
a^2 \int \sinh\eta\, d\eta\, d\theta\, 
(\chi_k^\pm(\lambda))^\dagger\,
\chi_{k'}^\pm(\lambda') = \delta_{kk'} \delta(\lambda-
\lambda')\, ,
\ee
etc.

The basis of spinors on $AdS_2\times S^2$ can be constructed
by taking the direct product of the spinors given in
\refb{ed2} and \refb{ed2a}. Let
$\psi_1$ denotes an eigenstate of $\not \hskip-4pt D_{S^2}$
with eigenvalue $i\zet_1=\pm ia^{-1}(l+1)$ and
$\psi_2$ denotes an eigenstate of $\not \hskip-4pt D_{AdS_2}$
with eigenvalue $i\zet_2=\pm ia^{-1}\lambda$.
Since $\sigma_3$ anti-commutes with 
$\not \hskip-4pt D_{S^2}$ and commutes with 
$\not \hskip-4pt D_{AdS_2}$, we
have, using \refb{ek1},
\ben \label{ek3}
\not \hskip-4pt D_{AdS_2\times S^2}\, \psi_1\otimes \psi_2
&=&  i\zet_1 \psi_1\otimes \psi_2 +
i\zet_2 \sigma_3 \, \psi_1\otimes \psi_2 \, , \nn
\not \hskip-4pt D_{AdS_2\times S^2} \, \sigma_3\,
\psi_1\otimes \psi_2
&=& i\zet_2  \, \psi_1\otimes \psi_2 
 -i\zet_1 \sigma_3 \, \psi_1\otimes \psi_2
\, .\nn
\een
Diagonalizing the $2\times 2$ matrix we see that
$\not \hskip-4pt D_{AdS_2\times S^2}$ has eigenvalues
$\pm i\sqrt{\zet_1^2 + \zet_2^2}$. Thus the square
of the eigenvalue of $\not \hskip-4pt D_{AdS_2\times S^2}$ is
given by the sum of squares of the eigenvalues of
$\not \hskip-4pt D_{AdS_2}$ and $\not \hskip-4pt D_{S^2}$,
and we have
\be \label{ednotsquare}
(\not \hskip-4pt D_{AdS_2\times S^2})^2
 \psi_1\otimes \psi_2 = -(\zet_1^2 + \zet_2^2)\,
  \psi_1\otimes \psi_2, \quad
(\not \hskip-4pt D_{AdS_2\times S^2})^2
\sigma_3 \psi_1\otimes \psi_2 = -(\zet_1^2 + \zet_2^2)\,
\sigma_3  \psi_1\otimes \psi_2\, .
  \ee

By introducing the `charge conjugation operator' 
\be \label{echargec}
\wt C=\sigma_2
\otimes \tau_1
\ee
and defining $\bar \psi =\psi^T \wt C$, we can express
the orthonormality relations \refb{edefchinorm}, \refb{eadsnorm} 
as
\be \label{enewnorm}
\int d^4 x \, \sqrt{\det g}\, \, \overline{\left(\chi^+_{l,m} \otimes 
\chi^+_k(\lambda)\right)} \, 
\left(\eta^+_{l',m'}\otimes \eta^-_{k'}(\lambda')\right)
= i \,
\delta_{l,l'}\delta_{m,m'}\delta_{k,k'}\delta(\lambda-\lambda')\, ,
\ee
etc. This is important since eventually we shall be dealing with fields
satisfying appropriate reality conditions for which $\bar\psi$ will be
defined as $\psi^T \wt C$.

In our analysis we shall also need to find a basis in which we can
expand the Rarita-Schwinger field $\Psi_\mu$. 
Let us denote by $\chi$ the spinor $\psi_1\otimes \psi_2$
where $\psi_1$ and $\psi_2$ are eigenstates of
$\not \hskip-4pt D_{S^2}$ and $\not \hskip-4pt D_{AdS_2}$ with
eigenvalues $i\zet_1$ and $i\zet_2$ respectively.
Then a (non-orthonormal set of) 
basis states for expanding $\Psi_\mu$ on $AdS_2\times S^2$
can
be chosen as follows:
\ben \label{ers5}
&&\Psi_\alpha =\gamma_\alpha \chi, \quad \Psi_m=0\, , \nn
&& \Psi_\alpha =0, \quad \Psi_m=\gamma_m \chi,  \nn
&&\Psi_\alpha =D_\alpha  \chi, \quad \Psi_m=0,  \nn
&& \Psi_\alpha =0, \quad \Psi_m= D_m 
\chi \, . 
\een
By including all possible eigenstates $\chi$ of 
$\not \hskip-4pt D_{S^2}$ and $\not \hskip-4pt D_{AdS_2}$
we shall generate the complete set of basis states for
expanding the Rarita-Schwinger field barring the subtleties
mentioned below.

The first subtlety arises due to the relations 
\be \label{eindep}
D_\alpha\chi^\pm_{0,0} = 
\pm {i\over 2} \, a^{-1}\, 
\gamma_\alpha\chi^\pm_{0,0}, \qquad
D_\alpha\eta^\pm_{0,0} = 
\pm {i\over 2} \, a^{-1}\, \gamma_\alpha\eta^\pm_{0,0}\, .
\ee
Thus if we take $\chi=\psi_1\otimes \psi_2$ 
where $\psi_1$ corresponds to any of the states
$\chi^\pm_{0,0}$ or $\eta^\pm_{0,0}$, and $\psi_2$ is any eigenstate of
$\not\hskip -4pt D_{AdS_2}$, then 
the basis vectors appearing in \refb{ers5} are not all independent, --
the modes in the third row of \refb{ers5} are related to those in
the first row.
The second point is that the modes given in \refb{ers5}
do not exhaust all the modes of the Rarita Schwinger
operator; there are some additional discrete modes of the form
\be \label{eadd1}
\xi_m^{(k)\pm}\equiv
\psi_1\otimes \left(D_m\pm {1\over 2a} \sigma_3
\gamma_m\right)\chi^\pm_k(i),
\qquad \wh\xi_m^{(k)\pm}\equiv
\psi_1\otimes \left(D_m\pm {1\over 2a}  \sigma_3
\gamma_m\right)
\eta^\pm_k(i), \quad k=1,\cdots \infty \, ,
\ee
where $\chi^\pm_k(\lambda)$ and $\eta^\pm_k(\lambda)$ 
have been defined
in \refb{ed2a}.
Since $\chi^\pm_k(i)$ and $\eta^\pm_k(i)$ are
not square integrable,
these states are not included in the set  given in \refb{ers5}.
However the modes described in \refb{eadd1} are square
integrable and hence they must be included among the eigenstates
of the Rarita-Schwinger operator.
These modes can be shown to satisfy the chirality projection
condition
\ben \label{eimp1}
\tau_3 \left(D_m\pm {1\over 2a}  \sigma_3
\gamma_m\right) \chi^\pm_k(i)
&=& - \left(D_m\pm {1\over 2a}  \sigma_3
\gamma_m\right) \chi^\pm_k(i), \nn 
\tau_3 \left(D_m\pm {1\over 2a}  \sigma_3
\gamma_m\right)\eta^\pm_k(i)
&=& \left(D_m\pm {1\over 2a}  \sigma_3
\gamma_m\right) \eta^\pm_k(i)\, .
\een

\sectiono{Some useful relations} \label{suse}

In this appendix we shall collect the results of some useful
integrals. Their derivation has been reviewed in
\cite{1005.3044,1106.0080}.
\ben\label{esa1}
&& \int_0^\infty d\lambda \, \lambda\, \tanh(\pi\lambda)\,
e^{-\bar s \lambda^2} \, \lambda^{2n}\cr
&=& {1\over 2} \bar s^{-1 - n} 
\Gamma(1+n) + 2 \sum_{m=0}^\infty \bar s^m 
{(2m+2n+1)!\over m!} \, (2\pi)^{-2(m+n+1)} \, (-1)^m\cr
&& \qquad \qquad \qquad \qquad \qquad \qquad
(2^{-2m -2n-1}-1)\, \zeta(2(m+n+1))\, ,
\een
\ben\label{esa2}
&& {\rm Im} \int_0^{e^{i\kappa}\times \infty} 
d\wt\lambda \, \wt\lambda\, \tan(\pi\wt\lambda)\,
e^{-\bar s \wt\lambda^2} \, \wt\lambda^{2n}\cr
&=& {1\over 2} \bar s^{-1 - n} 
\Gamma(1+n) + 2 \sum_{m=0}^\infty \bar s^m 
{(2m+2n+1)!\over m!} \, (2\pi)^{-2(m+n+1)} (-1)^{n+1}\cr
&& \qquad \qquad \qquad \qquad \qquad \qquad
(2^{-2m -2n-1}-1)\, \zeta(2(m+n+1))\, ,
\een
\ben\label{esa1a}
&& \int_0^\infty d\lambda \, \lambda\, \coth(\pi\lambda)\,
e^{-\bar s \lambda^2} \, \lambda^{2n}\cr
&=& {1\over 2} \bar s^{-1 - n} 
\Gamma(1+n) + 2 \sum_{m=0}^\infty \bar s^m 
{(2m+2n+1)!\over m!} \, (2\pi)^{-2(m+n+1)} \, (-1)^m\cr
&& \qquad \qquad \qquad \qquad \qquad \qquad
\zeta(2(m+n+1))\, ,
\een
\ben\label{esa2a}
&& {\rm Im} \int_0^{e^{i\kappa}\times \infty} 
d\wt\lambda \, \wt\lambda\, \cot(\pi\wt\lambda)\,
e^{-\bar s \wt\lambda^2} \, \wt\lambda^{2n}\cr
&=& {1\over 2} \bar s^{-1 - n} 
\Gamma(1+n) + 2 \sum_{m=0}^\infty \bar s^m 
{(2m+2n+1)!\over m!} \, (2\pi)^{-2(m+n+1)} (-1)^{n+1}\cr
&& \qquad \qquad \qquad \qquad \qquad \qquad
\zeta(2(m+n+1))\, .
\een

\sectiono{Symplectic transformation of the prepotential} \label{ssymplectic}

In general the coupling of the vector multiplet fields to
supergravity is determined by a prepotential $F(X^0,\cdots X^{n_V})$
where $F$ is a homogeneous function of degree 2 and $n_V$
is the number of vector multiplets. 
A general symplectic transformation takes the form
\be \label{egensym}
X^r \to M_{rs} X^s + N_{rs} F_s, \quad  
F_r = P_{rs} X^s
+ Q_{rs}  F_s\, , \quad 0\le r,s\le n_V\, ,
\ee
where $F_s=\p F/\p X^s$ and
$\pmatrix{M & N\cr P & Q}$ is an $Sp(2n_V+2)$ matrix
satisfying
\be \label{eun5}
M^T P - P^T M = 0, \quad N^T Q - Q^T N = 0, \quad
M^T Q - P^T N = I\, .
\ee
Our goal is to show that by a symplectic transformation we can
introduce new coordinates $Z^0,\cdots Z^{n_V}$ such that
at the attractor geometry $Z^k=0$ for $1\le k\le n_V$ and the
prepotential takes the form
\be \label{eun2}
\wh F = -{i\over 2} \left( (Z^0)^2 - \sum_{k=1}^{n_V} (Z^k)^2\right)+\cdots\, ,
\ee
where $\cdots$ denote terms which are cubic or higher
order in $Z^1,\cdots Z^{n_V}$. These higher order terms contain
information about the interactions of the theory and hence are
important in the full theory. But the quadratic terms in the fluctuations
about the black hole background are controlled by the terms up to
quadratic order in $Z^1,\cdots Z^{n_V}$, 
and hence for our analysis we can ignore the
effects of the cubic and higher order terms.

Since $Sp(2n_V+2)$ has $2(n_V+1)^2 + (n_V+1) 
= 2 n_V^2 + 5 n_V+3$ parameters, in the generic case
we can use them to introduce
new special coordinates $Y^0,Y^1,\cdots Y^{n_V}$
such that at the 
attractor value $Y^k=0$ for $k=1,\cdots n_V$. Since 
$Y^k$ are in general complex, this uses up
$2 n_V$ of the $2 n_V^2 + 5 n_V+3$ parameters.
We shall denote the new prepotential by 
$\check F$. 
If we expand $\check F$ around the point $Y^i=0$ the
expansion takes the form:
\be \label{eun1}
\check  F =  {i\over 2} A(Y^0)^2 + B_k Y^k Y^0  +
{i\over 2} C_{kl} Y^k Y^l + \cdots\, ,
\ee
for some complex 
constants $A$, $B_k$, $C_{kl}$. The $\cdots$ terms
are cubic and higher order in $Y^1,\cdots Y^{n_V}$ 
and as a result does
not affect the terms in the action quadratic in the
fluctuations. 
In order to arrive at the form \refb{eun2} we need to make
another set of symplectic transformations which
sets $A=1$, $B_k=0$ and $C_{kl}=-\delta_{kl}$.
This corresponds to $1+n_V+ n_V (n_V+1)/2$
complex constraints, \i.e.\ $n_V^2 + 3 n_V +2$ real
constraints and, in the generic case, can be achieved by
utilizing $n_V^2 + 3 n_V +2$ parameters of $Sp(2n_V+2)$.
Adding this to the $2n_V$ constraints which keep the
attractor values of $Z^k$ to be fixed at 0, we see that
we have $(n_V^2 + 5 n_V + 2)$ conditions. This is less
than the number of parameters 
$2 n_V^2 + 5 n_V+3$ of $Sp(2n_V)$ and hence
is achievable for a generic choice of the starting
prepotential. 

We shall now show how to find the
required symplectic transformation
explicitly in the case where the form of the
prepotential given in \refb{eun1} differs from the one in
\refb{eun2} by an infinitesimal amount, \i.e.\ when
\be \label{eun3}
A = -1 + \eps \tilde A, \quad B_k = \eps \, \tilde B_k, \quad
C_{kl} = \delta_{kl} + \eps\, \tilde C_{kl}\, ,
\ee
for an infinitesimal parameter $\eps$.
Now a general symplectic transformation relating the
variables $\vec Z$ and $\vec Y$ takes the form
\be \label{eun4}
Y^r = M_{rs} Z^s + N_{rs} \wh F_s, \quad  
\check F_r = P_{rs} Z^s
+ Q_{rs} \wh F_s\, , \quad 0\le r,s\le n\, ,
\ee
where $\pmatrix{M & N\cr P & Q}$ is an $Sp(2n_V+2)$ matrix
satisfying \refb{eun5}.
We choose the following infinitesimal $Sp(2n_V+2)$ matrices:
\ben \label{eun6}
&& M = I + \eps \wt M, \quad Q = I + \eps \wt Q, \quad P=
\eps \wt P, \quad
N = \eps \wt N, \quad \wt Q =-
\wt M^T, \quad \wt N = \wt N^T, \quad \wt P = \wt P^T\, , \nn
&& \wt M_{i0} = 0, \quad \wt N_{i0} = 0, \nn
&& 2 \wt M_{00} - i(\wt N_{00} + \wt P_{00}) = \wt A,
\quad \wt P_{0i} + i \wt M_{0i} = \wt B_i, \quad
- \wt M_{ij} - \wt M_{ji} - i \wt N_{ij} - i \wt P_{ij}
= \wt C_{ij}\, .
\een  
The first line ensures that the matrix $\pmatrix{M & N\cr
P & Q}$ describes an $Sp(2n_V+2)$ matrix to order
$\eps$. The second line
ensures that the attractor point $Y^i=0$ gets mapped to
$Z^i=0$ for $i=1,\cdots n_V$. Finally the last line ensures that
$\check F$ computed from \refb{eun4}, \refb{eun2} agrees with
\refb{eun1} to first order in $\eps$.

At the end of this process we are
still left with $n_V^2 + 1$ parameters of $Sp(2 n_V+2)$.
These transformations do not change the prepotential
but generate electric-magnetic duality rotation among
the Maxwell fields. 
For
example we can still make the symplectic transformation 
of the form
\be \label{esymp}
Z^0 \to \cos\alpha \, Z^0 + \sin\alpha \, F_0, \quad 
F_0\to -\sin\alpha\, Z^0 + \cos\alpha \, F_0\, ,
\ee
for some constant $\alpha$
without
changing the form of the prepotential. This induces an electric-magnetic
duality rotation among the electric and magnetic fields
$F^0_{\mu\nu}$ and $\wt F^0_{\mu\nu}$.

It is instructive to find the electric and magnetic charges 
$\{q_I, p^I\}$ and the near horizon electric field
$e^I$ ($0\le I\le n_V$) carried
by the black hole when the near horizon background is described
by $Z^k=0$ for $k=1,\cdots n_V$. For this we use the attractor
equations, derived for two derivative action in 
\cite{9508072,9602111,9602136} and for higher derivative terms in
\cite{9801081,9812082,9904005}. 
In the convention
of \cite{0603149} we have
\ben \label{eattract}
a^2 &=& {16\over w\bar w}\, , \nn
q_I &=& 4 i\left(\bar w^{-1} \overline{\wh F}_I - w^{-1} \wh 
F_I\right) \nn
p^I &=& 4 i\left(
\bar w^{-1} \bar Z^I - w^{-1} Z^I\right) \nn
 e^I &=& 4 \left( \bar w^{-1} \bar Z^I + w^{-1} Z^I\right)\, ,
 \quad 0\le I\le n_V\, ,
\een
where $a$ is the radii of $S^2$ and $AdS_2$ and $w$ is the
background value of an auxiliary anti-self-dual tensor field
$T^-_{\mu\nu}$: $T^-_{mn}=-i w\ve_{mn}$ for
$m,n\in AdS_2$. We shall choose
the gauge $Z^0=1$. Since $\wh F_k = i Z_k$ for $1\le k\le n_V$
it follows from \refb{eattract} that for $Z^k=0$, 
$p^k=q_k=e^k=0$ 
for $1\le k\le n_V$.
If we further choose $p^0=0$ with the help of the duality
rotation \refb{esymp},
then we see from \refb{eattract}
that in the $Z^0=1$ gauge $w$ must be real, and we have
\be \label{econse}
w = 4 a^{-1}, \quad q_0 = -8 w^{-1} = -2 a\, , \quad e^0 = 8 w^{-1}
= 2 a\, .
\ee
The near horizon electromagnetic fields are now given by
\ben \label{enearelec}
&& F^k_{\mu\nu} = 0 \quad \hbox{for} \quad 1\le k\le n_V,\quad
F^0_{\alpha\beta} = 0, \quad F^0_{mn} = -i e^0 \, a^{-2}\, \ve_{mn}
= -2 i a^{-1} \ve_{mn}\, , \nn
&& \qquad \qquad \qquad \qquad
\mu,\nu \in AdS_2\times S^2, \quad 
\alpha,\beta\in S^2, \quad m,n\in AdS_2\, .
\een

\end{document}